\batchmode
\makeatletter
\def\input@path{{\string"/Users/hari/Git/public/phd/papers/working/A unified Eulerian framework for multimaterial continuum mechanics/\string"}}
\makeatother
\documentclass[twoside,british,english,3p, number]{elsarticle}
\usepackage[T1]{fontenc}
\usepackage[latin9]{inputenc}
\usepackage{prettyref}
\usepackage{amsmath}
\usepackage{amssymb}
\usepackage{graphicx}
\usepackage{setspace}
\usepackage{wasysym}
\usepackage{nomencl}

\providecommand{\makenomenclature}{\makeglossary}
\makenomenclature

\makeatletter

\let\pr@chap=\pr@cha
\providecommand{\tabularnewline}{\\}

\journal{Journal of Computational Physics}

\usepackage{fancyhdr}
\usepackage{xcolor}
\usepackage{graphicx}

\usepackage{placeins}

\makeatother

\usepackage{babel}
\begin{document}

\begin{frontmatter}{}

\title{A unified Eulerian framework for multimaterial continuum mechanics}

\author[rvt]{Haran Jackson\corref{cor1}\corref{cor2}}

\ead{hj305@cam.ac.uk}

\author[rvt]{Nikos Nikiforakis}

\cortext[cor1]{Corresponding author}

\cortext[cor2]{Principal corresponding author}

\address[rvt]{Cavendish Laboratory, JJ Thomson Ave, Cambridge, UK, CB3 0HE}
\begin{abstract}
A framework for simulating the interactions between multiple different
continua is presented. Each constituent material is governed by the
same set of equations, differing only in terms of their equations
of state and strain dissipation functions. The interfaces between
any combination of fluids, solids, and vacuum are handled by a new
Riemann Ghost Fluid Method, which is agnostic to the type of material
on either side (depending only on the desired boundary conditions).

The Godunov-Peshkov-Romenski (GPR) model is used for modeling the
continua (having recently been used to solve a range of problems involving
Newtonian and non-Newtonian fluids, and elastic and elastoplastic
solids), and this study represents a novel approach for handling multimaterial
problems under this model.

The resulting framework is simple, yet capable of accurately reproducing
a wide range of different physical scenarios. It is demonstrated here
to accurately reproduce analytical results for known Riemann problems,
and to produce expected results in other cases, including some featuring
heat conduction across interfaces, and impact-induced deformation
and detonation of combustible materials. The framework thus has the
potential to streamline development of simulation software for scenarios
involving multiple materials and phases of matter, by reducing the
number of different systems of equations that require solvers, and
cutting down on the amount of theoretical work required to deal with
the interfaces between materials.
\end{abstract}

\end{frontmatter}{}

\global\long\def\dev{\operatorname{dev}}

\global\long\def\tr{\operatorname{tr}}

\section{Background}

\subsection{Multimaterial Models\label{subsec:Multiphase-Systems}}

This study concerns problems featuring immiscible materials. There
are many approaches available to solve these problems, broadly including
(but not limited to): Lagrangian and Arbitrary-Lagrangian-Eulerian
methods \citep{scardovelli_direct_1999,donea_arbitrary_1982}, volume
of fluid methods \citep{rider_reconstructing_1998,nichols_sola-vof:_1980},
diffuse interface methods \citep{saurel_multiphase_1999}, and level-set
methods (including the ghost fluid approach \citep{osher_level_2001,fedkiw_non-oscillatory_1999}).

Solids models tend to come in Lagrangian form, and often these are
combined with ALE forms for the fluid phases, so that the fluid meshes
may deform to match the deformation of the solid (see, for example,
Pin et al. \citep{Pin2007}). These schemes tend to be very accurate,
but like all Lagrangian schemes, they fail if the meshes become highly
contorted. Thus, adaptive remeshing is often necessary. Some authors
have coupled a Lagrangian solid scheme with an Eulerian fluid scheme,
but extra care must be taken when applying the boundary conditions
to the interface, which corresponds to the intersection of the Eulerian
and Lagrangian meshes (see Legay et al. \citep{Legay2006} for an
implementation using level sets, or Fedkiw \citep{Fedkiw2002} for
a GFM coupling). Some authors, such as Ryzhakov et al. \citep{Ryzhakov2010},
have found success in using the common Lagrangian formulations for
the solid, and a reformulated Lagrangian model for the fluid, implementing
the necessary adaptive remeshing. Yet another option is to model both
the fluid and the solid in an Eulerian framework, although this now
necessitates a level set method or volume of fluid method \citep{Hirt1981}
to track the interfaces. Also, these methods are more prone to losing
small-scale geometric features of the media, unless methods such as
AMR are employed to combat this \citep{Hou2012}.

Gavrilyuk, Favrie, et al have presented thermodynamically consistent
schemes where solid-fluid interfaces are modeled with the diffuse
interface method, with transverse velocities found using a ghost fluid
method \citep{favrie_solid-fluid_2009}. The fluid is governed by
the compressible Euler equations, and the solid by a conservative
hyperelastic model. This was later extended to encompass solids conforming
to the visco-plastic model of Maxwell type materials \citep{favrie_diffuse_2012},
and later still to an arbitrary number of interacting hyperelastic
solids and fluids governed by the compressible Euler equations \citep{ndanou_multi-solid_2015}.

In a recently submitted paper, Michael and Nikiforakis \citep{Michael}
(building on the work of Schoch et al. \citep{Schoch2013}) couple
various Eulerian models of reactive and inert fluids and solids by
use of a Riemann Ghost Fluid Method, with the ghost states calculated
using specialised mixed-material Riemann solvers for each interaction
(see \prettyref{sec:Ghost Fluid Methods}). Whilst these techniques
do not suffer from the mesh contortion issues inherent in Lagrangian
formulations of continuum mechanics, a fair amount of theoretical
work needs to be done to derive analytical relations describing the
interactions between every pair of models used.

If it were possible to describe all phases with the same Eulerian
model, this method could be used, with only one type of Riemann solver
needed to cope with any multiphase problem posed. The GPR model represents
such an opportunity. As will be seen, the model also includes terms
for heat conduction, which do not appear in the basic formulations
of many of the common models used in multiphase systems (e.g. the
Euler equations, or the Godunov-Romenski equations of solid mechanics).
Heat conduction is often ignored in multiphase modeling, but such
a framework based on the GPR model would almost unavoidably include
it.

At present, there is no way of dealing with material interfaces in
the GPR model, however. In this study, a modification of Barton's
\citep{Barton2010} application of Sambasivan and Udaykumar's Riemann
Ghost Fluid Method \citep{Sambasivan2009,Sambasivan2009a} is devised
for the GPR model, enabling the simulation of material interfaces.
This new method is tested on a variety of interface problems.

It is interesting to note that de Brauer et al \citep{de_brauer_cartesian_2016,brauer_cartesian_2017}
have presented a method for multimaterial modelling of a similar system
(including the distortion tensor of the GPR model, but excluding the
heat conduction terms). This method is based on level sets, similar
to the method presented in this paper. It should be noted, however,
that de Brauer et al do not apply their method to the modelling of
viscous flows.

The following two subsections outline the theory behind the GPR model
and ghost fluid methods. In Section 2 we explore the eigenstructure
of the GPR model, and use it to derive a Riemann solver for the Riemann
problem at the interfaces between different materials, which is able
to incorporate the boundary conditions that we wish to use. Section
3 presents results of fluid-fluid, solid-solid, fluid-solid, and solid-vacuum
problems, including some multidimensional cases, and some incorporating
heat conduction across the interfaces. Conclusions are drawn in section
4, along with discussion of potential limitations to the method presented
here, and ideas for further avenues of enquiry.

\subsection{The Model of Godunov, Peshkov and Romenski}

The GPR model takes the following form (see \citep{peshkov_hyperbolic_2016,dumbser_high_2015,boscheri_cell_2016,peshkov_theoretical_2019}):

\begin{subequations}

\begin{align}
\frac{\partial\rho}{\partial t}+\frac{\partial\left(\rho v_{k}\right)}{\partial x_{k}} & =0\label{eq:DensityEquation}\\
\frac{\partial\left(\rho v_{i}\right)}{\partial t}+\frac{\partial(\rho v_{i}v_{k}+p\delta_{ik}-\sigma_{ik})}{\partial x_{k}} & =0\label{eq:MomentumEquation}\\
\frac{\partial A_{ij}}{\partial t}+\frac{\partial\left(A_{ik}v_{k}\right)}{\partial x_{j}}+v_{k}\left(\frac{\partial A_{ij}}{\partial x_{k}}-\frac{\partial A_{ik}}{\partial x_{j}}\right) & =-\frac{\psi_{ij}}{\theta_{1}}\label{eq:DistortionEquation}\\
\frac{\partial\left(\rho J_{i}\right)}{\partial t}+\frac{\partial\left(\rho J_{i}v_{k}+T\delta_{ik}\right)}{\partial x_{k}} & =-\frac{\rho H_{i}}{\theta_{2}}\label{eq:ThermalEquation}\\
\frac{\partial\left(\rho E\right)}{\partial t}+\frac{\partial\left(\rho Ev_{k}+\left(p\delta_{ik}-\sigma_{ik}\right)v_{i}+q_{k}\right)}{\partial x_{k}} & =0\label{eq:EnergyEquation}
\end{align}

\end{subequations}

where $\rho$ is density, $\mathbf{v}$ is velocity, $\delta$ is
the Kronecker delta, $p$ is pressure, $\sigma$ is the sheer stress
tensor, $A$ is the distortion tensor, $\mathbf{J}$ is the thermal
impulse vector, $T$ is temperature, $E$ is total energy, and $\mathbf{q}$
is heat flux. $\psi=\frac{\partial E}{\partial A}$ and $\boldsymbol{H}=\frac{\partial E}{\partial\boldsymbol{J}}$,
and $\theta_{1}$ and $\theta_{2}$ are positive functions (given
below for the problems at hand). Additionally, we have:

\begin{equation}
\begin{cases}
p=\rho^{2}\left.\frac{\partial E}{\partial\rho}\right|_{s,A} & \sigma=-\rho A^{T}\left.\frac{\partial E}{\partial A}\right|_{\rho,s}\\
T=\left.\frac{\partial E}{\partial s}\right|_{\rho,A} & \boldsymbol{q}=T\frac{\partial E}{\partial\boldsymbol{J}}
\end{cases}\label{eq:compoundVars}
\end{equation}

where $s$ is the entropy of the system.

The GPR model represents the same set of equations as the model of
elastoplastic deformation originally proposed by Godunov and Romenski.
Peshkov and Romenski first subsequently proposed that these are the
equations of motion for an arbitrary continuum - not just a solid.
In doing so, they were able to apply the model to fluids too. Note
that material elements have not only finite size, but also internal
structure (encoded in the distortion), unlike in previous continuum
models.

The idea of $\tau_{1}$ - the strain dissipation time - has its roots
in Frenkel's ``particle settled life time'' (see \citep{frenkel_kinetic_1947}).
$\tau_{1}$ represents a continuous analogue of Frenkel's object.
It can be thought of as the characteristic time taken for a particle
to move by a distance roughly the same as the particle's size. Thus,
the typical time taken for a material element to rearrange with its
neighbors is characterized by $\tau_{1}$ . As long as a continuum
description is appropriate for the material at hand, it is thus that
the GPR model seeks to describe all three major phases of matter.
For example, we have the following relations:

\begin{equation}
\tau_{1}=\begin{cases}
\infty & elastic\,solids\\
0 & inviscid\,fluids
\end{cases}
\end{equation}

The equation governing $\boldsymbol{J}$ - and its contribution to
the system's total energy - are derived from Romenski's model of hyperbolic
heat transfer, (see \citep{malyshev_hyperbolic_1986,romenski_hyperbolic_1989}).
These concepts were later implemented in \citep{romenski_conservative_2007,romenski_conservative_2010}.
The entropy flux is the derivative of the specific internal energy
with respect to $\boldsymbol{J}$, and it is in this way that $\mathbf{J}$
is defined (as the variable conjugate to the entropy flux). As remarked
by Romenski, it is more convenient to evolve $\boldsymbol{J}$ and
$E$ than \textbf{$\mathbf{q}$} or the entropy flux, and thus the
equations take the form given here. Similarly to $\tau_{1}$, $\tau_{2}$
is a relaxation time, characterizing the average speed of relaxation
of thermal impulse due to heat exchange between neighboring particles.

$E$ must be specified to close the system. The energy contains contributions
from the micro-, meso-, and macro-scale:

\begin{equation}
E=E_{1}\left(\rho,s\right)+E_{2}\left(\rho,s,A,\boldsymbol{J}\right)+E_{3}\left(\boldsymbol{v}\right)\label{eq:EnergyDefinition}
\end{equation}

$E_{3}$ is the usual specific kinetic energy per unit mass:

\begin{equation}
E_{3}=\frac{1}{2}\left\Vert \boldsymbol{v}\right\Vert ^{2}
\end{equation}

$E_{2}$ takes the quadratic form:

\begin{equation}
E_{2}=\frac{c_{s}\left(\rho,s\right)^{2}}{4}\left\Vert \dev\left(G\right)\right\Vert _{F}^{2}+\frac{c_{t}\left(\rho,s\right)^{2}}{2}\left\Vert \boldsymbol{J}\right\Vert ^{2}
\end{equation}

where 

\begin{equation}
\dev\left(G\right)=G-\frac{1}{3}\tr\left(G\right)I
\end{equation}

$\left\Vert \cdot\right\Vert _{F}$ is the Frobenius norm\footnote{The Frobenius norm is defined by: $\left\Vert X\right\Vert _{F}=\sqrt{\sum_{i,j}\left|X_{ij}\right|^{2}}$}
and $G=A^{T}A$ is the Gramian matrix of the distortion\footnote{$G$ is known as the \textit{Finger tensor} in the solid mechanics
community}, and $\dev\left(G\right)$ is the deviator (trace-free part) of $G$.
$c_{s}$ is the characteristic velocity of transverse perturbations.
$c_{t}$ is related to the characteristic velocity of propagation
of heat waves\footnote{Note that \citep{dumbser_high_2015} denotes this variable by $\alpha$,
which is avoided here due to a clash with a parameter of one of the
equations of state used.}:

\begin{equation}
c_{h}=\frac{c_{t}}{\rho}\sqrt{\frac{T}{c_{v}}}
\end{equation}

In previous studies (e.g. \citep{dumbser_high_2015,boscheri_cell_2016}),
$c_{t}$ has been taken to be constant, as it will be in this study.

In this study, $E_{1}$ is taken to be one of the following forms:

1. The stiffened gas EOS:

\begin{equation}
E_{1}=\frac{p+\gamma p_{\infty}}{\rho\left(\gamma-1\right)}
\end{equation}

2. The shock Mie-Gruneisen EOS:

\begin{equation}
E_{1}=\frac{p_{ref}}{2}\left(\frac{1}{\rho_{0}}-\frac{1}{\rho}\right)+\frac{p-p_{ref}}{\Gamma_{0}\rho_{0}}
\end{equation}

where

\begin{equation}
p_{ref}=\frac{c_{0}^{2}\left(\frac{1}{\rho_{0}}-\frac{1}{\rho}\right)}{\left(\frac{1}{\rho_{0}}-s\left(\frac{1}{\rho_{0}}-\frac{1}{\rho}\right)\right)^{2}}
\end{equation}

3. The Godunov-Romenski EOS (see \citep{barton_exact_2009}):

\begin{equation}
E_{1}=\frac{p}{\rho\gamma}+\frac{c_{0}^{2}}{2\alpha^{2}}\left(\left(\frac{\rho}{\rho_{0}}\right)^{\alpha}-1\right)\left(\left(1-\frac{2\alpha}{\gamma}\right)\left(\frac{\rho}{\rho_{0}}\right)^{\alpha}-1\right)
\end{equation}

with

\begin{equation}
c_{s}=b_{0}\left(\frac{\rho}{\rho_{0}}\right)^{\beta}
\end{equation}

4. The JWL EOS:

\begin{equation}
E_{1}=\left(\frac{A}{\rho_{0}R_{1}}e^{-\frac{R_{1}\rho_{0}}{\rho}}+\frac{B}{\rho_{0}R_{2}}e^{-\frac{R_{2}\rho_{0}}{\rho}}\right)+\frac{p-\left(Ae^{-\frac{R_{1}\rho_{0}}{\rho}}+Be^{-\frac{R_{2}\rho_{0}}{\rho}}\right)}{\rho\Gamma_{0}}
\end{equation}

Tabulated equations of state are common-place in the field (see \citep{buyukcizmeci_tabulated_2014,levashov_tabular_2007,peterson_global_2012}
for a range of use cases). There is no a priori reason why they cannot
be used for $E_{1}$ under the framework presented in this study,
in the same manner as other hydrodynamic systems (e.g. see \citep{boettger_tabular_2012,hempert_simulation_2017}).
This is out of scope of this paper, however, and presents an avenue
of future research. Note that $E_{1},c_{s},c_{t}$ are permitted to
depend upon $\rho,p$ instead of $\rho,s$ (as is the case in this
study), or indeed $\rho,T$ if the material requires it (such as materials
whose shear modulus depend on temperature).

The following forms are taken:

\begin{subequations}

\begin{align}
\theta_{1} & =\frac{\tau_{1}c_{s}^{2}}{3\left|A\right|^{\frac{5}{3}}}\qquad\tau_{1}=\begin{cases}
\frac{6\mu}{\rho_{0}c_{s}^{2}} & Newtonian\thinspace fluids\\
\frac{6\mu^{\frac{1}{n}}}{\rho_{0}c_{s}^{2}}\left|\frac{1}{\boldsymbol{\sigma}}\right|^{\frac{1-n}{n}} & power\thinspace law\thinspace fluids\\
\tau_{0}\left(\frac{\sigma_{Y}}{\left\Vert \dev\left(\sigma\right)\right\Vert _{F}}\right)^{n} & elastoplastic\thinspace solids
\end{cases}\\
\theta_{2} & =\tau_{2}c_{t}^{2}\frac{\rho T_{0}}{\rho_{0}T}\qquad\tau_{2}=\frac{\rho_{0}\kappa}{T_{0}c_{t}^{2}}
\end{align}

\end{subequations}

The justification of these choices is that classical Navier\textendash Stokes\textendash Fourier
theory is recovered in the stiff limit $\tau_{1},\tau_{2}\rightarrow0$
(see \citep{dumbser_high_2015}). The rules for power-law fluids and
elastoplastic solids are based on material from \citep{jackson_numerical_2019}
and \citep{barton_eulerian_2011}, respectively.

Finally, it is straightforward to verify that as a consequence of
\eqref{eq:compoundVars} we have the following relations:

\begin{subequations}

\begin{align}
\sigma & =-\rho c_{s}^{2}G\dev\left(G\right)\\
\boldsymbol{q} & =c_{t}^{2}T\boldsymbol{J}
\end{align}

\end{subequations}

and

\begin{subequations}

\begin{align}
-\frac{\psi}{\theta_{1}(\tau_{1})} & =-\frac{3}{\tau_{1}}\left(\det A\right)^{\frac{5}{3}}A\dev\left(G\right)\\
-\frac{\rho\boldsymbol{H}}{\theta_{2}\left(\tau_{2}\right)} & =-\frac{T\rho_{0}}{T_{0}\tau_{2}}\boldsymbol{J}
\end{align}

\end{subequations}

Finally, the following constraint holds (see \citep{peshkov_hyperbolic_2016}):

\begin{equation}
\det\left(A\right)=\frac{\rho}{\rho_{0}}\label{eq:detA constraint}
\end{equation}

See \citep{jackson_numerical_2019} for an interpretation of the physical
meaning of the relaxation times $\tau_{1},\tau_{2}$ and thermal impulse
vector $\boldsymbol{J}$.

\subsection{Ghost Fluid Methods\label{sec:Ghost Fluid Methods}}

Ghost fluid methods, combined with level set methods, are used to
model the evolution of interfaces between different materials. They
are detailed here, as it is with such a method that this study proposes
to model the interfaces between different materials described by the
GPR model.

\subsubsection{Level Set Methods}

Given a scalar function $f$ on $\mathbb{R}^{n}$, the level set of
$f$ at level $c$ is defined as:

\begin{equation}
\Gamma_{c}=\left\{ x:f\left(x\right)=c\right\} 
\end{equation}

Given normal direction speed $v$, $f$ is advected according to the
level set equation \citep{Osher2002}:

\begin{equation}
\frac{\partial f}{\partial t}=v\left|\nabla f\right|\label{eq:LevelSetEqn}
\end{equation}

The advection of a point in a fluid with velocity $v$ can be modeled
by taking $f=\left|\mathbf{x-x_{0}}\right|$ where $\mathbf{x_{0}}$
is the position of the point at time $t=0$, and tracking $\Gamma_{0}$.
\eqref{eq:LevelSetEqn} is solved by an appropriate numerical method.
The numerical methods used in this study are described in \prettyref{chap:A Riemann Ghost Fluid Method for the GPR Model}.
$f$ will usually have to be renormalized at every time step, to avoid
unwanted distortions such as becoming a multivalued function.

\subsubsection{The Original Ghost Fluid Method}

\begin{figure}
\begin{centering}
\includegraphics[height=0.25\textheight]{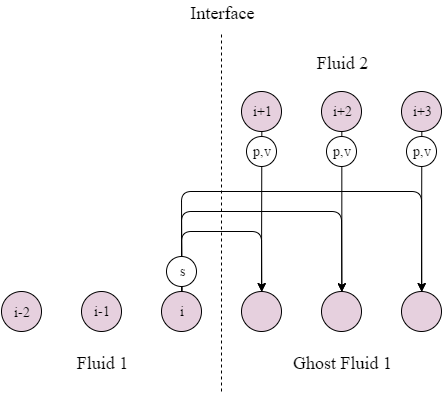}
\par\end{centering}
\caption{\label{fig:OriginalGFMNoFix}The Original Ghost Fluid Method}
\medskip{}
\end{figure}

The Original Ghost Fluid Method of Fedkiw et al. \citep{Fedkiw1999}
(an adaptation of the work of Glimm et al. \citep{Glimm1981}) is
a numerical method for the Euler equations for simulating interfaces
between multiple materials. The primitive variables for the Euler
equations in 1D are given by $\boldsymbol{P}=\left(\begin{array}{ccc}
\rho & v & p\end{array}\right)^{T}$.

Suppose the interface between two fluids is modeled on spatial domain
$\left[0,1\right]$, divided into $N$ cells with width $\Delta x=\frac{1}{N}$.
Let the time step be $\Delta t$ and let $\boldsymbol{P_{i}^{n}}$
be the set of primitive variables in cell $i$ at time $t_{n}=n\Delta t$.
Let the level set function $f$ have root $x_{n}$ where $x_{n}\in\left[\text{\ensuremath{\left(i+\frac{1}{2}\right)}}\Delta x,\text{\ensuremath{\left(i+\frac{3}{2}\right)}}\Delta x\right]$.
Thus, at time $t_{n}$ the interface lies between the cells with primitive
variables $\boldsymbol{P_{i}^{n},P_{i+1}^{n}}$. Define two sets of
primitive variables:

\begin{equation}
\boldsymbol{P_{j}^{\left(1\right)}}=\begin{cases}
\boldsymbol{P_{j}^{n}} & j\leq i\\
\left(\begin{array}{ccc}
\rho\left(s_{i}^{n},p_{j}^{n},\gamma_{i}^{n}\right) & v_{j}^{n} & p_{j}^{n}\end{array}\right) & j>i
\end{cases}
\end{equation}

\begin{equation}
\boldsymbol{P_{j}^{\left(2\right)}}=\begin{cases}
\boldsymbol{P_{j}^{n}} & j\geq i+1\\
\left(\begin{array}{ccc}
\rho\left(s_{i+1}^{n},p_{j}^{n},\gamma_{i+1}^{n}\right) & v_{j}^{n} & p_{j}^{n}\end{array}\right) & j<i+1
\end{cases}
\end{equation}

where:

\begin{equation}
\rho\left(s,p,\gamma\right)=\left(\frac{p}{s}\right)^{\frac{1}{\gamma}}
\end{equation}

All cells in $\boldsymbol{P^{\left(1\right)}}$ to the left of the
interface have the same state variables as those of $\boldsymbol{P^{n}}$.
All cells to the right have the same pressure and velocity as their
counterparts in $\boldsymbol{P^{n}}$, but the same entropy as $\boldsymbol{P_{i}^{n}}$.
This affects their density. The situation is analogous for $\boldsymbol{P^{\left(2\right)}}$.
This is demonstrated in \prettyref{fig:OriginalGFMNoFix}.

$\boldsymbol{P^{\left(1\right)}},\boldsymbol{P^{\left(2\right)}}$
are stepped forward by time step $\Delta t$ using a standard Eulerian
method. $f$ is advected using \eqref{eq:LevelSetEqn}, taking the
velocity in each cell to be that of $\boldsymbol{P^{n}}$. Now let
$f\left(x_{n+1}\right)=0$ where $x_{n+1}\in\left[\left(k+\frac{1}{2}\right)\Delta x,\left(k+\frac{3}{2}\right)\Delta x\right]$
for some $k$. Define:

\begin{equation}
\boldsymbol{P_{j}^{n+1}}=\begin{cases}
\boldsymbol{P_{j}^{\left(1\right)}} & j\leq k\\
\boldsymbol{P_{j}^{\left(2\right)}} & j>k
\end{cases}
\end{equation}

The rationale behind the original GFM is that in most applications,
pressure and velocity are continuous across the interface, and thus
the ghost cells may take the real pressure and velocity values. Entropy
is generally discontinuous at a contact discontinuity, resulting in
large truncation errors if a standard finite difference scheme is
used to solve the system. Thus, entropy is extrapolated as a constant
from the interface boundary cell into the ghost region.

\begin{figure}
\begin{centering}
\includegraphics[height=0.25\textheight]{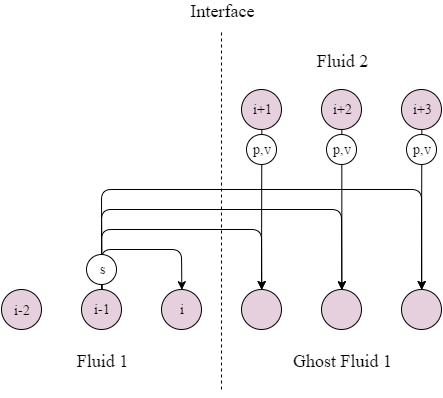}
\par\end{centering}
\caption{\label{fig:OriginalGFMIsobaricFix}The Original Ghost Fluid Method,
with the isobaric fix}
\medskip{}
\end{figure}

Fedkiw et al. advised to use the \textit{isobaric fix} technique.
This involves setting the entropy of cell $i$, and all cells in the
right ghost region, to that of cell $i-1$, and setting the entropy
of cell $i+1$, and all cells in the left ghost region, to that of
cell $i+2$. This is demonstrated in \prettyref{fig:OriginalGFMIsobaricFix}.

Effectively, the ghost regions behave like they are composed of the
same fluid as the regions they extend (as they have the same entropy),
facilitating calculation of the next time step, but they have the
same pressure and velocity profiles as the real fluids they replace,
meaning the boundary conditions at the interface are upheld.

\begin{figure}
\begin{centering}
\includegraphics[height=0.2\textheight]{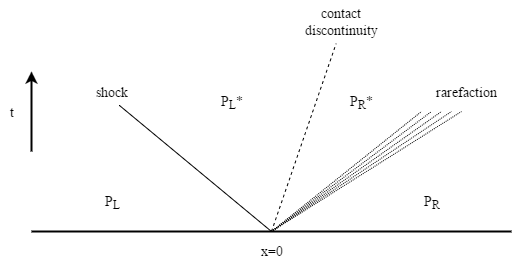}
\par\end{centering}
\caption{\label{fig:RiemannProblem}The qualitative structure of the solution
to the Riemann Problem, showing the different possible types of waves}
\medskip{}
\end{figure}

\subsubsection{The Riemann Ghost Fluid Method}

The Riemann Problem in its general form is the solution of the following
initial value problem. Given a set of variables $\boldsymbol{P}$
dependent on space and time, and a hyperbolic set of equations which
govern their spatio-temporal evolution, $\boldsymbol{P}\left(x,t\right)$
is sought for $t>0$, given the initial condition:

\begin{equation}
\boldsymbol{P}\left(x,0\right)=\begin{cases}
\boldsymbol{P_{L}} & x<0\\
\boldsymbol{P_{R}} & x>0
\end{cases}
\end{equation}

This problem is denoted by $RP\left(\boldsymbol{P_{L}},\boldsymbol{P_{R}}\right)$.
Exact solvers exist for the Riemann Problem for various sets of governing
equations, such as the Euler equations \citep{Toro2009}, the equations
of non-linear elasticity \citep{Barton2009}, or the shallow water
equations \citep{Alcrudo2001}, among others. There also exist approximate
solvers for general conservative \citep{Miller2004,Liu1975} or non-conservative
\citep{Dumbser2016} hyperbolic systems of PDEs. The references given
here form a very small sample of the work that has been done in this
area.

The solution of the Riemann Problem usually takes the form of a set
of waves, between which $\boldsymbol{P}$ is constant. The waves can
either be a contact discontinuity (across which pressure and velocity
are continuous), a shock (across which all variables may be discontinuous),
or a rarefaction (along which the variables vary continuously between
their values on either side of the wave). The number and form of the
waves are determined by the governing equations and the initial conditions.
The states of the variables either side of the contact discontinuity
in the middle are known as the \textit{star states}. This qualitative
description is depicted in \prettyref{fig:RiemannProblem}.

Liu et al. \citep{Liu2003} demonstrated that the original GFM fails
to resolve strong shocks at material interfaces. This is because the
method effectively solves two separate single-fluid Riemann problems.
The waves present in these Riemann problems do not necessarily correspond
to those in the real Riemann problem across the interface. The Riemann
Ghost Fluid Method of Sambasivan et al. \citep{Sambasivan2009} aims
to rectify this.

\begin{figure}
\begin{centering}
\includegraphics[height=0.25\textheight]{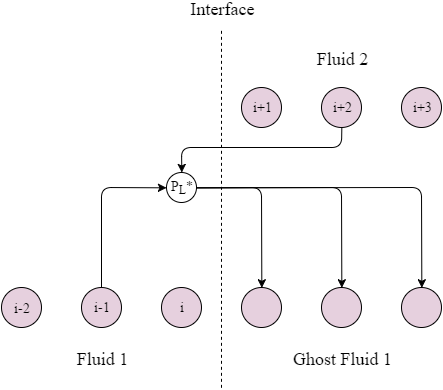}
\par\end{centering}
\caption{\label{fig:RiemannGFM}The Riemann Ghost Fluid Method}
\medskip{}
\end{figure}

Given $\boldsymbol{P^{n}}$ and $x_{n}\in\left[\left(i+\frac{1}{2}\right)\Delta x,\left(i+\frac{3}{2}\right)\Delta x\right]$,
the ghost cells for fluid 1 are populated with the left star state
of $RP\left(\boldsymbol{P_{i-1}^{n}},\boldsymbol{P_{i+2}^{n}}\right)$,
and the ghost cells for fluid 2 are populated with the right star
state. $RP\left(\boldsymbol{P_{i-1}^{n}},\boldsymbol{P_{i+2}^{n}}\right)$
is solved rather than $RP\left(\boldsymbol{P_{i}^{n}},\boldsymbol{P_{i+1}^{n}}\right)$,
as $\boldsymbol{P_{i}^{n}},\boldsymbol{P_{i+1}^{n}}$ often contain
errors generated by the fact that they lie on the material interface.
$\boldsymbol{P^{n+1}}$ is then generated as before from the newly
formed $\boldsymbol{P^{\left(1\right)}},\boldsymbol{P^{\left(2\right)}}$.
This process is demonstrated in \prettyref{fig:RiemannGFM}.

\subsection{Finite Volume Scheme}

The presented system is solved with a finite volume scheme in this
study, using a regularized square meshing. Once the values of the
ghost cells have been calculated, we require a numerical method to
calculate the value of the grid at the next time step. In this study,
the finite volume method presented in \citep{jackson_fast_2017,jackson_numerical_2019}
was used. It is outlined here for completeness.

Note that \eqref{eq:DensityEquation}, \eqref{eq:MomentumEquation},
\eqref{eq:DistortionEquation}, \eqref{eq:ThermalEquation}, \eqref{eq:EnergyEquation}
can be written in the following form:

\begin{equation}
\frac{\partial\boldsymbol{Q}}{\partial t}+\boldsymbol{\nabla}\cdot\boldsymbol{F}\left(\boldsymbol{Q}\right)+\boldsymbol{B}\left(\boldsymbol{Q}\right)\cdot\nabla\boldsymbol{Q}=\boldsymbol{S}\left(\boldsymbol{Q}\right)
\end{equation}

As described in \citep{toro_reimann_2009}, a method to solve a system
such as this is to solve the following subsystems:

\begin{subequations}

\begin{align}
\frac{\partial\boldsymbol{Q}}{\partial t}+\boldsymbol{\nabla}\cdot\boldsymbol{F}\left(\boldsymbol{Q}\right)+\boldsymbol{B}\left(\boldsymbol{Q}\right)\cdot\nabla\boldsymbol{Q} & =\boldsymbol{0}\label{eq:HomogeneousSubsystem}\\
\frac{d\boldsymbol{Q}}{dt} & =\boldsymbol{S}\left(\boldsymbol{Q}\right)\label{eq:ODESubsystem}
\end{align}

\end{subequations}

Let $H^{\delta t},S^{\delta t}$ be the operators that take data $\boldsymbol{Q}\left(x,t\right)$
to $\boldsymbol{Q}\left(x,t+\delta t\right)$ under systems \eqref{eq:HomogeneousSubsystem}
and \eqref{eq:ODESubsystem} respectively. A second-order scheme (in
time) for solving the full set of PDEs over time step $\left[0,\Delta t\right]$
is obtained by calculating $\boldsymbol{Q_{\Delta t}}$ using a Strang
splitting:

\begin{equation}
\boldsymbol{Q_{\Delta t}}=S^{\frac{\Delta t}{2}}H^{\Delta t}S^{\frac{\Delta t}{2}}\boldsymbol{Q_{0}}
\end{equation}

In this study, the homogeneous subsystem is be solved using a WENO
reconstruction of the data, followed by a finite volume update, and
the temporal ODEs will be solved with appropriate ODE solvers. The
WENO method was chosen due to the arbitrarily high-order spatial reconstructions
it is able to produce.

Noting that $\frac{d\rho}{dt}=0$ over the ODE time step, the operator
$S$ entails solving the following systems:

\begin{subequations}

\begin{align}
\frac{dA}{dt} & =\frac{-3}{\tau_{1}}\left(\det A\right)^{\frac{5}{3}}A\dev\left(G\right)\label{eq:DistortionODE}\\
\frac{d\boldsymbol{J}}{dt} & =-\frac{1}{\tau_{2}}\frac{T\rho_{0}}{T_{0}\rho}\boldsymbol{J}\label{eq:ThermalODE}
\end{align}

\end{subequations}

These systems are solved separately, and thus he second-order Strang
splitting becomes:

\begin{equation}
\boldsymbol{Q_{\Delta t}}=D^{\frac{\Delta t}{2}}T^{\frac{\Delta t}{2}}H^{\Delta t}T^{\frac{\Delta t}{2}}D^{\frac{\Delta t}{2}}\boldsymbol{Q_{0}}
\end{equation}

where $D^{\delta t},T^{\delta t}$ are the operators solving the distortion
and thermal impulse ODEs respectively, over time step $\delta t$.

The constraint \eqref{eq:detA constraint} is enforced by rescaling
the singular values of the distortion all by the same factor at each
timestep, to ensure that $\det A=\frac{\rho}{\rho_{0}}$.

\subsubsection{The Homogeneous System}

A WENO reconstruction of the cell-averaged data is performed at the
start of the time step (as described in \citep{dumbser_ader-weno_2013}).
Focusing on a single cell $C_{i}$ at time $t_{n}$, we have $\boldsymbol{w^{n}}\left(\boldsymbol{x}\right)=\boldsymbol{w^{n}}_{p}\Psi_{p}\left(\boldsymbol{\chi}\left(\boldsymbol{x}\right)\right)$
in $C_{i}$ where $\Psi_{p}$ is a tensor product of basis functions
in each of the spatial dimensions. The flux in $C$ is approximated
by $\boldsymbol{F}\left(\boldsymbol{x}\right)\approx\boldsymbol{F}\left(\boldsymbol{w}_{p}\right)\Psi_{p}\left(\boldsymbol{\chi}\left(\boldsymbol{x}\right)\right)$.
$\boldsymbol{w}_{p}$ are stepped forwards half a time step using
the update formula:

\begin{align}
\frac{\boldsymbol{w_{p}^{n+\frac{1}{2}}}-\boldsymbol{w_{p}^{n}}}{\Delta t/2}= & -\boldsymbol{F}\left(\boldsymbol{w_{k}^{n}}\right)\cdot\nabla\Psi_{k}\left(\boldsymbol{\chi_{p}}\right)\\
 & -\boldsymbol{B}\left(\boldsymbol{w_{p}^{n}}\right)\cdot\left(\boldsymbol{w_{k}^{n}}\nabla\Psi_{k}\left(\boldsymbol{\chi_{p}}\right)\right)\nonumber 
\end{align}

i.e.

\begin{equation}
\boldsymbol{w_{p}^{n+\frac{1}{2}}}=\boldsymbol{w_{p}^{n}}-\frac{\Delta t}{2\Delta x}\left(\begin{array}{c}
\boldsymbol{F}\left(\boldsymbol{w_{k}^{n}}\right)\cdot\nabla\Psi_{k}\left(\boldsymbol{\chi_{p}}\right)\\
+\boldsymbol{B}\left(\boldsymbol{w_{p}^{n}}\right)\cdot\left(\boldsymbol{w_{k}^{n}}\nabla\Psi_{k}\left(\boldsymbol{\chi_{p}}\right)\right)
\end{array}\right)\label{eq:WENO half step}
\end{equation}

where $\boldsymbol{\chi_{p}}$ is the node corresponding to $\Psi_{p}$.

Integrating \eqref{eq:HomogeneousSubsystem} over $C$ gives:

\begin{equation}
\boldsymbol{Q_{i}^{n+1}}=\boldsymbol{Q_{i}^{n}}-\Delta t_{n}\left(\boldsymbol{P_{i}^{n+\frac{1}{2}}}+\boldsymbol{D_{i}^{n+\frac{1}{2}}}\right)
\end{equation}

where

\begin{subequations}

\begin{align}
\boldsymbol{Q_{i}^{n}} & =\frac{1}{V}\int_{C}\boldsymbol{Q}\left(\boldsymbol{x},t_{n}\right)d\boldsymbol{x}\\
\boldsymbol{P_{i}^{n+\frac{1}{2}}} & =\frac{1}{V}\int_{C}\boldsymbol{B}\left(\boldsymbol{Q}\left(\boldsymbol{x},t_{n+\frac{1}{2}}\right)\right)\cdot\nabla\boldsymbol{Q}\left(\boldsymbol{x},t_{n+\frac{1}{2}}\right)d\boldsymbol{x}\\
\boldsymbol{D_{i}^{n+\frac{1}{2}}} & =\frac{1}{V}\varoint_{\partial C}\boldsymbol{\mathcal{D}}\left(\boldsymbol{Q^{-}}\left(\boldsymbol{s},t_{n+\frac{1}{2}}\right),\boldsymbol{Q^{+}}\left(\boldsymbol{s},t_{n+\frac{1}{2}}\right)\right)d\boldsymbol{s}
\end{align}

\end{subequations}

where $V$ is the volume of $C$ and $\boldsymbol{Q^{-},Q^{+}}$ are
the interior and exterior extrapolated states at the boundary of $C$,
respectively.

Note that \eqref{eq:HomogeneousSubsystem} can be rewritten as:

\begin{equation}
\frac{\partial\boldsymbol{Q}}{\partial t}+\boldsymbol{M}\left(\boldsymbol{Q}\right)\cdot\nabla\boldsymbol{Q}=\boldsymbol{0}
\end{equation}

where $\boldsymbol{M}=\frac{\partial\boldsymbol{F}}{\partial\boldsymbol{Q}}+\boldsymbol{B}$.
Let $\boldsymbol{n}$ be the normal to the boundary at point $\boldsymbol{s}\in\partial C$.
For the GPR model, $\hat{M}=\boldsymbol{M}\left(\boldsymbol{Q}\left(\boldsymbol{s}\right)\right)\cdot\boldsymbol{n}$
is a diagonalizable matrix with decomposition $\hat{M}=\hat{R}\hat{\Lambda}\hat{R}^{-1}$
where the columns of $\hat{R}$ are the right eigenvectors and $\hat{\Lambda}$
is the diagonal matrix of eigenvalues. Define also $\boldsymbol{\hat{F}}=\boldsymbol{F}\cdot\boldsymbol{n}$
and $\hat{B}=\boldsymbol{B}\cdot\boldsymbol{n}$. Using these definitions,
the interface terms arising in the FV formula have the following form:

\begin{align}
\boldsymbol{\mathcal{D}}\left(\boldsymbol{Q^{-}},\boldsymbol{Q^{+}}\right) & =\frac{1}{2}\left(\boldsymbol{\hat{F}}\left(\boldsymbol{Q^{+}}\right)+\boldsymbol{\hat{F}}\left(\boldsymbol{Q^{-}}\right)\right)\\
 & +\frac{1}{2}\left(\tilde{B}\left(\boldsymbol{Q^{+}}-\boldsymbol{Q^{-}}\right)+\tilde{M}\left(\boldsymbol{Q^{+}}-\boldsymbol{Q^{-}}\right)\right)\nonumber 
\end{align}

$\tilde{M}$ is chosen to either correspond to a Rusanov/Lax-Friedrichs
flux (see \citep{toro_reimann_2009}):

\begin{equation}
\tilde{M}=\max\left(\max\left|\hat{\Lambda}\left(\boldsymbol{Q^{+}}\right)\right|,\max\left|\hat{\Lambda}\left(\boldsymbol{Q^{-}}\right)\right|\right)
\end{equation}

or a Roe flux (see \citep{dumbser_simple_2011}):
\begin{equation}
\hat{M}=\left|\int_{0}^{1}M\left(\boldsymbol{\boldsymbol{q^{-}}}+z\left(\boldsymbol{q^{+}}-\boldsymbol{q^{-}}\right)\right)dz\right|
\end{equation}

or a simplified Osher\textendash Solomon flux (see \citep{dumbser_simple_2011,dumbser_universal_2011}):

\begin{equation}
\tilde{M}=\int_{0}^{1}\left|\hat{M}\left(\boldsymbol{Q^{-}}+z\left(\boldsymbol{Q^{+}}-\boldsymbol{Q^{-}}\right)\right)\right|dz
\end{equation}

where

\begin{equation}
\left|\hat{M}\right|=\hat{R}\left|\hat{\Lambda}\right|\hat{R}^{-1}
\end{equation}

$\tilde{B}$ takes the following form:

\begin{equation}
\tilde{B}=\int_{0}^{1}\hat{B}\left(\boldsymbol{Q^{-}}+z\left(\boldsymbol{Q^{+}}-\boldsymbol{Q^{-}}\right)\right)dz
\end{equation}

$\boldsymbol{P_{i}^{n+\frac{1}{2}}},\boldsymbol{D_{i}^{n+\frac{1}{2}}}$
are calculated using an $N+1$-point Gauss-Legendre quadrature, replacing
$\boldsymbol{Q}\left(\boldsymbol{x},t_{n+\frac{1}{2}}\right)$ with
$\boldsymbol{w^{n+\frac{1}{2}}}\left(\boldsymbol{x}\right)$.

\subsubsection{The Thermal Impulse ODEs\label{subsec:The-Thermal-Impulse-ODEs}}

In \citep{jackson_fast_2017} it was shown that \eqref{eq:ThermalODE}
has the following analytical solution:

\begin{equation}
\boldsymbol{J}\left(t\right)=\boldsymbol{J}\left(0\right)\sqrt{\frac{1}{e^{at}-\frac{b}{a}\left(e^{at}-1\right)\left\Vert \boldsymbol{J}\left(0\right)\right\Vert ^{2}}}
\end{equation}

where

\begin{subequations}

\begin{align}
a & =\frac{2\rho_{0}}{\tau_{2}T_{0}\rho c_{v}}\left(E-E_{2}^{\left(A\right)}\left(A\right)-E_{3}\left(\boldsymbol{v}\right)\right)\\
b & =\frac{\rho_{0}c_{t}^{2}}{\tau_{2}T_{0}\rho c_{v}}
\end{align}

\end{subequations}

\subsubsection{The Distortion ODEs\label{subsec:The-Distortion-ODEs}}

Take the following singular value decomposition:

\begin{equation}
A=U\Sigma V^{T}
\end{equation}

Denote the singular values of $A$ by $a_{1},a_{2},a_{3}$ and define:

\begin{equation}
x_{i}=\frac{a_{i}^{2}}{\left(\frac{\rho}{\rho_{0}}\right)^{\frac{2}{3}}}
\end{equation}

Define also:

\begin{subequations}

\begin{align}
m_{0} & =\frac{x_{1}+x_{2}+x_{3}}{3}\\
u_{0} & =\frac{\left(x_{1}-x_{2}\right)^{2}+\left(x_{2}-x_{3}\right)^{2}+\left(x_{3}-x_{1}\right)^{2}}{3}
\end{align}

\end{subequations}

In \citep{jackson_fast_2017} it was shown that for Newtonian fluids,
after timestep $\Delta t$, $x_{i}$ become:

\begin{equation}
x_{i}=\frac{\sqrt{6u_{\Delta t}}}{3}\cos\left(\frac{\theta}{3}\right)+m_{\Delta t}
\end{equation}

where

\begin{subequations}

\begin{align}
\theta & =\tan^{-1}\left(\frac{\sqrt{6u_{\Delta t}^{3}-81\Delta^{2}}}{9\Delta}\right)\\
\Delta & =-2m_{\Delta t}^{3}+m_{\Delta t}u_{\Delta t}+2\\
m_{\Delta t} & =1+\frac{e^{-9\tau}}{3}\left(ae^{3\tau}-b\right)\\
u_{\Delta t} & =e^{-9\tau}\left(2ae^{3\tau}-3b\right)\\
a & =9m_{0}-u_{0}-9\\
b & =6m_{0}-u_{0}-6\\
\tau & =\frac{2}{\tau_{1}}\left(\frac{\rho}{\rho_{0}}\right)^{\frac{7}{3}}\Delta t
\end{align}

\end{subequations}

These new values for $x_{i}$ are used to calculate the value of $A$
after timestep $\text{\ensuremath{\Delta t}}$.

In \citep{jackson_numerical_2019} it was shown that for elastoplastic
solids, the same procedure can be undertaken, where now:

\begin{equation}
\tau=\frac{2\lambda}{nc}\log\left(\frac{nc}{\tau_{0}\lambda}\left(\frac{\rho}{\rho_{0}}\right)^{\frac{4n+7}{3}}\left(\frac{\sqrt{c}}{6}\frac{\rho c_{s}^{2}}{\sigma_{0}}\right)^{n}\Delta t+1\right)
\end{equation}

where

\begin{subequations}

\begin{align}
c & =108a-324b+108a^{2}-396ab+297b^{2}\\
 & -24\left(a^{2}b-2ab^{2}+b^{3}\right)-4\left(a-b\right)^{4}\\
\lambda & =18a-36b+9a^{2}-\frac{132ab}{5}+\frac{33b^{2}}{2}\\
 & -\frac{8a^{2}b}{7}+2ab^{2}-\frac{8b^{3}}{9}-\frac{a^{4}}{6}\\
 & +\frac{16a^{3}b}{27}-\frac{4a^{2}b^{2}}{5}+\frac{16ab^{3}}{33}-\frac{b^{4}}{9}
\end{align}

\end{subequations}

\subsubsection{Time Step}

Let $\Lambda_{i}^{n}$ be the set of eigenvalues of the GPR system
evaluated at $\boldsymbol{Q_{i}^{n}}$ (given explicitly in \prettyref{sec:Eigenstructure of the HPR Model}).
$C_{cfl}<1$ is a constant (usually taken to be $0.9$, unless the
problem being simulated is particularly demanding, requiring a lower
value). The eigenvalues determine the speed of propagation of information
in the solution to the Riemann Problem at the cell interfaces, and
the time step is chosen to ensure that the characteristics do not
enter into other cells between $t_{n}$ and $t_{n+1}$:

\begin{equation}
\Delta t_{n}=\frac{C_{cfl}\cdot\Delta x}{\max_{i}\left|\Lambda_{i}^{n}\right|}
\end{equation}

\section{A Riemann Ghost Fluid Method for the GPR Model\label{chap:A Riemann Ghost Fluid Method for the GPR Model}}

\subsection{Eigenstructure of the GPR Model\label{sec:Eigenstructure of the HPR Model}}

Take a hyperbolic system of the following form, noting that the GPR
model takes this form:

\begin{equation}
\frac{\partial\boldsymbol{P}}{\partial t}+M\frac{\partial\boldsymbol{P}}{\partial x}=\boldsymbol{S}
\end{equation}

Let $\left\{ \boldsymbol{l_{i}}\right\} $ be the set of left eigenvectors
of the system matrix $M$, in other words:

\begin{equation}
\boldsymbol{l_{i}}^{T}M=\lambda_{i}\boldsymbol{l_{i}}^{T}
\end{equation}

Along characteristics corresponding to $\lambda_{i}$, we have:

\begin{align}
\boldsymbol{l_{i}^{T}}\left(\frac{\partial\boldsymbol{P}}{\partial t}+M\frac{\partial\boldsymbol{P}}{\partial x}\right) & =\boldsymbol{l_{i}^{T}}\left(\frac{\partial\boldsymbol{P}}{\partial t}+\frac{dx}{dt}\frac{\partial\boldsymbol{P}}{\partial x}\right)\label{eq:characteristics}\\
 & =\boldsymbol{l_{i}^{T}}\frac{d\boldsymbol{P}}{dt}=\boldsymbol{l_{i}^{T}}\boldsymbol{S}\nonumber 
\end{align}

The second line of this equation is crucial to Riemann Ghost Fluid
Method presented in this study. As such, we must now investigate the
eigenvalues and eigenvectors of $M$.

\subsubsection{Eigenvalues}

Considering the primitive system matrix \prettyref{eq:M1}, it is
clear that the eigenvalues of the GPR system in the first spatial
axis consist of $v_{1}$ repeated 8 times, along with the roots of:

\begin{equation}
\left|\begin{array}{cc}
\left(v_{1}-\lambda\right)I & \Xi_{2}\\
\Xi_{1} & \left(v_{1}-\lambda\right)I
\end{array}\right|=0
\end{equation}

where

\begin{equation}
\Xi_{1}=-\frac{1}{\rho}\left(\begin{array}{ccccc}
\frac{\partial\sigma_{11}}{\partial\rho} & -1 & \frac{\partial\sigma_{11}}{\partial A_{11}} & \frac{\partial\sigma_{11}}{\partial A_{21}} & \frac{\partial\sigma_{11}}{\partial A_{31}}\\
\frac{\partial\sigma_{21}}{\partial\rho} & 0 & \frac{\partial\sigma_{21}}{\partial A_{11}} & \frac{\partial\sigma_{21}}{\partial A_{21}} & \frac{\partial\sigma_{21}}{\partial A_{31}}\\
\frac{\partial\sigma_{31}}{\partial\rho} & 0 & \frac{\partial\sigma_{31}}{\partial A_{11}} & \frac{\partial\sigma_{31}}{\partial A_{21}} & \frac{\partial\sigma_{31}}{\partial A_{31}}\\
-T_{\rho} & -T_{p} & 0 & 0 & 0
\end{array}\right)
\end{equation}

\begin{equation}
\Xi_{2}=\left(\begin{array}{cccc}
\rho & 0 & 0 & 0\\
\left(\rho c_{0}^{2}+\sigma_{11}-\rho\frac{\partial\sigma_{11}}{\partial\rho}\right) & \left(\sigma_{21}-\rho\frac{\partial\sigma_{21}}{\partial\rho}\right) & \left(\sigma_{31}-\rho\frac{\partial\sigma_{31}}{\partial\rho}\right) & \frac{\rho c_{h}^{2}}{T_{p}}\\
A_{11} & A_{12} & A_{13} & 0\\
A_{21} & A_{22} & A_{23} & 0\\
A_{31} & A_{32} & A_{33} & 0
\end{array}\right)
\end{equation}

By the properties of block matrices\footnote{If $A$ is \foreignlanguage{british}{invertible}, $\det\left(\begin{array}{cc}
A & B\\
C & D
\end{array}\right)=\det\left(A\right)\det\left(D-CA^{-1}B\right)$}, the remaining eigenvalues are $v_{1}$ and the roots of $\left|\left(v_{1}-\lambda\right)^{2}I-\Xi_{1}\Xi_{2}\right|=0$.
Thus, $\lambda_{i}=v_{1}\pm\sqrt{\tilde{\lambda_{i}}}$ where the
$\tilde{\lambda_{i}}$ are the eigenvalues of the following matrix:

\begin{equation}
\Xi=\Xi_{1}\Xi_{2}=\left(\begin{array}{cccc}
\Omega_{11}^{1}+\left(c_{0}^{2}+\frac{\sigma_{11}}{\rho}-\frac{\partial\sigma_{11}}{\partial\rho}\right) & \Omega_{12}^{1}+\left(\frac{\sigma_{21}}{\rho}-\frac{\partial\sigma_{21}}{\partial\rho}\right) & \Omega_{13}^{1}+\left(\frac{\sigma_{31}}{\rho}-\frac{\partial\sigma_{31}}{\partial\rho}\right) & \frac{c_{h}^{2}}{T_{p}}\\
\Omega_{21}^{1} & \Omega_{22}^{1} & \Omega_{23}^{1} & 0\\
\Omega_{31}^{1} & \Omega_{32}^{1} & \Omega_{33}^{1} & 0\\
T_{\rho}+T_{p}\left(c_{0}^{2}+\frac{\sigma_{11}}{\rho}-\frac{\partial\sigma_{11}}{\partial\rho}\right) & T_{p}\left(\frac{\sigma_{21}}{\rho}-\frac{\partial\sigma_{21}}{\partial\rho}\right) & T_{p}\left(\frac{\sigma_{31}}{\rho}-\frac{\partial\sigma_{31}}{\partial\rho}\right) & c_{h}^{2}
\end{array}\right)
\end{equation}

where $\Omega$ is given shortly. Similar results hold for the other
two spatial directions. In general it is not possible to express the
eigenvalues of $\Xi$ in terms of the eigenvalues of its submatrices.
Note, however, that if $c_{t}=0$ then one of the eigenvalues is $0$
and the remaining eigenvalues can be found analytically, using the
form given in the appendix of \citep{dumbser_high_2015}.

It is straightforward to verify the following:

\begin{equation}
\frac{\partial\sigma_{ij}}{\partial A_{mn}}=-c_{s}^{2}\rho\left(\begin{array}{c}
\delta_{in}\left(A\dev\left(G\right)\right)_{mj}+\delta_{jn}\left(A\dev\left(G\right)\right)_{mi}\\
+A_{mi}G_{jn}+A_{mj}G_{in}-\frac{2}{3}G_{ij}A_{mn}
\end{array}\right)\label{eq:dsigma/dA}
\end{equation}

The quantity $\Omega$ is named here the \textit{acoustic tensor},
due to its similarity to the acoustic tensor described in \citep{barton_exact_2009}:

{\small{}
\begin{align}
\Omega_{ij}^{d} & =-\frac{1}{\rho}\frac{\partial\sigma_{id}}{\partial A_{kd}}A_{kj}-\frac{\sigma_{id}}{\rho}\delta_{dj}\\
 & =c_{s}^{2}\left(\begin{array}{c}
\delta_{id}\left(G\dev\left(G\right)\right)_{dj}+\left(G\dev\left(G\right)\right)_{id}\delta_{dj}\\
+\left(G\dev\left(G\right)\right)_{ij}+G_{ij}G_{dd}+\frac{1}{3}G_{dj}G_{id}
\end{array}\right)\nonumber \\
 & =c_{s}^{2}\left(E^{d}G\dev\left(G\right)+G\dev\left(G\right)E^{d}+G\dev\left(G\right)+G_{dd}G+\frac{1}{3}G_{d}G_{d}^{T}\right)\nonumber 
\end{align}
}{\small\par}

where $E_{ij}^{d}=\delta_{id}\delta_{jd}$.

\subsubsection{Eigenvectors (with Heat Conduction)}

By hyperbolicity of the system, $\Xi$ can be expressed as:

\begin{equation}
\Xi=Q^{-1}D^{2}Q
\end{equation}

where $D$ is a diagonal matrix with positive diagonal entries. The
eigenvectors corresponding to $\lambda_{i}=v_{1}\pm\sqrt{\tilde{\lambda_{i}}}$
take the form $\left(\begin{array}{cccc}
\hat{u} & 0_{6} & \tilde{u} & 0_{2}\end{array}\right)^{T}$ where $\boldsymbol{\hat{u}}\in\mathbb{R}^{5},\boldsymbol{\tilde{u}}\in\mathbb{R}^{4}$
satisfy:

\begin{equation}
\left(\begin{array}{cc}
v_{1}I & \Xi_{2}\\
\Xi_{1} & v_{1}I
\end{array}\right)\left(\begin{array}{c}
\boldsymbol{\hat{u}}\\
\boldsymbol{\tilde{u}}
\end{array}\right)=\left(v_{1}\pm\sqrt{\tilde{\lambda_{i}}}\right)\left(\begin{array}{c}
\boldsymbol{\hat{u}}\\
\boldsymbol{\tilde{u}}
\end{array}\right)
\end{equation}

Thus, $\Xi_{2}\boldsymbol{\tilde{u}}=\pm\sqrt{\tilde{\lambda_{i}}}\boldsymbol{\hat{u}}$
and $\Xi_{1}\boldsymbol{\hat{u}}=\pm\sqrt{\tilde{\lambda_{i}}}\boldsymbol{\tilde{u}}$.
Combining these results, $\Xi\boldsymbol{\tilde{u}}=\tilde{\lambda_{i}}\boldsymbol{\tilde{u}}$.
Thus, $\boldsymbol{\tilde{u}}$ is a right eigenvector of $\Xi$ and,
taking the form $Q^{-1}\boldsymbol{e_{i}}$ for some $i=1\ldots4$.

The four eigenvectors corresponding to eigenvalues of the form $v_{1}+\sqrt{\tilde{\lambda_{i}}}$
are columns 1-4 of matrix $R$ in \eqref{eq:Right-Eigenvectors-1}.
Those corresponding to eigenvalues of the form $v_{1}-\sqrt{\tilde{\lambda_{i}}}$
are columns 5-8. By inspection (using the system matrix \eqref{eq:M1}),
it can be verified that the remaining 9 eigenvectors (corresponding
to eigenvalue $v_{1}$) are the remaining columns.

Note that the index $d$ appearing in these representations should
be taken as $1,2,3$ for eigenvectors in directions $x,y,z$, respectively.
$0_{m,n}$ is defined to be the 0-matrix of shape $\left(m,n\right)$
and $I_{n}$ the identity matrix of size $n$.

\begin{equation}
R=\left\{ \left(\begin{array}{cc}
\frac{1}{2}\Xi_{2}\left(D^{2}Q\right)^{-1} & \frac{1}{2}\Xi_{2}\left(D^{2}Q\right)^{-1}\\
0_{6,4} & 0_{6,4}\\
\frac{1}{2}\left(DQ\right)^{-1} & -\frac{1}{2}\left(DQ\right)^{-1}\\
0_{2,4} & 0_{2,4}
\end{array}\right),\left(\begin{array}{c}
-cT_{p}\\
cT_{\rho}\\
c\Pi_{d}^{-1}\boldsymbol{w}\\
0_{12,1}
\end{array}\right),\left(\begin{array}{cc}
0_{2,3} & 0_{2,3}\\
-\Pi_{1}^{-1}\Pi_{2} & -\Pi_{1}^{-1}\Pi_{3}\\
I_{3} & 0_{3,3}\\
0_{3,3} & I_{3}\\
0_{6,3} & 0_{6,3}
\end{array}\right),\left(\begin{array}{c}
0_{15,2}\\
I_{2}
\end{array}\right)\right\} \label{eq:Right-Eigenvectors-1}
\end{equation}

where

\begin{subequations}

\begin{align}
\left(\Pi_{k}\right)_{ij} & =\frac{\partial\sigma_{id}}{\partial A_{jk}}\\
\boldsymbol{w} & =T_{p}\frac{\partial\boldsymbol{\sigma_{d}}}{\partial\rho}+T_{\rho}\boldsymbol{e_{d}}\\
c & =\frac{1}{\boldsymbol{e_{d}^{T}}\left(\Pi_{d}A\right)^{-1}\boldsymbol{w}+\frac{T_{p}}{\rho}}
\end{align}

\end{subequations}

A similar analysis yields the left eigenvectors as the rows of \eqref{eq:Left-Eigenvectors-1}.

\begin{equation}
L=\left\{ \begin{array}{c}
\left(\begin{array}{ccccc}
Q\Xi_{1} & -\frac{1}{\rho}Q_{:,1:3}\Pi_{2} & -\frac{1}{\rho}Q_{:,1:3}\Pi_{3} & DQ & 0_{4,2}\\
Q\Xi_{1} & -\frac{1}{\rho}Q_{:,1:3}\Pi_{2} & -\frac{1}{\rho}Q_{:,1:3}\Pi_{3} & -DQ & 0_{4,2}
\end{array}\right)\\
\left(\begin{array}{cccccc}
-\frac{1}{\rho} & 0 & \boldsymbol{e_{d}^{T}}A^{-1} & \boldsymbol{e_{d}^{T}}A^{-1}\Pi_{1}^{-1}\Pi_{2} & \boldsymbol{e_{d}^{T}}A^{-1}\Pi_{1}^{-1}\Pi_{3} & 0_{1,6}\end{array}\right)\\
\left(\begin{array}{cccc}
0_{3,5} & I_{3} & 0_{3,3} & 0_{3,6}\\
0_{3,5} & 0_{3,3} & I_{3} & 0_{3,6}
\end{array}\right)\\
\left(\begin{array}{cc}
0_{2,15} & I_{2}\end{array}\right)
\end{array}\right\} \label{eq:Left-Eigenvectors-1}
\end{equation}

\subsubsection{Eigenvectors (without Heat Conduction)}

If the system does not include the heat conduction terms, the eigenstructure
of the system matrix changes. $\Xi_{1},\Xi_{2},\Xi$ now take the
following values:

\begin{equation}
\Xi_{1}=-\frac{1}{\rho}\left(\begin{array}{ccccc}
\frac{\partial\sigma_{11}}{\partial\rho} & -1 & \frac{\partial\sigma_{11}}{\partial A_{11}} & \frac{\partial\sigma_{11}}{\partial A_{21}} & \frac{\partial\sigma_{11}}{\partial A_{31}}\\
\frac{\partial\sigma_{21}}{\partial\rho} & 0 & \frac{\partial\sigma_{21}}{\partial A_{11}} & \frac{\partial\sigma_{21}}{\partial A_{21}} & \frac{\partial\sigma_{21}}{\partial A_{31}}\\
\frac{\partial\sigma_{31}}{\partial\rho} & 0 & \frac{\partial\sigma_{31}}{\partial A_{11}} & \frac{\partial\sigma_{31}}{\partial A_{21}} & \frac{\partial\sigma_{31}}{\partial A_{31}}
\end{array}\right)
\end{equation}

\begin{equation}
\Xi_{2}=\left(\begin{array}{ccc}
\rho & 0 & 0\\
\left(\rho c_{0}^{2}+\sigma_{11}-\rho\frac{\partial\sigma_{11}}{\partial\rho}\right) & \left(\sigma_{21}-\rho\frac{\partial\sigma_{21}}{\partial\rho}\right) & \left(\sigma_{31}-\rho\frac{\partial\sigma_{31}}{\partial\rho}\right)\\
A_{11} & A_{12} & A_{13}\\
A_{21} & A_{22} & A_{23}\\
A_{31} & A_{32} & A_{33}
\end{array}\right)
\end{equation}

\begin{equation}
\Xi=\Xi_{1}\Xi_{2}=\left(\begin{array}{ccc}
\Omega_{11}^{1}+\left(c_{0}^{2}+\frac{\sigma_{11}}{\rho}-\frac{\partial\sigma_{11}}{\partial\rho}\right) & \Omega_{12}^{1}+\left(\frac{\sigma_{21}}{\rho}-\frac{\partial\sigma_{21}}{\partial\rho}\right) & \Omega_{13}^{1}+\left(\frac{\sigma_{31}}{\rho}-\frac{\partial\sigma_{31}}{\partial\rho}\right)\\
\Omega_{21}^{1} & \Omega_{22}^{1} & \Omega_{23}^{1}\\
\Omega_{31}^{1} & \Omega_{32}^{1} & \Omega_{33}^{1}
\end{array}\right)
\end{equation}

Using the eigendecomposition $\Xi=Q^{-1}D^{2}Q$ as before, we have:

\begin{equation}
R=\left\{ \left(\begin{array}{cc}
\frac{1}{2}\Xi_{2}\left(D^{2}Q\right)^{-1} & \frac{1}{2}\Xi_{2}\left(D^{2}Q\right)^{-1}\\
0_{6,3} & 0_{6,3}\\
\frac{1}{2}\left(DQ\right)^{-1} & -\frac{1}{2}\left(DQ\right)^{-1}
\end{array}\right),\left(\begin{array}{cc}
1 & 0\\
0 & 1\\
-\Pi_{1}^{-1}\frac{\partial\boldsymbol{\sigma_{1}}}{\partial\rho} & \Pi_{1}^{-1}\boldsymbol{e_{1}}\\
\boldsymbol{0_{9}} & \boldsymbol{0_{9}}
\end{array}\right),\left(\begin{array}{cc}
0_{2,3} & 0_{2,3}\\
-\Pi_{1}^{-1}\Pi_{2} & -\Pi_{1}^{-1}\Pi_{3}\\
I_{3} & 0_{3,3}\\
0_{3,3} & I_{3}\\
0_{3,3} & 0_{3,3}
\end{array}\right)\right\} \label{eq:Right-Eigenvectors-Non-Thermal}
\end{equation}

By considering their products with the first 8 columns of $R$, two
of the left eigenvectors corresponding the the 7th and 8th right eigenvectors
must come in the form of the rows of the following matrix:

\begin{equation}
\left(\begin{array}{cccc}
W & X & Y & Z\end{array}\right)\label{eq:WXYZ}
\end{equation}

where $W\in\mathbb{R}^{2,5}$ and $X,Y,Z\in\mathbb{R}^{2,3}$, and:

\begin{subequations}

\begin{align}
W\Xi_{2}\left(D^{2}Q\right)^{-1}+Z\left(DQ\right)^{-1} & =0\\
W\Xi_{2}\left(D^{2}Q\right)^{-1}-Z\left(DQ\right)^{-1} & =0\\
W\left(\begin{array}{c}
0_{2,3}\\
-\Pi_{1}^{-1}\Pi_{2}
\end{array}\right)+X & =0\\
W\left(\begin{array}{c}
0_{2,3}\\
-\Pi_{1}^{-1}\Pi_{3}
\end{array}\right)+Y & =0
\end{align}

\end{subequations}

Solving this system for $X,Y,Z$:

\begin{subequations}

\begin{align}
Z & =0\\
X & =W_{:,3:5}\Pi_{1}^{-1}\Pi_{2}\\
Y & =W_{:,3:5}\Pi_{1}^{-1}\Pi_{3}
\end{align}

\end{subequations}

Define:

\begin{align}
\aleph & \equiv\left(\begin{array}{ccccc}
\left(\Xi_{2}\right)_{11} & \left(\Xi_{2}\right)_{12} & \left(\Xi_{2}\right)_{13} & 1 & 0\\
\left(\Xi_{2}\right)_{21} & \left(\Xi_{2}\right)_{22} & \left(\Xi_{2}\right)_{23} & 0 & 1\\
\left(\Xi_{2}\right)_{31} & \left(\Xi_{2}\right)_{32} & \left(\Xi_{2}\right)_{33} & C_{11} & C_{12}\\
\left(\Xi_{2}\right)_{41} & \left(\Xi_{2}\right)_{42} & \left(\Xi_{2}\right)_{43} & C_{21} & C_{22}\\
\left(\Xi_{2}\right)_{51} & \left(\Xi_{2}\right)_{52} & \left(\Xi_{2}\right)_{53} & C_{31} & C_{32}
\end{array}\right)\\
 & =\left(\begin{array}{ccccc}
B_{11} & B_{12} & B_{13} & 1 & 0\\
B_{21} & B_{22} & B_{23} & 0 & 1\\
A_{11} & A_{12} & A_{13} & C_{11} & C_{12}\\
A_{21} & A_{22} & A_{23} & C_{21} & C_{22}\\
A_{31} & A_{32} & A_{33} & C_{31} & C_{32}
\end{array}\right)\nonumber 
\end{align}

where

\begin{subequations}

\begin{align}
B & =\left(\begin{array}{ccc}
\rho & 0 & 0\\
\left(\rho c_{0}^{2}+\sigma_{11}-\rho\frac{\partial\sigma_{11}}{\partial\rho}\right) & \left(\sigma_{21}-\rho\frac{\partial\sigma_{21}}{\partial\rho}\right) & \left(\sigma_{31}-\rho\frac{\partial\sigma_{31}}{\partial\rho}\right)
\end{array}\right)\\
C & =\Pi_{1}^{-1}\left(\begin{array}{cc}
-\frac{\partial\boldsymbol{\sigma_{1}}}{\partial\rho} & \boldsymbol{e_{1}}\end{array}\right)
\end{align}

\end{subequations}

By the properties of block matrices:

\begin{equation}
\aleph^{-1}=\left(\begin{array}{cc}
-A^{-1}C\left(I-BA^{-1}C\right)^{-1} & A^{-1}\left(I+C\left(I-BA^{-1}C\right)^{-1}BA^{-1}\right)\\
\left(I-BA^{-1}C\right)^{-1} & -\left(I-BA^{-1}C\right)^{-1}BA^{-1}
\end{array}\right)
\end{equation}

By the orthonormality of eigenvectors, we must have:

\begin{equation}
W\aleph=\left(\begin{array}{ccccc}
0 & 0 & 0 & 1 & 0\\
0 & 0 & 0 & 0 & 1
\end{array}\right)
\end{equation}

Thus, it is straightforward to confirm that:

\begin{equation}
W=\left(\begin{array}{cc}
\left(I-BA^{-1}C\right)^{-1} & -\left(I-BA^{-1}C\right)^{-1}BA^{-1}\end{array}\right)
\end{equation}
\begin{subequations}

Thus, we have:

\begin{align}
W & =\left(I-BA^{-1}C\right)^{-1}\left(\begin{array}{cc}
I_{2} & -BA^{-1}\end{array}\right)\\
X & =-\left(I-BA^{-1}C\right)^{-1}BA^{-1}\Pi_{1}^{-1}\Pi_{2}\\
Y & =-\left(I-BA^{-1}C\right)^{-1}BA^{-1}\Pi_{1}^{-1}\Pi_{3}
\end{align}

\end{subequations}

Finally, combining the preceding results with \eqref{eq:WXYZ}, we
have:

\begin{equation}
L=\left\{ \begin{array}{c}
\left(\begin{array}{cccc}
Q\Xi_{1} & -\frac{1}{\rho}Q\Pi_{2} & -\frac{1}{\rho}Q\Pi_{3} & DQ\\
Q\Xi_{1} & -\frac{1}{\rho}Q\Pi_{2} & -\frac{1}{\rho}Q\Pi_{3} & -DQ
\end{array}\right)\\
\left(I_{2}-BA^{-1}C\right)^{-1}\left(\begin{array}{ccccc}
I_{2} & -BA^{-1} & -BA^{-1}\Pi_{1}^{-1}\Pi_{2} & -BA^{-1}\Pi_{1}^{-1}\Pi_{3} & 0_{2,3}\end{array}\right)\\
\left(\begin{array}{cccc}
0_{3,5} & I_{3} & 0_{3,3} & 0_{3,3}\\
0_{3,5} & 0_{3,3} & I_{3} & 0_{3,3}
\end{array}\right)
\end{array}\right\} \label{eq:Left-Eigenvectors-Non-Thermal}
\end{equation}

\subsection{Solving the Riemann Problem\label{sec:SolvingTheRiemannProblem}}

Barton et al. have presented an RGFM for the equations of non-linear
elasticity \citep{barton_conservative_2011,barton_eulerian_2010}.
Owing to the similarity of the structure of the non-linear elasticity
equations to those of the GPR model (differing only in the presence
of source terms, the form of the shear stress tensor, and possibly
the EOS), their method is built upon here. The resulting method is
named the \textit{GPR-RGFM}.

\begin{figure}
\begin{centering}
\includegraphics[height=0.25\textheight]{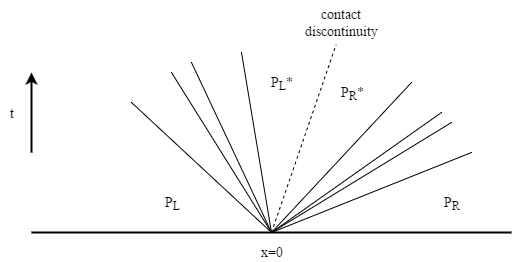}
\par\end{centering}
\caption{\label{fig:GPR_RP}The Riemann Problem for the GPR model, assuming
all waves are distinct}
\medskip{}
\end{figure}

The Riemann Problem of the GPR model takes the form demonstrated in
\prettyref{fig:GPR_RP}. Assuming all waves are distinct, there are
four waves on either side of the contact discontinuity. On each side,
one wave corresponds to the thermal impulse (manifesting as a heat
wave) and the other three correspond to the distortion components
in the axis in which the Riemann Problem is considered (manifesting
as two shear waves and one longitudinal pressure wave). It is important
to note that - owing to the source terms - the star states are not
constant in the spacetime region in which they reside, so the method
presented here produces only an approximation to them.

The method is presented here along the first spatial axis. It can
easily be adapted along any axis by taking the components of all relevant
vector quantities (velocity, distortion, and thermal impulse) in the
direction normal to the interface.

Denote the vector of primitive variables by $\boldsymbol{P}$. Take
the set of left eigenvectors $L$ \eqref{eq:Left-Eigenvectors-1}
with eigenvalues $\text{\ensuremath{\left\{  \lambda_{i}\right\} } }$.
From \eqref{eq:characteristics}, we have the standard set of relations
along characteristics (curves along which $\frac{dx}{dt}=\lambda_{i}$):

\begin{equation}
L\cdot d\boldsymbol{P}=dt\cdot L\cdot\boldsymbol{S}\label{eq:L.dP=00003Ddt.L.S}
\end{equation}

In what follows, we enact an operator splitting of the two processes
present in the system \eqref{eq:L.dP=00003Ddt.L.S}:

\begin{subequations}

\begin{align}
L\cdot d\boldsymbol{P} & =\boldsymbol{0}\label{eq:L.dP=00003D0}\\
\frac{d\boldsymbol{P}}{dt} & =\boldsymbol{S}\label{eq:dP/dt=00003DS}
\end{align}

\end{subequations}

\textbf{$\boldsymbol{P^{*K}}$} is now sought, where $K=L$ or $K=R$,
denoting the left or right sides of the interface, respectively. Take
the following linearization:

\begin{equation}
d\boldsymbol{P^{K}}\approx\boldsymbol{P^{*K}}-\boldsymbol{P^{K}}
\end{equation}

\begin{figure}
\begin{centering}
\includegraphics[height=0.25\textheight]{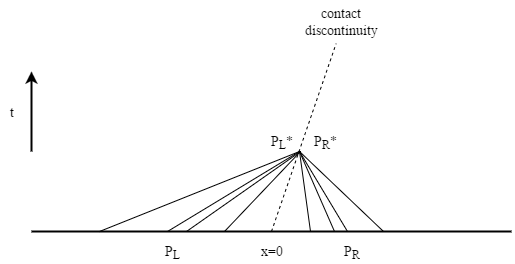}
\par\end{centering}
\caption{\label{fig:HPR_RP2}Different sets of characteristic curves, traveling
from their respective initial points to the star region}
\medskip{}
\end{figure}

13 relations from \eqref{eq:L.dP=00003D0} are taken: 4 regarding
the 4 sets of characteristics traveling into the contact discontinuity
from side $K$ (with speeds greater or less than $v$ for $K=L$ or
$K=R$, respectively), and 9 relating to the contact discontinuity
itself. This is demonstrated in \prettyref{fig:HPR_RP2}. 4 more relations
must be derived to solve the system for $\boldsymbol{P^{*K}}$.

Define the total stress tensor as:

\begin{equation}
\Sigma\equiv pI-\sigma
\end{equation}

The values of $\Sigma,T$ under variables $\boldsymbol{P^{*K}}$ are
obtained by expanding the following Taylor series:

\begin{subequations}

\begin{align}
\Sigma^{*} & =\Sigma+\left(\rho^{*}-\rho\right)\frac{\partial\Sigma}{\partial\rho}+\left(p^{*}-p\right)\frac{\partial\Sigma}{\partial p}+\left(A_{mn}^{*}-A_{mn}\right)\frac{\partial\Sigma}{\partial A_{mn}}+O\left(d\boldsymbol{P}^{2}\right)\label{eq:q* Taylor Series}\\
T^{*} & =T+\left(\rho^{*}-\rho\right)\frac{\partial T}{\partial\rho}+\left(p^{*}-p\right)\frac{\partial T}{\partial p}+O\left(d\boldsymbol{P}^{2}\right)\label{eq:sigma* Taylor Series}
\end{align}

\end{subequations}

\nomenclature[g]{$\Sigma$}{Total stress tensor}

Thus, we have:

\begin{subequations}

\begin{align}
\Sigma^{*}-\Sigma & \approx\left(p^{*}-p\right)I-\left(\rho^{*}-\rho\right)\frac{\partial\sigma}{\partial\rho}-\left(A_{mn}^{*}-A_{mn}\right)\frac{\partial\sigma}{\partial A_{mn}}\label{eq:SigmaCond}\\
T^{*}-T & \approx\left(\rho^{*}-\rho\right)\frac{\partial T}{\partial\rho}+\left(p^{*}-p\right)\frac{\partial T}{\partial p}\label{eq:TempCond}
\end{align}

\end{subequations}

These are the extra required relations. Thus we have:

\begin{equation}
\hat{L}^{K}\cdot\left(\boldsymbol{P^{*K}}-\boldsymbol{P^{K}}\right)=\boldsymbol{c^{K}}\label{eq:L(P*-P)=00003Dc}
\end{equation}

where $\hat{L}^{K}$ takes the form found in \eqref{eq:Left-Conditions},
with $\xi=-1$ for $K=R$ and $\xi=1$ for $K=L$, and:

\begin{equation}
\boldsymbol{c^{K}}=\left(\begin{array}{c}
\boldsymbol{\Sigma_{1}^{*K}}-\boldsymbol{\Sigma_{1}^{K}}\\
T^{*K}-T^{K}\\
\boldsymbol{0}
\end{array}\right)
\end{equation}

The inverse of $\hat{L}^{K}$ takes the form found in \eqref{eq:Right-Conditions}.

$\hat{L}^{K},\hat{L}^{K}{}^{-1}$ are evaluated at $\boldsymbol{P^{K}}$.
It remains to find expressions for $\Sigma^{*}$ and $T^{*}$ in terms
of $\boldsymbol{P^{L}},\boldsymbol{P^{R}}$ to close the system. The
obtained values depend on the boundary conditions chosen, as explained
below.

\subsubsection{Boundary Conditions}

\paragraph{Stick Boundary Conditions}

The following boundary conditions are used:

\begin{subequations}

\begin{align}
\boldsymbol{\Sigma_{1}^{*L}} & =\boldsymbol{\Sigma_{1}^{*R}}\\
T^{*L} & =T^{*R}\\
\boldsymbol{v^{*L}} & =\boldsymbol{v^{*R}}\\
q_{1}^{*L} & =q_{1}^{*R}
\end{align}

\end{subequations}

Taking the relevant rows of $\boldsymbol{P^{*K}}=\boldsymbol{P^{K}}+\hat{L}^{K}{}^{-1}\boldsymbol{c^{K}}$:

\begin{equation}
\left(\begin{array}{c}
\boldsymbol{v^{*}}\\
J_{1}^{*}
\end{array}\right)=\left(\begin{array}{c}
\boldsymbol{v^{K}}\\
J_{1}^{K}
\end{array}\right)+Y^{K}\left(\left(\begin{array}{c}
\boldsymbol{\Sigma_{1}^{*}}\\
T^{*}
\end{array}\right)-\left(\begin{array}{c}
\boldsymbol{\Sigma_{1}^{K}}\\
T^{K}
\end{array}\right)\right)
\end{equation}

Thus:

{\small{}
\begin{equation}
\left(\begin{array}{c}
\boldsymbol{\Sigma_{1}^{*}}\\
T^{*}
\end{array}\right)=\left(Y^{L}-Y^{R}\right)^{-1}\left(\begin{array}{c}
\left(\begin{array}{c}
\boldsymbol{v^{R}}\\
J_{1}^{R}
\end{array}\right)-\left(\begin{array}{c}
\boldsymbol{v^{L}}\\
J_{1}^{L}
\end{array}\right)+Y^{L}\left(\begin{array}{c}
\boldsymbol{\Sigma_{1}^{L}}\\
T^{L}
\end{array}\right)-Y^{R}\left(\begin{array}{c}
\boldsymbol{\Sigma_{1}^{R}}\\
T^{R}
\end{array}\right)\end{array}\right)
\end{equation}
}{\small\par}

\paragraph{Slip Boundary Conditions}

The following boundary conditions are used:

\begin{subequations}

\begin{align}
\Sigma_{11}^{*L} & =\Sigma_{11}^{*R}\\
\Sigma_{12}^{*L},\Sigma_{12}^{*R} & =0\\
\Sigma_{13}^{*L},\Sigma_{13}^{*R} & =0\\
T^{*L} & =T^{*R}\\
v_{1}^{*L} & =v_{1}^{*R}\\
q_{1}^{*L} & =q_{1}^{*R}
\end{align}

\end{subequations}

Taking the relevant rows of $\boldsymbol{P^{*K}}=\boldsymbol{P^{K}}+\hat{L}^{K}{}^{-1}\boldsymbol{c^{K}}$:

\begin{equation}
\left(\begin{array}{c}
v_{1}^{*}\\
J_{1}^{*}
\end{array}\right)=\left(\begin{array}{c}
v_{1}^{K}\\
J_{1}^{K}
\end{array}\right)+\tilde{Y}^{K}\left(\left(\begin{array}{c}
\Sigma_{11}^{*}\\
0\\
0\\
T^{*}
\end{array}\right)-\left(\begin{array}{c}
\Sigma_{11}^{K}\\
\Sigma_{12}^{K}\\
\Sigma_{13}^{K}\\
T^{K}
\end{array}\right)\right)
\end{equation}

where

\begin{equation}
\tilde{Y}^{K}=\left(\begin{array}{c}
\boldsymbol{Y_{1}^{K}}\\
\boldsymbol{Y_{4}^{K}}
\end{array}\right)
\end{equation}

Thus:

{\small{}
\begin{equation}
\left(\begin{array}{c}
\Sigma_{11}^{*}\\
T^{*}
\end{array}\right)=\left(\hat{Y}^{L}-\hat{Y}^{R}\right)^{-1}\left(\begin{array}{c}
\left(\begin{array}{c}
v_{1}^{R}\\
J_{1}^{R}
\end{array}\right)-\left(\begin{array}{c}
v_{1}^{K}\\
J_{1}^{L}
\end{array}\right)+Y^{L}\left(\begin{array}{c}
\boldsymbol{\Sigma_{1}^{L}}\\
T^{L}
\end{array}\right)-Y^{R}\left(\begin{array}{c}
\boldsymbol{\Sigma_{1}^{R}}\\
T^{R}
\end{array}\right)\end{array}\right)
\end{equation}
}{\small\par}

where

\begin{equation}
\hat{Y}^{K}=\left(\begin{array}{cc}
Y_{11}^{K} & Y_{14}^{K}\\
Y_{41}^{K} & Y_{44}^{K}
\end{array}\right)
\end{equation}

\paragraph{Vacuum Boundary Conditions}

The following boundary conditions are used:

\begin{subequations}

\begin{align}
\boldsymbol{\Sigma_{1}^{*}} & =\boldsymbol{0}\\
q_{1}^{*} & =0
\end{align}

\end{subequations}

Taking the relevant row of $\boldsymbol{P^{*K}}=\boldsymbol{P^{K}}+\hat{L}^{K}{}^{-1}\boldsymbol{c^{K}}$:

\begin{equation}
J_{1}^{*}=J_{1}^{K}+\boldsymbol{Y_{4}^{K}}\cdot\left(\left(\begin{array}{c}
\boldsymbol{0}\\
T^{*}
\end{array}\right)-\left(\begin{array}{c}
\boldsymbol{\Sigma_{1}^{K}}\\
T^{K}
\end{array}\right)\right)
\end{equation}

As $q_{1}^{*}=0$ implies that $J_{1}^{*}=0$, we have:

\begin{equation}
T^{*}=\frac{1}{Y_{44}^{K}}\left(\boldsymbol{Y_{4}^{K}}\cdot\left(\begin{array}{c}
\boldsymbol{\Sigma_{1}^{K}}\\
T^{K}
\end{array}\right)-J_{1}^{K}\right)=T^{K}+\frac{\boldsymbol{Y_{4,:3}^{K}}\cdot\boldsymbol{\Sigma_{1}^{K}}-J_{1}^{K}}{Y_{44}^{K}}
\end{equation}

\paragraph{Iteration}

\eqref{eq:L(P*-P)=00003Dc} is solved for $\boldsymbol{P^{*K}}$,
which is taken to be the star state if the following conditions are
satisfied:

\begin{subequations}

\begin{align}
\frac{\left|\boldsymbol{\Sigma_{1}^{*L}}-\boldsymbol{\Sigma_{1}^{*R}}\right|}{\min\left(\rho_{0}^{L},\rho_{0}^{R}\right)\times\min\left(c_{s}^{L},c_{s}^{R}\right)^{2}} & <TOL\\
\frac{\left|v_{1}^{L}-v_{1}^{R}\right|}{\min\left(c_{s}^{L},c_{s}^{R}\right)} & <TOL\\
\frac{\left|q_{1}^{L}-q_{1}^{R}\right|}{\min\left(\tilde{q}^{L},\tilde{q}^{R}\right)} & <TOL\\
\frac{\left|T^{L}-T^{R}\right|}{\min\left(T_{0}^{L},T_{0}^{R}\right)} & <TOL
\end{align}

\end{subequations}

where

\begin{equation}
\tilde{q}=\frac{c_{t}^{2}}{\rho_{0}}\sqrt{\frac{T_{0}^{3}}{c_{V}}}
\end{equation}

These convergence criteria are chosen so that the variables required
to be less than $TOL$ are dimensionless. At every iteration, \eqref{eq:dP/dt=00003DS}
is solved using the ODE solvers presented in \citep{jackson_fast_2017,jackson_numerical_2019}.

\subsubsection{Linear Conditions with Heat Conduction}

We now obtain $\hat{L}^{K}$ and its inverse in order to solve \eqref{eq:L(P*-P)=00003Dc}.
Replacing the first four lines of \eqref{eq:Left-Eigenvectors-1}
with the conditions \eqref{eq:SigmaCond}, \eqref{eq:TempCond}, we
have:

\begin{equation}
\hat{L}^{K}=\left\{ \begin{array}{c}
\left(\begin{array}{cccccc}
-\frac{\partial\boldsymbol{\sigma_{d}}}{\partial\rho} & \boldsymbol{e_{d}} & -\Pi_{1} & -\Pi_{2} & -\Pi_{3} & 0_{3,6}\\
\frac{\partial T}{\partial\rho} & \frac{\partial T}{\partial p} & 0_{1,3} & 0_{1,3} & 0_{1,3} & 0_{1,6}
\end{array}\right)\\
\left(\begin{array}{ccccc}
Q\Xi_{1} & -\frac{1}{\rho}Q_{:,1:3}\Pi_{2} & -\frac{1}{\rho}Q_{:,1:3}\Pi_{3} & \xi DQ & 0_{4,2}\end{array}\right)\\
\left(\begin{array}{cccccc}
-\frac{1}{\rho} & 0 & \boldsymbol{e_{d}^{T}}A^{-1} & \boldsymbol{e_{d}^{T}}A^{-1}\Pi_{1}^{-1}\Pi_{2} & \boldsymbol{e_{d}^{T}}A^{-1}\Pi_{1}^{-1}\Pi_{3} & 0_{1,6}\end{array}\right)\\
\left(\begin{array}{cccc}
0_{3,5} & I_{3} & 0_{3,3} & 0_{3,6}\\
0_{3,5} & 0_{3,3} & I_{3} & 0_{3,6}
\end{array}\right)\\
\left(\begin{array}{cc}
0_{2,15} & I_{2}\end{array}\right)
\end{array}\right\} \label{eq:Left-Conditions}
\end{equation}

Thus, the inverse of the left-eigenvector matrix becomes:

\begin{equation}
\hat{L}^{K}{}^{-1}=\left\{ \left(\begin{array}{c}
X^{-1}I_{5,4}\\
0_{6,4}\\
Y\\
0_{2,4}
\end{array}\right),\left(\begin{array}{c}
0_{11,4}\\
\xi\left(DQ\right)^{-1}\\
0_{2,4}
\end{array}\right),\left(\begin{array}{c}
-cT_{p}\\
cT_{\rho}\\
c\Pi_{d}^{-1}\boldsymbol{w}\\
0_{12,1}
\end{array}\right),\left(\begin{array}{cc}
0_{2,3} & 0_{2,3}\\
-\Pi_{1}^{-1}\Pi_{2} & -\Pi_{1}^{-1}\Pi_{3}\\
I_{3} & 0_{3,3}\\
0_{3,3} & I_{3}\\
0_{6,3} & 0_{6,3}
\end{array}\right),\left(\begin{array}{c}
0_{15,2}\\
I_{2}
\end{array}\right)\right\} \label{eq:Right-Conditions}
\end{equation}

where:

\begin{subequations}

\begin{equation}
X=\left(\begin{array}{ccccc}
\tilde{B}_{11} & \tilde{B}_{12} & \left(-\Pi_{1}\right)_{11} & \left(-\Pi_{1}\right)_{12} & \left(-\Pi_{1}\right)_{13}\\
\tilde{B}_{21} & \tilde{B}_{22} & \left(-\Pi_{1}\right)_{21} & \left(-\Pi_{1}\right)_{22} & \left(-\Pi_{1}\right)_{23}\\
\tilde{B}_{31} & \tilde{B}_{32} & \left(-\Pi_{1}\right)_{31} & \left(-\Pi_{1}\right)_{32} & \left(-\Pi_{1}\right)_{33}\\
\tilde{D}_{11} & \tilde{D}_{12} & \tilde{C}_{11} & \tilde{C}_{12} & \tilde{C}_{13}\\
\tilde{D}_{21} & \tilde{D}_{22} & \tilde{C}_{21} & \tilde{C}_{22} & \tilde{C}_{23}
\end{array}\right)
\end{equation}

\begin{equation}
Y=-\xi Q^{-1}D^{-1}Q\Xi_{1}X^{-1}I_{5,4}
\end{equation}

\end{subequations}

and also:

\begin{subequations}

\begin{equation}
\tilde{B}=\left(\begin{array}{cc}
-\frac{\partial\boldsymbol{\sigma_{d}}}{\partial\rho} & \boldsymbol{e_{d}}\end{array}\right)
\end{equation}

\begin{equation}
\tilde{C}=\left(\begin{array}{ccc}
0 & 0 & 0\\
A_{d1}^{-1} & A_{d2}^{-1} & A_{d3}^{-1}
\end{array}\right)
\end{equation}

\begin{equation}
\tilde{D}=\left(\begin{array}{cc}
\frac{\partial T}{\partial\rho} & \frac{\partial T}{\partial p}\\
-\frac{1}{\rho} & 0
\end{array}\right)
\end{equation}

\end{subequations}

By inversion of block matrices\footnote{$\left(\begin{array}{cc}
A & B\\
C & D
\end{array}\right)^{-1}=\left(\begin{array}{cc}
\left(A-BD^{-1}C\right)^{-1} & -\left(A-BD^{-1}C\right)^{-1}BD^{-1}\\
-D^{-1}C\left(A-BD^{-1}C\right)^{-1} & D^{-1}\left(I+C\left(A-BD^{-1}C\right)^{-1}BD^{-1}\right)
\end{array}\right)$}:

\begin{equation}
X^{-1}=\left(\begin{array}{cc}
\tilde{D}^{-1}\tilde{C}Z^{-1} & \tilde{D}^{-1}\left(I-\tilde{C}Z^{-1}\tilde{B}\tilde{D}^{-1}\right)\\
-Z^{-1} & Z^{-1}\tilde{B}\tilde{D}^{-1}
\end{array}\right)
\end{equation}

where

\begin{equation}
Z=\Pi_{1}+\frac{\rho}{T_{p}}\left(T_{p}\frac{\partial\boldsymbol{\sigma_{d}}}{\partial\rho}+T_{\rho}\boldsymbol{e_{d}}\right)\boldsymbol{e_{d}^{T}}A^{-1}
\end{equation}

\subsubsection{Linear Conditions without Heat Conduction}

If the heat conduction terms are dropped from the GPR model, the eigenstructure
of the system changes, along with the solution of the linear conditions.
$\Xi$ retains the same definition, but is now a $3\times3$ matrix
(comprising the top-left corner of $\Xi$ under heat conduction).
Thus, $Q,D$ are also $3\times3$ matrices. Taking the eigenvectors
\eqref{eq:Left-Eigenvectors-Non-Thermal}, the linear conditions become:

\begin{equation}
\hat{L}^{K}=\left\{ \begin{array}{c}
\left(\begin{array}{ccccc}
-\frac{\partial\boldsymbol{\sigma_{d}}}{\partial\rho} & \boldsymbol{e_{d}} & -\Pi_{1} & -\Pi_{2} & -\Pi_{3}\end{array}\right)\\
\left(\begin{array}{cccc}
Q\Xi_{1} & -\frac{1}{\rho}Q\Pi_{2} & -\frac{1}{\rho}Q\Pi_{3} & \xi DQ\end{array}\right)\\
\left(I-BA^{-1}C\right)^{-1}\left(\begin{array}{ccccc}
I_{2} & -BA^{-1} & -BA^{-1}\Pi_{1}^{-1}\Pi_{2} & -BA^{-1}\Pi_{1}^{-1}\Pi_{3} & 0_{2,3}\end{array}\right)\\
\left(\begin{array}{cccc}
0_{3,5} & I_{3} & 0_{3,3} & 0_{3,3}\\
0_{3,5} & 0_{3,3} & I_{3} & 0_{3,3}
\end{array}\right)
\end{array}\right\} \label{eq:Left-Conditions-Non-Thermal}
\end{equation}

\begin{equation}
\hat{L}^{K}{}^{-1}=\left\{ \left(\begin{array}{c}
X^{-1}I_{5,4}\\
0_{6,3}\\
Y
\end{array}\right),\left(\begin{array}{c}
0_{11,3}\\
\xi\left(DQ\right)^{-1}
\end{array}\right),\left(\begin{array}{cc}
1 & 0\\
0 & 1\\
-\Pi_{1}^{-1}\frac{\partial\boldsymbol{\sigma_{1}}}{\partial\rho} & \Pi_{1}^{-1}\boldsymbol{e_{1}}\\
\boldsymbol{0_{9}} & \boldsymbol{0_{9}}
\end{array}\right),\left(\begin{array}{cc}
0_{2,3} & 0_{2,3}\\
-\Pi_{1}^{-1}\Pi_{2} & -\Pi_{1}^{-1}\Pi_{3}\\
I_{3} & 0_{3,3}\\
0_{3,3} & I_{3}\\
0_{3,3} & 0_{3,3}
\end{array}\right)\right\} \label{eq:Right-Conditions-Non-Thermal}
\end{equation}

where:

\begin{subequations}

\begin{equation}
X=\left(\begin{array}{ccccc}
\tilde{B}_{11} & \tilde{B}_{12} & \left(-\Pi_{1}\right)_{11} & \left(-\Pi_{1}\right)_{12} & \left(-\Pi_{1}\right)_{13}\\
\tilde{B}_{21} & \tilde{B}_{22} & \left(-\Pi_{1}\right)_{21} & \left(-\Pi_{1}\right)_{22} & \left(-\Pi_{1}\right)_{23}\\
\tilde{B}_{31} & \tilde{B}_{32} & \left(-\Pi_{1}\right)_{31} & \left(-\Pi_{1}\right)_{32} & \left(-\Pi_{1}\right)_{33}\\
\Delta_{11}^{-1} & \Delta_{12}^{-1} & \left(-\Delta^{-1}BA^{-1}\right)_{11} & \left(-\Delta^{-1}BA^{-1}\right)_{12} & \left(-\Delta^{-1}BA^{-1}\right)_{13}\\
\Delta_{21}^{-1} & \Delta_{22}^{-1} & \left(-\Delta^{-1}BA^{-1}\right)_{21} & \left(-\Delta^{-1}BA^{-1}\right)_{22} & \left(-\Delta^{-1}BA^{-1}\right)_{23}
\end{array}\right)
\end{equation}

\begin{equation}
Y=-\xi Q^{-1}D^{-1}Q\Xi_{1}X^{-1}I_{5,4}
\end{equation}

\end{subequations}

where

\begin{subequations}

\begin{align}
\Delta & =I-BA^{-1}C\\
\tilde{B} & =\left(\begin{array}{cc}
-\frac{\partial\boldsymbol{\sigma_{1}}}{\partial\rho} & \boldsymbol{e_{1}}\end{array}\right)\\
B & =\left(\begin{array}{ccc}
\rho & 0 & 0\\
\left(\rho c_{0}^{2}+\sigma_{11}-\rho\frac{\partial\sigma_{11}}{\partial\rho}\right) & \left(\sigma_{21}-\rho\frac{\partial\sigma_{21}}{\partial\rho}\right) & \left(\sigma_{31}-\rho\frac{\partial\sigma_{31}}{\partial\rho}\right)
\end{array}\right)
\end{align}

\end{subequations}

By inversion of block matrices:

\begin{equation}
X^{-1}=\left(\begin{array}{cc}
-BA^{-1}\tilde{Z} & \left(I+BA^{-1}\tilde{Z}\tilde{B}\right)\left(I-BA^{-1}\Pi_{1}^{-1}\tilde{B}\right)\\
-\tilde{Z} & \tilde{Z}\tilde{B}\left(I-BA^{-1}\Pi_{1}^{-1}\tilde{B}\right)
\end{array}\right)
\end{equation}

where

\begin{equation}
\tilde{Z}=\left(\Pi_{1}-\tilde{B}BA^{-1}\right)^{-1}
\end{equation}

\section{Results}

The GPR-RGFM is now assessed. The first fives tests in this chapter
are Riemann problems that have appeared elsewhere in the literature.
Reference solutions to these problems have been calculated by various
methods, as described for each test individually. The sixth test is
new; it assess the ability of the GPR-RGFM to correctly model heat
conduction across interfaces. The last two tests are well-known two-dimensional
problems, to demonstrate the applicability of the method to multiple
dimensions. The stiffened gas EOS parameters for three commonly-used
fluids are given in \prettyref{tab:fluid params}.

\begin{table*}[h]
\begin{centering}
\bigskip{}
\begin{tabular}{|c|c|c|c|c|c|c|c|c|}
\hline 
 &
$\rho_{0}$ &
$c_{v}$ &
\textbf{$\gamma$} &
$p_{\infty}$ &
$c_{s}$ &
\textbf{$c_{t}$} &
$\mu$ &
$P_{r}$\tabularnewline
\hline 
\hline 
Air &
1.18 &
718 &
1.4 &
- &
55 &
50 &
$1.85\times10^{-5}$ &
0.714\tabularnewline
\hline 
Helium &
0.163 &
3127 &
5/3 &
- &
55 &
50 &
$1.99\times10^{-5}$ &
0.688\tabularnewline
\hline 
Water &
997 &
950 &
4.4 &
$6\times10^{8}$ &
1 &
1 &
$10^{-3}$ &
7\tabularnewline
\hline 
\end{tabular}
\par\end{centering}
\caption{\label{tab:fluid params}EOS parameters for different fluids (using
SI units)}
\medskip{}
\end{table*}

\subsection{Helium Bubble\label{sec:Helium-Bubble}}

The interface between two different gases is now modeled. As in Test
B of Wang et al. \citep{wang_thermodynamically_2004}, a bubble of
helium - surrounded by air - initially occupies the region $x\in\left[0.4,0.6\right]$.
A shock front in the air, initially at $x=0.05$, travels towards
the helium bubble. The initial conditions are given in \prettyref{tab:HeliumBubble}
and the EOS parameters for each material are given in \prettyref{tab:fluid params}.

\begin{figure*}[p]
\begin{centering}
\includegraphics[width=0.5\textwidth]{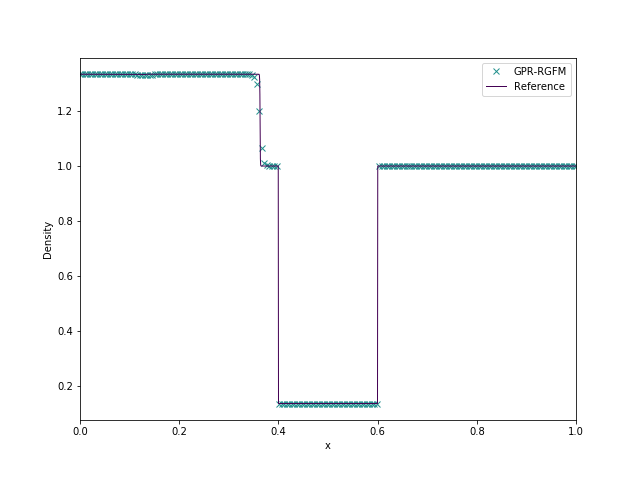}\includegraphics[width=0.5\textwidth]{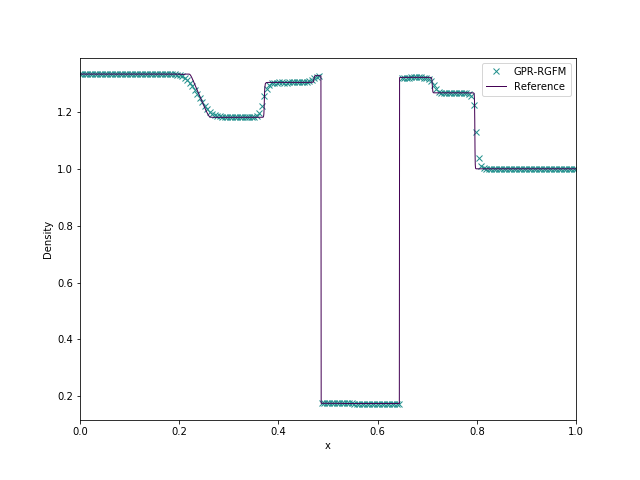}
\par\end{centering}
\begin{centering}
\includegraphics[width=0.5\textwidth]{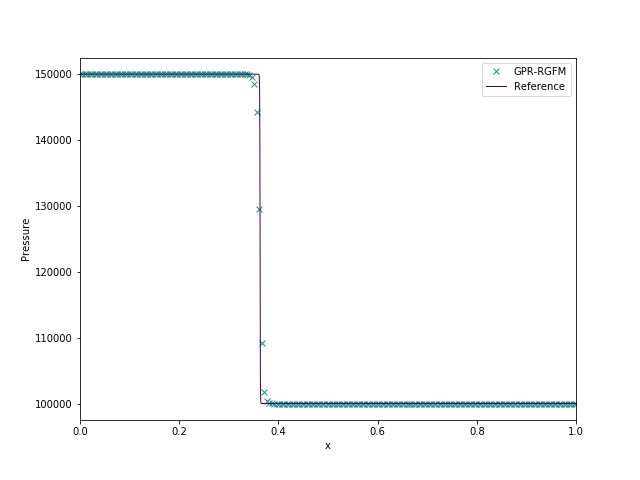}\includegraphics[width=0.5\textwidth]{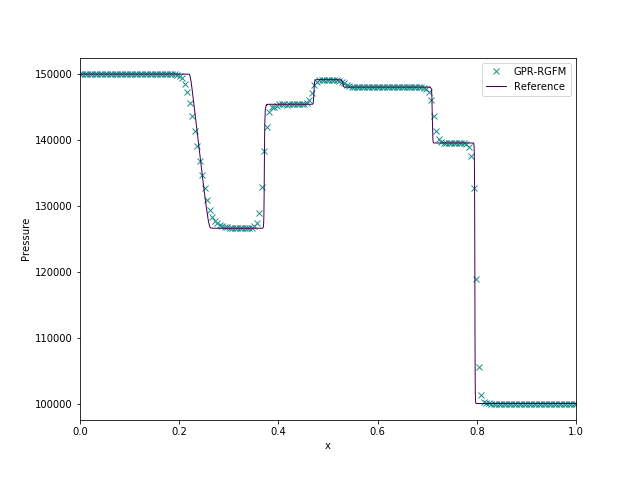}
\par\end{centering}
\begin{centering}
\includegraphics[width=0.5\textwidth]{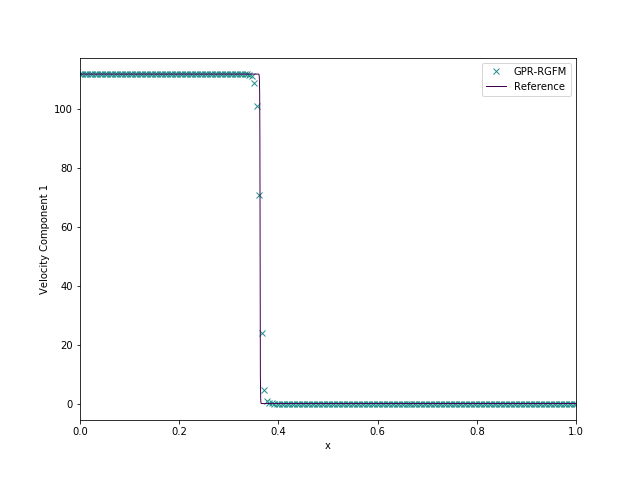}\includegraphics[width=0.5\textwidth]{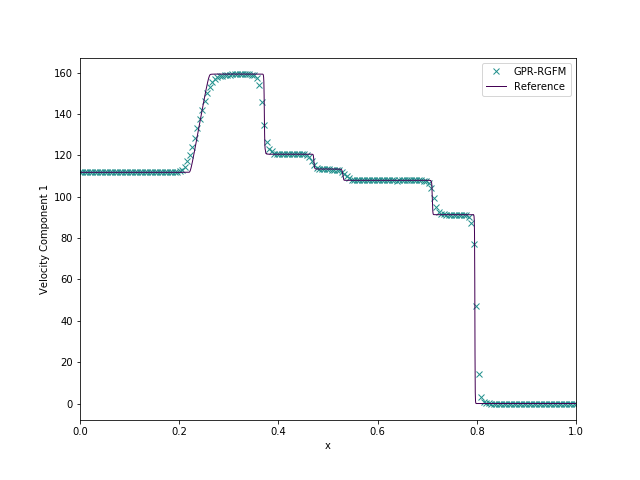}
\par\end{centering}
\caption{\label{fig:Helium-Bubble}Density, pressure, and velocity for the
helium bubble test with GPR-RGFM at times $t=7\times10^{-4}$ (left)
and $t=14\times10^{-4}$ (right)}
\end{figure*}

200 cells are used. Reference solutions are computed using the exact
solver for mixed ideal gas Riemann problems under the Euler equations
(presented in \citep{toro_reimann_2009}). The results for times $t=7\times10^{-4}$
and $t=14\times10^{-4}$ are displayed in \prettyref{fig:Helium-Bubble}.
In the former, the shock is about to hit the helium bubble (corresponding
to the region of low density). In the latter, the shock has traveled
through the helium bubble, compressing it slightly, and the bubble
itself has moved almost 0.1 spatial units to the right. There is good
correspondence with the results in \citep{wang_thermodynamically_2004}
and the sharp discontinuity in density is maintained.

\begin{table*}[h]
\begin{centering}
\bigskip{}
\begin{tabular}{|c|c|c|c|c|c|}
\hline 
 &
$\rho$ &
$p$ &
\textbf{$\boldsymbol{v}$} &
$A$ &
\textbf{$\boldsymbol{J}$}\tabularnewline
\hline 
\hline 
$x<0.05$ &
$1.3333$ &
$1.5\times10^{5}$ &
$\left(\begin{array}{ccc}
35.35\sqrt{10} & 0 & 0\end{array}\right)$ &
$\left(\frac{1.3333}{1.18}\right)^{\frac{1}{3}}I_{3}$ &
$\boldsymbol{0}$\tabularnewline
\hline 
$0.05\leq x<0.4$ &
$1$ &
$10^{5}$ &
$\boldsymbol{0}$ &
$\left(\frac{1}{1.18}\right)^{\frac{1}{3}}I_{3}$ &
$\boldsymbol{0}$\tabularnewline
\hline 
$0.4\leq x<0.6$ &
$0.1379$ &
$10^{5}$ &
$\boldsymbol{0}$ &
$\left(\frac{0.1379}{0.163}\right)^{\frac{1}{3}}I_{3}$ &
$\boldsymbol{0}$\tabularnewline
\hline 
$0.6\leq x\leq1$ &
$1$ &
$10^{5}$ &
$\boldsymbol{0}$ &
$\left(\frac{1}{1.18}\right)^{\frac{1}{3}}I_{3}$ &
$\boldsymbol{0}$\tabularnewline
\hline 
\end{tabular}
\par\end{centering}
\caption{\label{tab:HeliumBubble}Initial conditions for the helium bubble
test}
\medskip{}
\end{table*}

\subsection{Water-Air Shock Tube\label{subsec:Water-Air-Shock-Tube}}

This test comprises an interface between water and air, with initial
data taken from Chinnayya et al. \citep{chinnayya_modelling_2004}
(see \prettyref{tab:WaterAirShockTube}). The aim of this test is
to evaluate the ability of the GPR-RGFM at capturing interfaces between
qualitatively different fluids. The water is initially at high pressure,
and the air at atmospheric pressure. Due to the large difference in
state variables and qualitative characteristics of the two fluids,
this is an example of a test with which the original GFM for the Euler
equations does not perform well.

\begin{figure*}[p]
\begin{centering}
\includegraphics[width=0.5\textwidth]{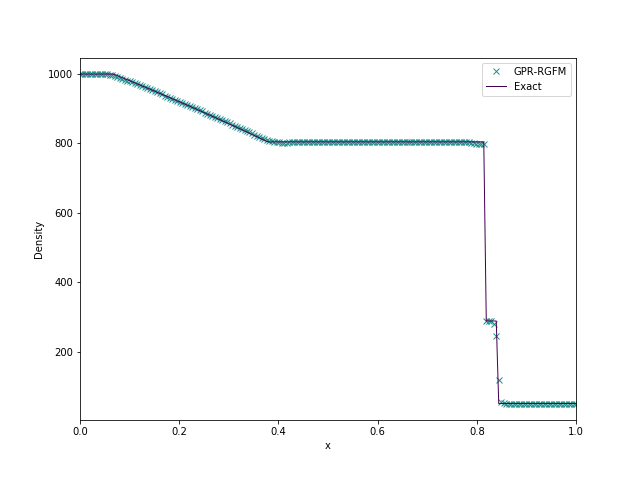}\includegraphics[width=0.5\textwidth]{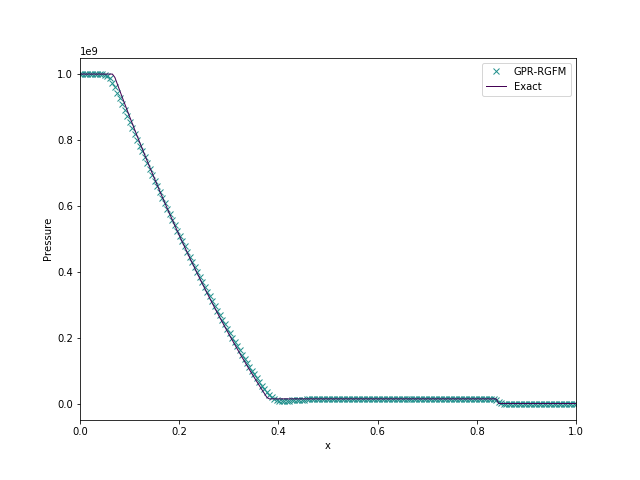}
\par\end{centering}
\begin{centering}
\includegraphics[width=0.5\textwidth]{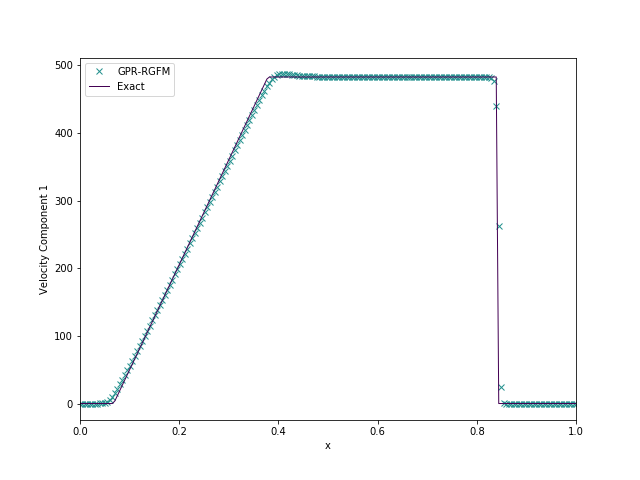}\includegraphics[width=0.5\textwidth]{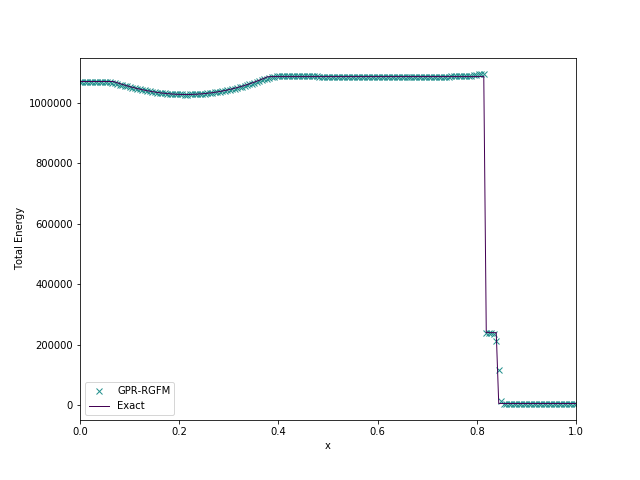}
\par\end{centering}
\caption{\label{fig:Water-Air}Density, pressure, velocity, and internal energy
for the water-air shock tube test with GPR-RGFM}
\end{figure*}

The results using the GPR-RGFM with 200 cells are shown in \prettyref{fig:Water-Air},
along with the exact solution to the Euler equations (computed using
the extension to the stiffened gas equations of the exact Riemann
solver presented in \citep{toro_reimann_2009}). As can be seen, the
material interface is captured well, with the correct intermediate
density found by the numerical method.

\begin{table}
\begin{centering}
\bigskip{}
\begin{tabular}{|c|c|c|c|c|c|}
\hline 
 &
$\rho$ &
$p$ &
\textbf{$\boldsymbol{v}$} &
$A$ &
\textbf{$\boldsymbol{J}$}\tabularnewline
\hline 
\hline 
$0\leq x<0.7$ &
$1000$ &
$10^{9}$ &
$\boldsymbol{0}$ &
$I_{3}$ &
$\boldsymbol{0}$\tabularnewline
\hline 
$0.7\leq x\leq1$ &
$50$ &
$10^{5}$ &
$\boldsymbol{0}$ &
$\sqrt[3]{50}\cdot I_{3}$ &
$\boldsymbol{0}$\tabularnewline
\hline 
\end{tabular}
\par\end{centering}
\caption{\label{tab:WaterAirShockTube}Initial conditions for the water-air
shock tube test}
\medskip{}
\end{table}

\subsection{PBX9404-Copper Shock Tube\label{subsec:PBX9404-Copper-Shock-Tube}}

This test is taken from \citep{barton_conservative_2011}, with the
aim of testing the ability of the GPR-RGFM to model interfaces between
fluids and solids. High pressure, reacted PBX9404 is in contact with
copper at position $x=0.5$ on domain $x\in\left[0,1\right]$, with
both materials initially at rest. The pressure of the PBX is initially
$18.9GPa$, and the entropy of the copper is initially $0$. The PBX
follows an ideal gas EOS, with parameters $\rho_{0}=1840$, $\gamma=2.85$,
$c_{s}=1$, $\mu=10^{-2}$. The copper follows the Godunov-Romenski
EOS, with parameters $\rho_{0}=8930$, $c_{v}=390$, $T_{0}=300$,
$c_{0}=3939$, $\alpha=1$, $\beta=3$, $\gamma=2$, $b_{0}=2141$.
The test is run until time $t=0.5\times10^{-6}$, using 500 cells.

The exact solution to this test is calculated using the iterative
solver described in \citep{barton_exact_2009}. The error in the wavespeeds
is calculated from the residual error in the traction and velocities
across the central contact, as required by the Rankine\textendash Hugoniot
conditions and boundary conditions. The wavespeeds are found by iteratively
reducing the residual using the Newton\textendash Raphson method.

\begin{figure*}[p]
\begin{centering}
\includegraphics[width=0.5\textwidth]{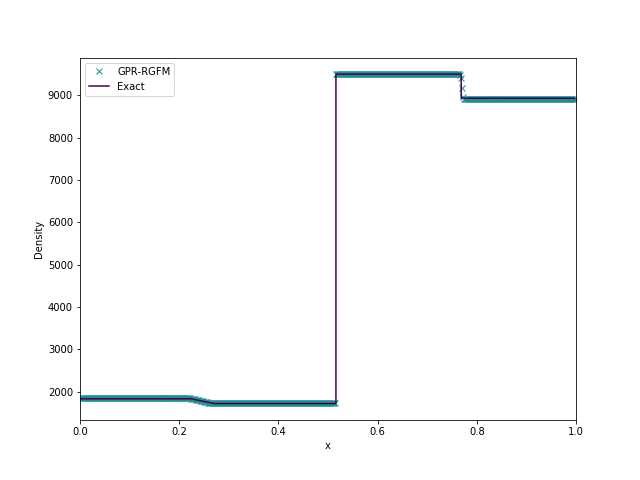}
\par\end{centering}
\begin{centering}
\includegraphics[width=0.5\textwidth]{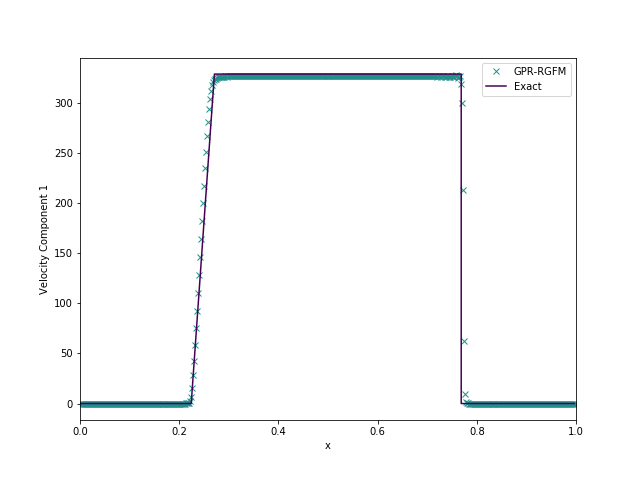}
\par\end{centering}
\begin{centering}
\includegraphics[width=0.5\textwidth]{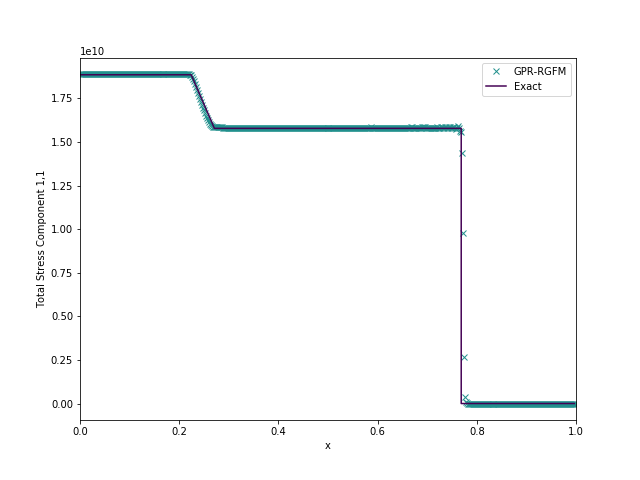}
\par\end{centering}
\caption{\label{fig:Copper-PBX}Density, velocity, and total stress for the
Copper-PBX test with GPR-RGFM}
\end{figure*}

Plots for density, velocity, and total stress are given in \prettyref{fig:Copper-PBX}.
As can be seen, the GPR-RGFM is able to reproduce the solution to
high fidelity, with a perfectly sharp discontinuity in the density,
and a very well resolved discontinuity in the total stress.

\subsection{Aluminium in Vacuum\label{subsec:Aluminium-in-Vacuum}}

This test is taken from \citep{barton_eulerian_2010}. The initial
conditions of the test consist of a slab of aluminium, initially with
velocity $\left(\begin{array}{ccc}
2 & 0 & 0.1\end{array}\right)$, meeting a vacuum at point $x=0.5$, on the domain $x\in\left[0,1\right]$.
The distortion of the aluminium is initially given by:

\begin{equation}
A=\left(\begin{array}{ccc}
1 & 0 & 0\\
-0.01 & 0.95 & 0.02\\
-0.015 & 0 & 0.9
\end{array}\right)^{-1}
\end{equation}

The initial density of the aluminium is thus given as $\rho=\rho_{0}\det\left(A\right)$.
The aluminium is modeled using the Godunov-Romenski EOS, with parameters
$\rho_{0}=2.71$, $c_{v}=9\times10^{-4}$, $T_{0}=300$, $c_{0}=5.037$,
$\alpha=1$, $\beta=3.577$, $\gamma=2.088$, \textbf{$b_{0}=3.16$},
$c_{t}=2$, $\kappa=204$.

\begin{figure*}[p]
\begin{centering}
\includegraphics[width=0.5\textwidth]{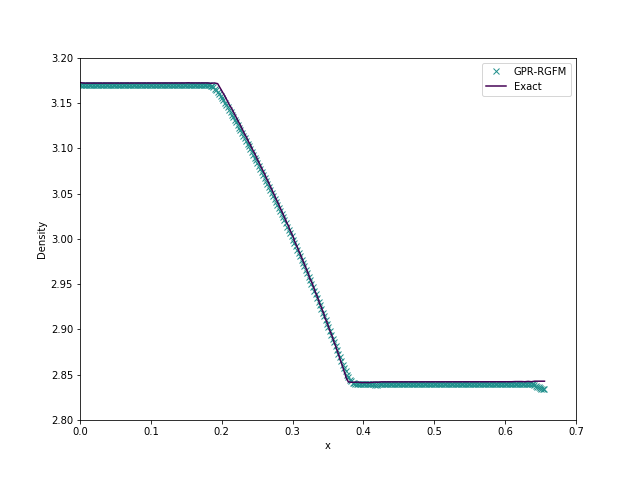}
\par\end{centering}
\begin{centering}
\includegraphics[width=0.5\textwidth]{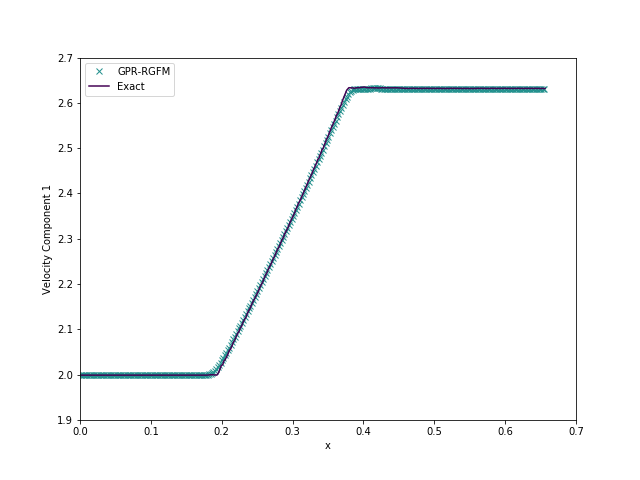}\includegraphics[width=0.5\textwidth]{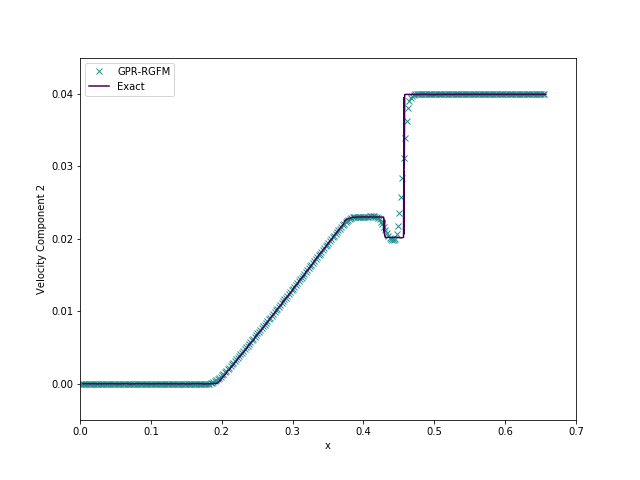}
\par\end{centering}
\begin{centering}
\includegraphics[width=0.5\textwidth]{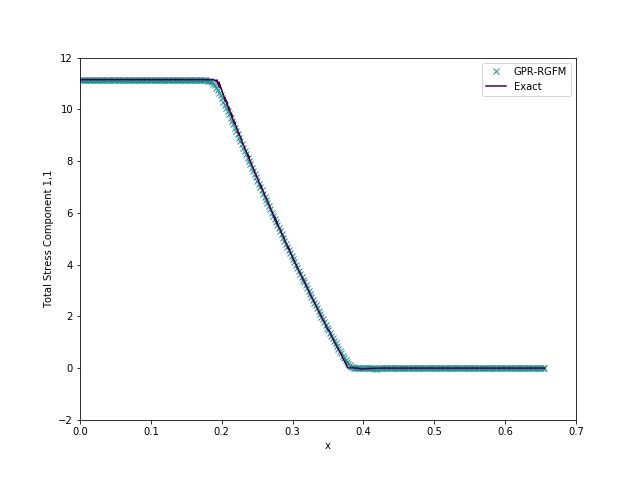}\includegraphics[width=0.5\textwidth]{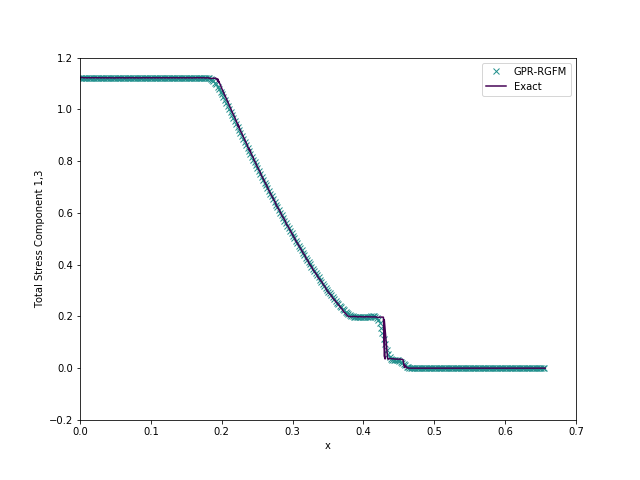}
\par\end{centering}
\caption{\label{fig:Aluminium-in-Vacuum}Density, velocity, and total stress
for the aluminium-vacuum test with GPR-RGFM, not including thermal
conduction}
\end{figure*}

\begin{figure*}[p]
\begin{centering}
\includegraphics[width=0.5\textwidth]{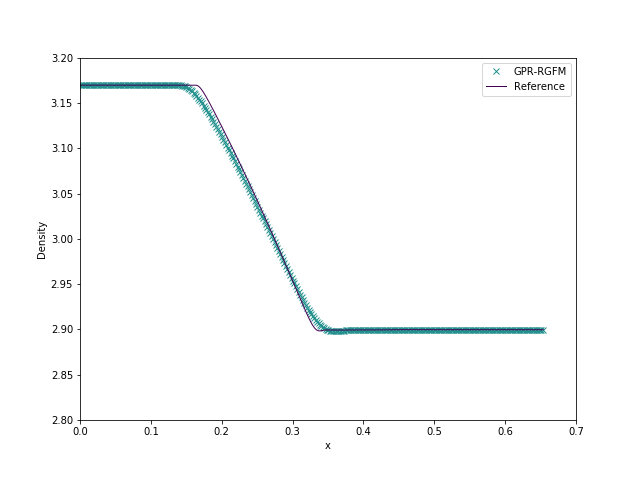}
\par\end{centering}
\begin{centering}
\includegraphics[width=0.5\textwidth]{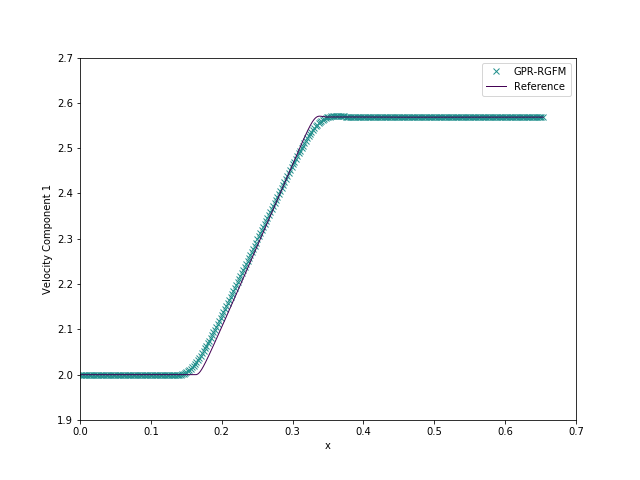}
\par\end{centering}
\begin{centering}
\includegraphics[width=0.5\textwidth]{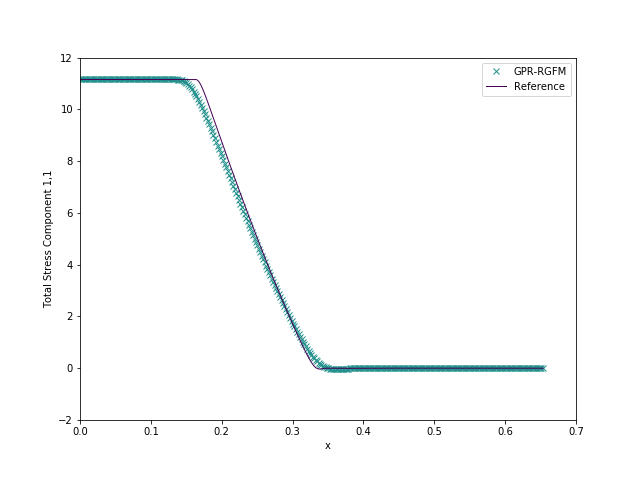}
\par\end{centering}
\caption{\label{fig:Aluminium-in-Vacuum-Thermal}Density, velocity, and total
stress for the aluminium-vacuum test with GPR-RGFM, including thermal
conduction}
\end{figure*}

The test was run until time $t=0.06$, using 500 cells. The results
of solving this problem with the GPR-RGFM, not including thermal conduction
(as in \citep{barton_eulerian_2010}), are given in \prettyref{fig:Aluminium-in-Vacuum}.
The results of solving the problem, including thermal conduction,
are given in \prettyref{fig:Aluminium-in-Vacuum-Thermal}. The exact
solutions are calculated using the iterative method presented in \citep{barton_exact_2009},
as described in the previous test.

As can be seen, in both cases, the GPR-RGFM is able to accurately
capture the longitudinal wave and the two transverse shock waves that
propagate to the left side of the initial point of contact. Note that
at $t=0.06$, the vacuum occupies the region $\left[\sim0.65,1\right]$.
As this region is empty, the plots in the aforementioned figures are
shown over the interval $\left[0,0.7\right]$, to give greater resolution
to the region of interest.

Without thermal conduction, the interface suffers from a ``heating
error'' of the same kind discussed in \citep{barton_eulerian_2010},
manifesting itself as a slight undershoot in the density of the metal
at the interface. Note that by incorporating thermal conduction into
the numerical method, this heating error completely disappears, without
the use of an entropy fix (as in \citep{barton_eulerian_2010}). It
must be noted that, in this case, the waves in the state variables
now appear to be slightly more diffused than the reference solution.
This is the expected effect of incorporating the phenomenon of thermal
conduction into this physical problem.


\subsection{Heat Conduction in a Gas\label{sec:HeatConductionInAGas}}

This test is based on the Heat Conduction in a Gas Test of Dumbser
et al. \citep{dumbser_high_2015}. Two ideal gases at different temperatures
are initially in contact at position $x=0$. The initial conditions
for this problem are given in \prettyref{tab:HeatConduction-1}.

\begin{table}
\begin{centering}
\bigskip{}
\begin{tabular}{|c|c|c|c|c|c|}
\hline 
 &
$\rho$ &
$p$ &
\textbf{$\boldsymbol{v}$} &
$A$ &
\textbf{$\boldsymbol{J}$}\tabularnewline
\hline 
\hline 
$x<0$ &
$2$ &
$1$ &
$\boldsymbol{0}$ &
$\sqrt[3]{2}\cdot I_{3}$ &
$\boldsymbol{0}$\tabularnewline
\hline 
$x\geq0$ &
$0.5$ &
$1$ &
$\boldsymbol{0}$ &
$\frac{1}{\sqrt[3]{2}}\cdot I_{3}$ &
$\boldsymbol{0}$\tabularnewline
\hline 
\end{tabular}
\par\end{centering}
\caption{\label{tab:HeatConduction-1}Initial conditions for the heat conduction
test}
\medskip{}
\end{table}

\begin{figure*}[p]
\begin{centering}
\includegraphics[width=0.5\textwidth]{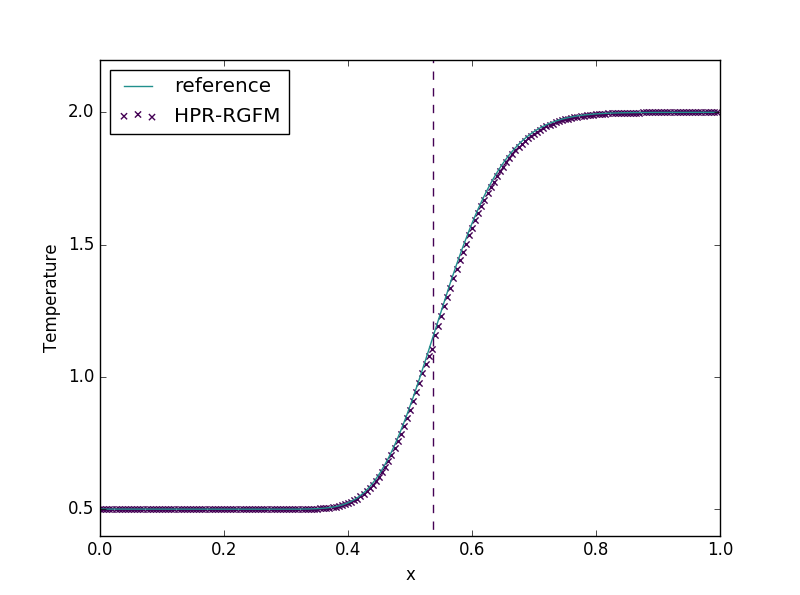}
\par\end{centering}
\begin{centering}
\includegraphics[width=0.5\textwidth]{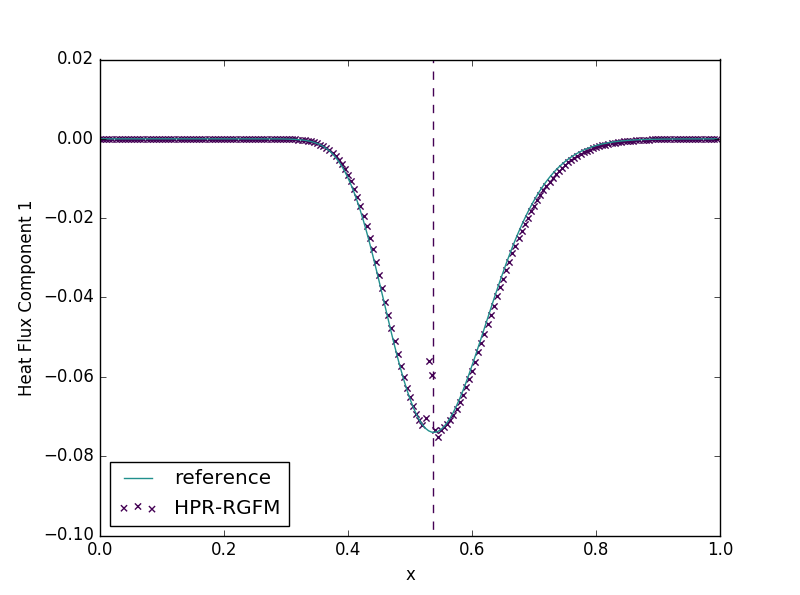}
\par\end{centering}
\begin{centering}
\includegraphics[width=0.5\textwidth]{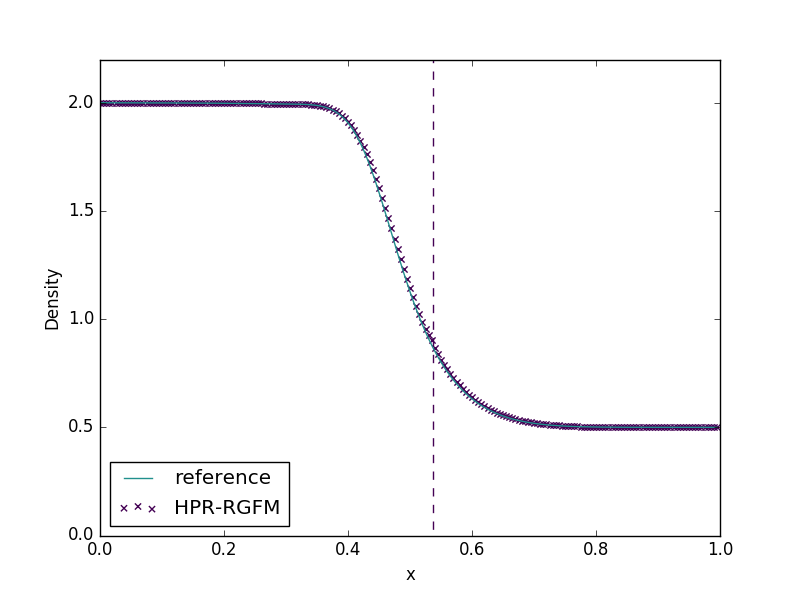}
\par\end{centering}
\caption{\label{fig:HeatFluxInAGas}Temperature, heat flux, and density for
the intermaterial heat conduction test with GPR-RGFM}
\end{figure*}

The material parameters are taken to be: $\gamma=1.4$, $c_{v}=2.5$,
$\rho_{0}=1$, $p_{0}=1$, $c_{s}=1$, $c_{t}=1$, $\mu=10^{-2}$,
$\kappa=10^{-2}$. An interface is initially placed between the two
volumes of air at $x=0.5$. The final time is taken to be $t=1$,
and 200 cells are used. Results are displayed in \prettyref{fig:HeatFluxInAGas},
using the results from \citep{dumbser_high_2015} as a reference.
The material interface is denoted by a dashed vertical line.

The temperature curve generated using the GPR-RGFM matches very well
the reference solution. The interface has moved to $x=0.53756$, as
is to be expected, as the cooler gas on the left expands as it heats
up, and the hotter gas on the right contracts as it cools. Initially,
the mass of the left volume is 1 and the right volume is 0.25. At
$t=1$, these masses are 0.9997 and 0.2503, respectively. Thus, mass
on either side is conserved to a good approximation. Although the
GPR-RGFM results for the heat flux match the reference solution well
over most of the domain, there are aberrations in a small region around
the interface. Although this doesn't affect the temperature curve,
these discrepancies are undesirable, and possible methods to rectify
them are discussed in \prettyref{sec:Discussion}.

\subsection{Taylor Bar}

This follows a similar form to that found in \citep{boscheri_cell_2016,maire_nominally_2013}.
A bar of aluminium of dimensions $100\times500$ travels towards a
solid wall at speed $0.015$. The surrounding environment is a vacuum.
The aluminium bar is modelled by the shock Mie-Gruneisen equation
of state, with parameters $\rho_{0}=2.785,$ $c_{v}=9\times10^{-4}$,
$c_{0}=0.533$, $\Gamma_{0}=2$, $s=1.338$. The aluminium also follows
a plasticity law with parameters $b_{0}=0.305$, $\sigma_{Y}=0.003$,
$\tau_{0}=1$, $n=20$. The domain has dimensions $300\times510$,
with $\Delta x,\Delta y=1$.

\begin{figure}
\begin{centering}
\includegraphics[width=0.5\textwidth]{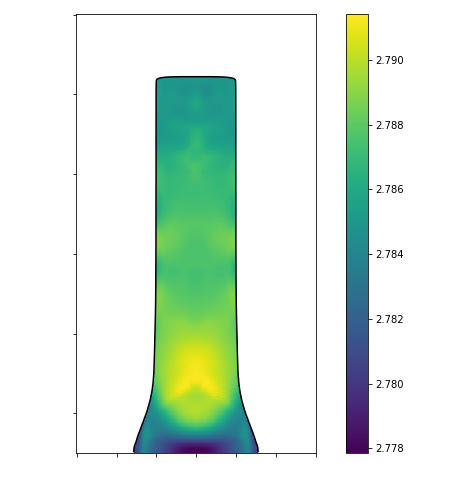}\includegraphics[width=0.5\textwidth]{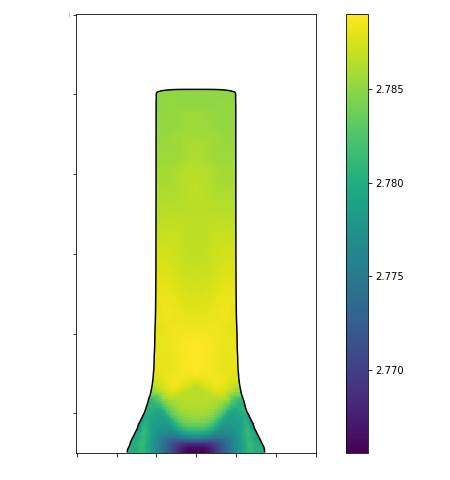}
\par\end{centering}
\begin{centering}
\includegraphics[width=0.5\textwidth]{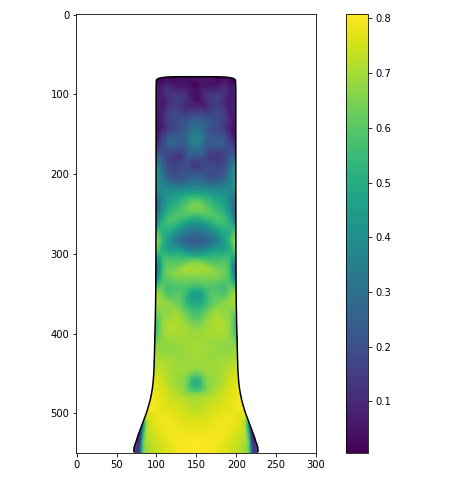}\includegraphics[width=0.5\textwidth]{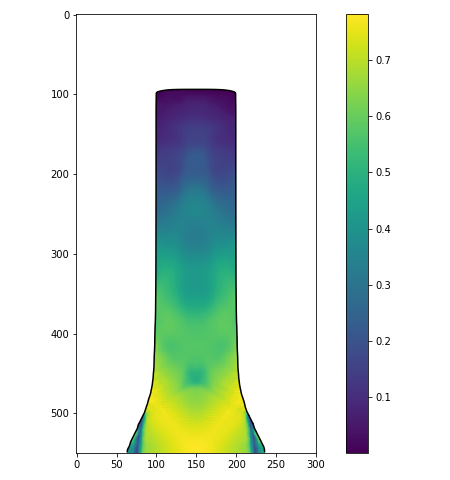}
\par\end{centering}
\caption{\label{fig:The-Taylor-bar}Density (top) and plastic deformation (bottom)
for the Taylor bar test, at times $t=0.0025$ (left) and $t=0.005$
(right)}
\end{figure}

The density and plastic deformation of the bar at times $t=0.0025$
and $t=0.005$ are shown in \prettyref{fig:The-Taylor-bar}. Unfortunately
there are no experimental results for this test, but the reader is
asked to note the good agreement here with the results found in \citep{maire_nominally_2013}.
In that study, the boundary between the bar and the vacuum is captured
using a Lagrangian scheme, and it is reassuring that the same behaviour
is captured here with a characteristically different numerical method.

\subsection{Aluminum Plate Impact}

This test follows the form found in Michael \& Nikiforakis \citep{michael_multi-physics_2018}
(based on the original formulation found in \citep{howell_free_2002}).
An aluminum projectile impacts upon an aluminum plate at speed $400$.
The domain is $\left[0,0.03\right]\times\left[0,0.04\right]$, with
the projectile initially occupying $\left[0.001,0.006\right]\times\left[0.014,0.026\right]$,
and the plate occupying $\left[0.006,0.028\right]\times\left[0.003,0.037\right]$.
We have $\Delta x,\Delta y=10^{-4}$. The surroundings are taken to
be a vacuum. The aluminium follows a Godunov-Romenski EOS with parameters
$\rho_{0}=2710$, $c_{v}=900$, $T_{0}=300$, $c_{0}=5037$, $\alpha=1$,
$\beta=3.577$, $\gamma=2.088$, $b_{0}=3160$, $\sigma_{Y}=4\times10^{8}$,
$\tau_{0}=1$, $n=100$. Gauges are placed initially at $x=0.0078125$,
$0.0114375$, $0.0150625$, $0.0186875$, $0.0223125$ to measure
the state variables over time, and these gauges are permitted to move
with the local velocity of the material. The test is run until time
$t=5\times10^{-6}$.

\begin{figure}
\begin{centering}
\includegraphics[width=0.5\textwidth]{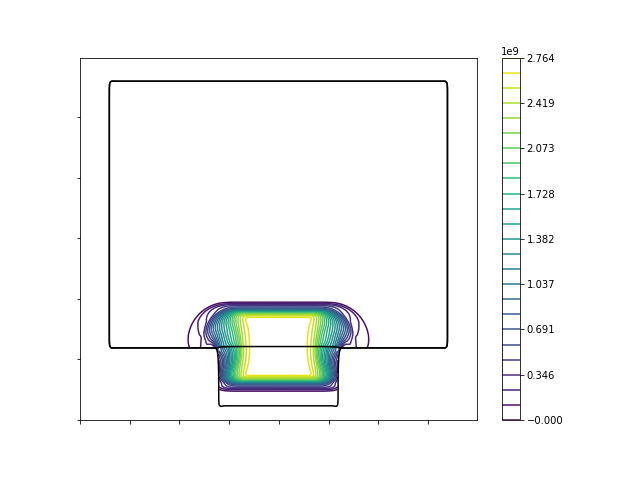}\includegraphics[width=0.5\textwidth]{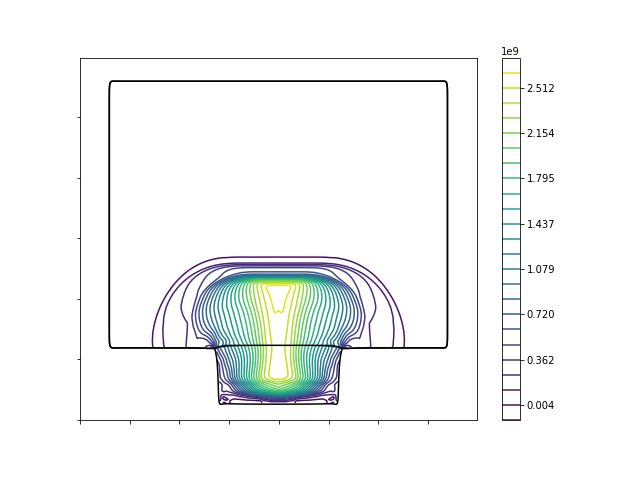}
\par\end{centering}
\begin{centering}
\includegraphics[width=0.5\textwidth]{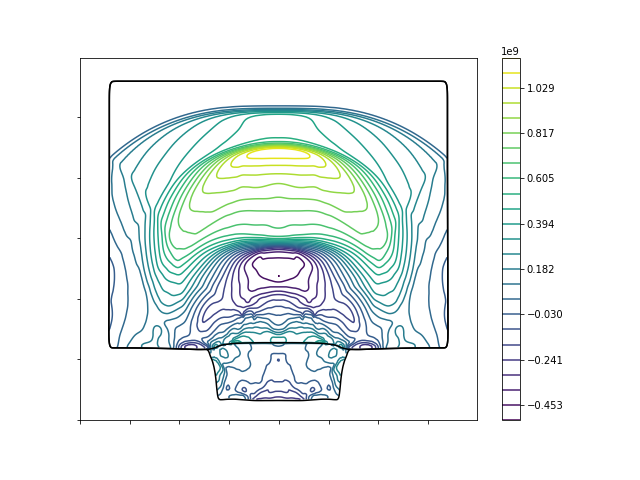}\includegraphics[width=0.5\textwidth]{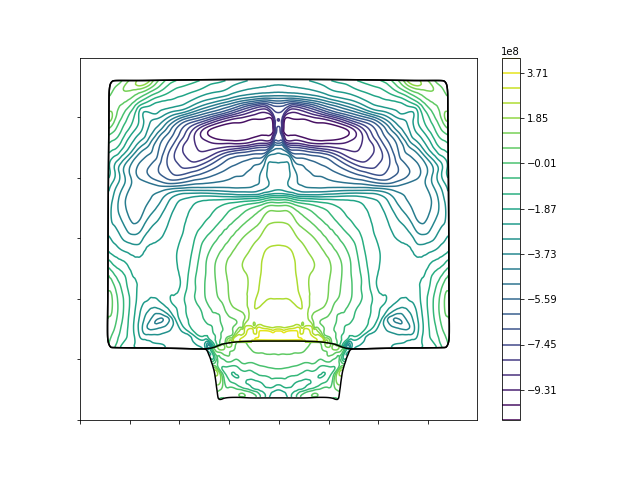}
\par\end{centering}
\caption{\label{fig:The-aluminium-plate}Pressure contour plots for the aluminium
plate impact test, at times $0.5\mu s$, $1\mu s$, $3\mu s$, $5\mu s$}
\end{figure}

The pressure contours throughout the aluminium at various times are
shown in \prettyref{fig:The-aluminium-plate}. Despite relying on
a slightly different plasticity model to that found in \citep{michael_multi-physics_2018},
it can be seen that these plots are in very good agreement with those
found in the aforementioned publication. Note that release waves can
be seen on the sides of the plate, in agreement with Michael \& Nikiforakis.

\begin{figure}
\begin{centering}
\includegraphics[width=0.5\textwidth]{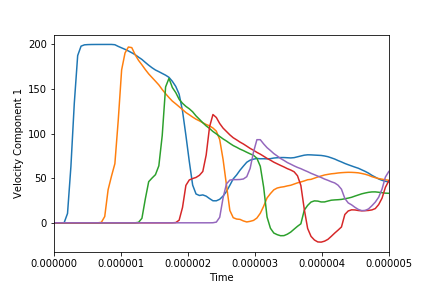}\includegraphics[width=0.5\textwidth]{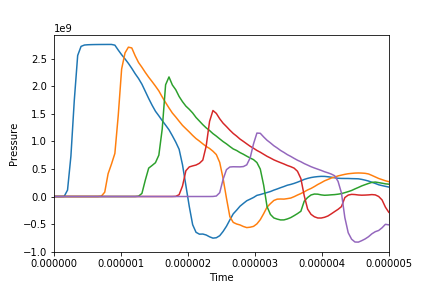}
\par\end{centering}
\begin{centering}
\includegraphics[width=0.5\textwidth]{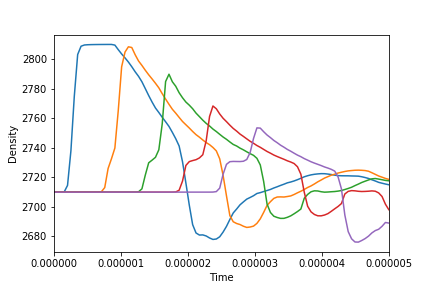}\includegraphics[width=0.5\textwidth]{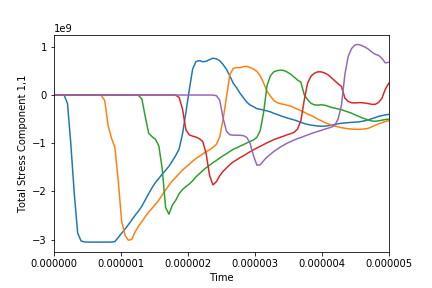}
\par\end{centering}
\caption{\label{fig:The-aluminium-plate-2}$x$-velocity, pressure, density,
and total stress over time, as measured by the various gauges of the
aluminium plate impact test}
\end{figure}

Plots over time of the $x$-velocity, pressure, density, and total
stress - as measured by the gauges - are given in \prettyref{fig:The-aluminium-plate-2}.
Note the good agreement between these plots and those found in \citep{michael_multi-physics_2018,howell_free_2002},
both in terms of their qualitative shape, and the arrival times of
the waves that they represent. One can clearly see the separation
between the elastic precursor wave and the trailing plastic wave in
the impacted plate, and the subsequent return waves that are generated
once these waves reach the end of the plate. This implies that the
GPR-RGFM has correctly captured the aluminium-vacuum interface. 

\subsection{Confined C4 Detonation without Backplate}

This test is a variation of that found in \citep{michael_multi-physics_2018}.
A steel bar of length 0.03 and width 0.018 impacts upon a steel plate
of depth 0.003, which is covering a region of depth 0.009 composed
of C4 . The bar is initially traveling with speed 700. The system
is surrounded by air.

The steel is modeled using a shock Mie-Gruneisen EOS, with parameters
$\rho_{0}=7870$, $c_{v}=134$, $c_{0}=4569$, $\Gamma_{0}=2.17$,
$s=1.49$, $c_{s}=3235$, $\sigma_{Y}=0.53\times10^{9}$, $\tau_{0}=1$,
$n=10$. The C4 is modeled using a JWL EOS, with parameters $\rho_{0}=1601$,
$c_{v}=2.487\times10^{6}/1601$, $\Gamma_{0}=0.8938$, $A=7.781\times10^{13}$,
$B=-5.031\times10^{9}$, $R_{1}=11.3$, $R_{2}=1.13$, $c_{s}=1487$.
The air is modeled using an ideal gas EOS, with parameters $\rho_{0}=1.18$,
$c_{v}=718$, $\gamma=1.4$, $c_{s}=50$, $\mu=1.85\times10^{-5}$.

The reaction of the C4 is captured using the ignition and growth model
\citep{lee_phenomenological_1980}, where total energy $E$ is modified
to include the term:

\begin{equation}
E_{r}\left(\lambda\right)=-Q\left(1-\lambda\right)
\end{equation}

with $\lambda$ being the volume fraction of unreacted material, governed
by the dynamical equation:

\begin{subequations}

\begin{align}
\frac{\partial\left(\rho\lambda\right)}{\partial t}+\frac{\partial\left(\rho\lambda v_{k}\right)}{\partial x_{k}}= & -\rho K\\
K= & I\lambda^{b}\left(\frac{\rho}{\rho_{0}}-1-a\right)^{x}H\left(\phi_{I}-\phi\right)\\
 & +G_{1}\lambda^{c}\phi^{d}p^{y}H\left(\phi_{G1}-\phi\right)\nonumber 
\end{align}

\end{subequations}

In this test, the parameters are taken to be $Q=9\times10^{9}/1601$,
$I=4\times10^{6}$, $G_{1}=1.4\times10^{-20}$, $a=0.0367$, $b=2/3$,
$c=2/3$, $d=1/3$, $x=7$, $y=2$, $\phi_{I}=0.022$, $\phi_{G1}=1$.

\begin{figure}
\begin{centering}
\includegraphics[width=0.5\textwidth]{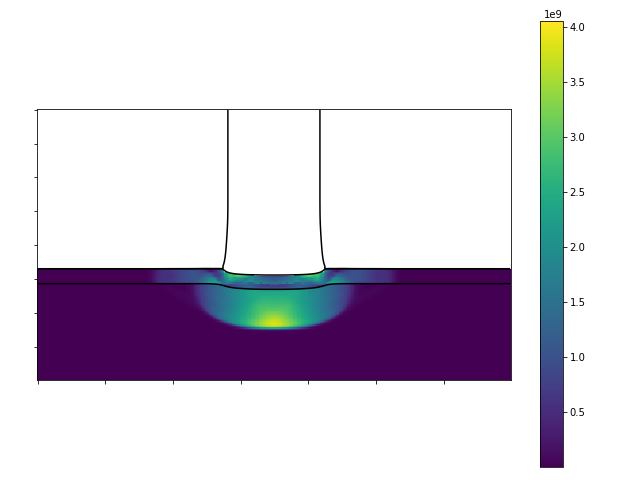}\includegraphics[width=0.5\textwidth]{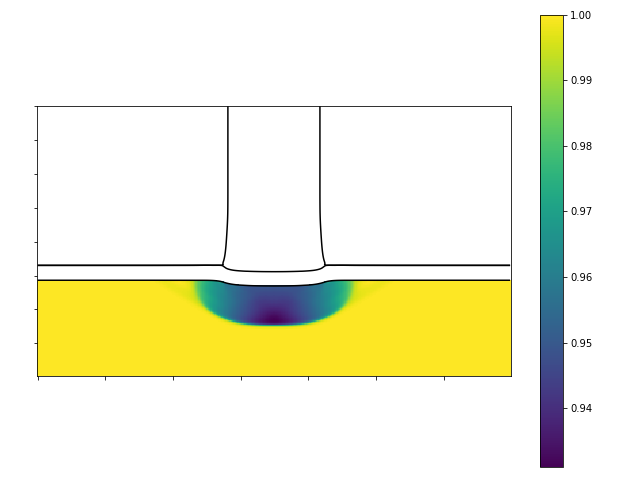}
\par\end{centering}
\begin{centering}
\includegraphics[width=0.5\textwidth]{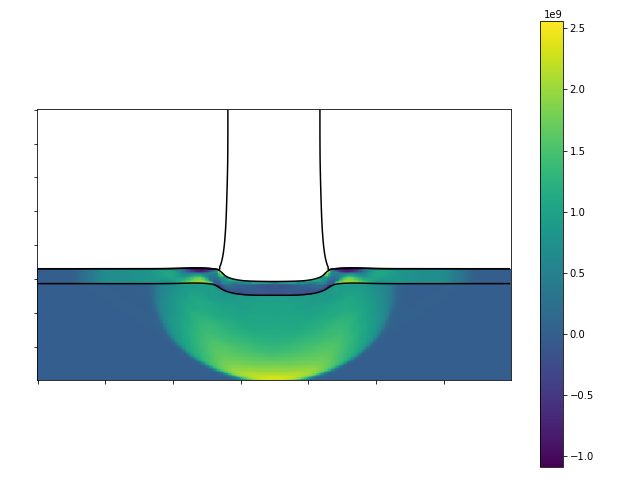}\includegraphics[width=0.5\textwidth]{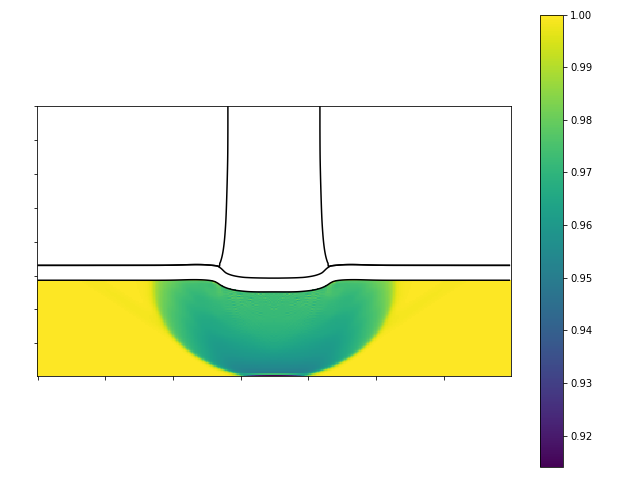}
\par\end{centering}
\caption{\label{fig:confined-detonation-no-backplate}Pressure (left) and reactant
concentration (right) for the confined detonation test (without back
plate), at times $2.4\mu s$ (top) and $4.9\mu s$ (bottom)}
\end{figure}

\prettyref{fig:confined-detonation-no-backplate} displays the resulting
pressure and C4 concentration at times $t=2.4\times10^{-6}$ and $t=4.9\times10^{-6}$.
As can be seen, the kinetic energy of the bar is correctly transmitted
to the steel plate, with the plate deforming in a manner qualitatively
identical to that found in \citep{michael_multi-physics_2018}. This
energy is in turn transmitted to the C4, leading to an exothermic
reaction and a symmetrical wavefront that travels through the material.
The C4 concentration is depleted to 0.93 at time $t=2.4\times10^{-6}$
and to 0.915 at time $t=4.9\times10^{-6}$.

\subsection{Confined C4 Detonation}

\begin{figure}
\begin{centering}
\includegraphics[width=0.5\textwidth]{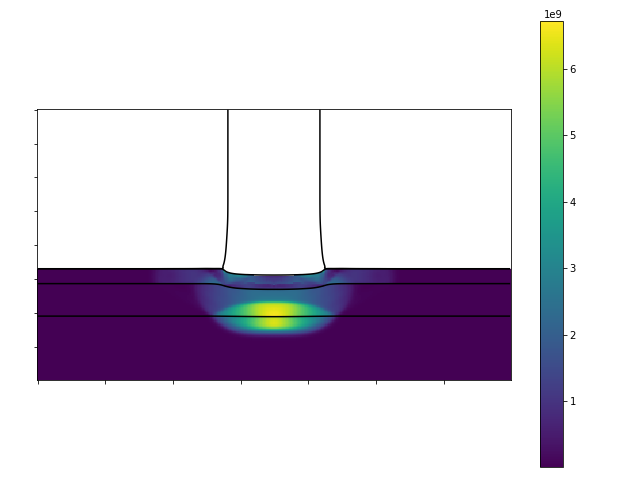}\includegraphics[width=0.5\textwidth]{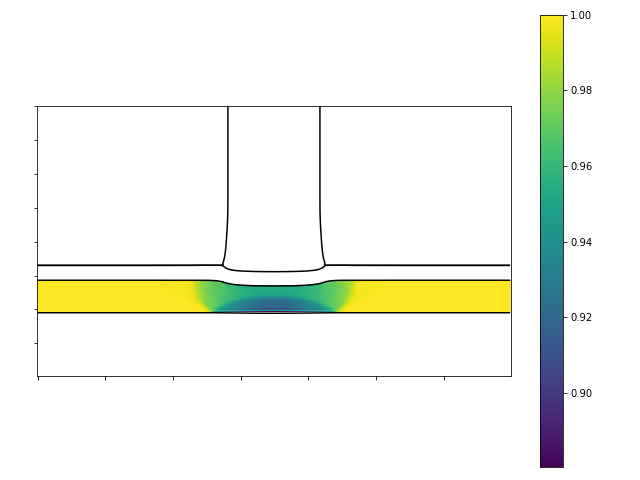}
\par\end{centering}
\begin{centering}
\includegraphics[width=0.5\textwidth]{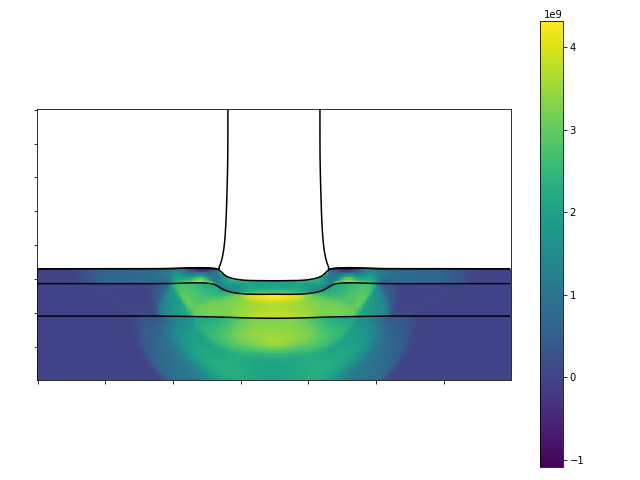}\includegraphics[width=0.5\textwidth]{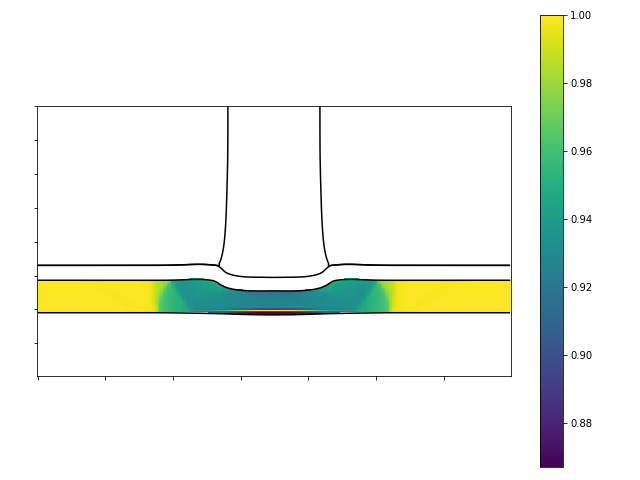}
\par\end{centering}
\caption{\label{fig:confined-detonation}Pressure (left) and reactant concentration
(right) for the confined detonation test, at times $2.4\mu s$ (top)
and $4.9\mu s$ (bottom)}
\end{figure}

This test is identical to the previous test, except a steel plate
of depth 0.003 is now placed behind the C4, so that the explosive
is entirely confined. As can be seen from \prettyref{fig:confined-detonation},
the kinetic energy of the bar is once again correctly transmitted
to the steel plate and C4, with the same deformation occurring in
the first steel plate. At time $t=2.4\times10^{-6}$ we see the wave
in the C4 both partially rebounding off the backplate back into the
reactant, and partially traveling on through the backplate. At the
earlier time, the reactant concentration has been depleted to 0.88,
and at the later time to 0.865. This corroborates the results of \citep{michael_multi-physics_2018},
in that the presence of the backplate accelerates the reactive processes.

\subsection{Confined C4 Detonation with Air Gap}

This problem is designed to test the ability of the framework presented
in this paper to capture the interaction of widely varying media.
It is identical to the previous problem, except an air gap is now
placed between the first steel plate and the C4. The air has the same
EOS parameters as the surrounding air.

\begin{figure}
\begin{centering}
\includegraphics[width=0.5\textwidth]{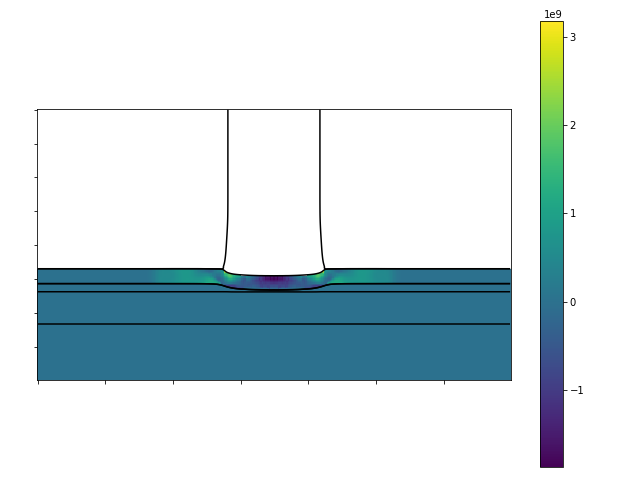}\includegraphics[width=0.5\textwidth]{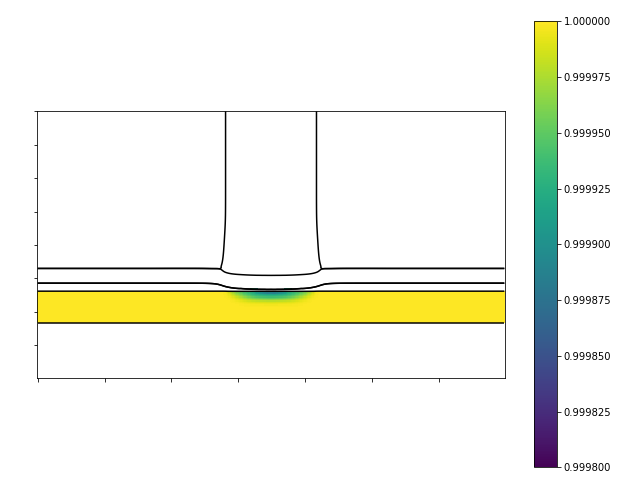}
\par\end{centering}
\begin{centering}
\includegraphics[width=0.5\textwidth]{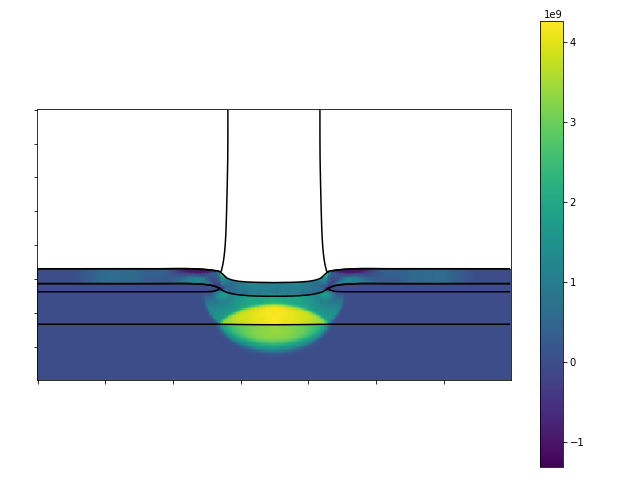}\includegraphics[width=0.5\textwidth]{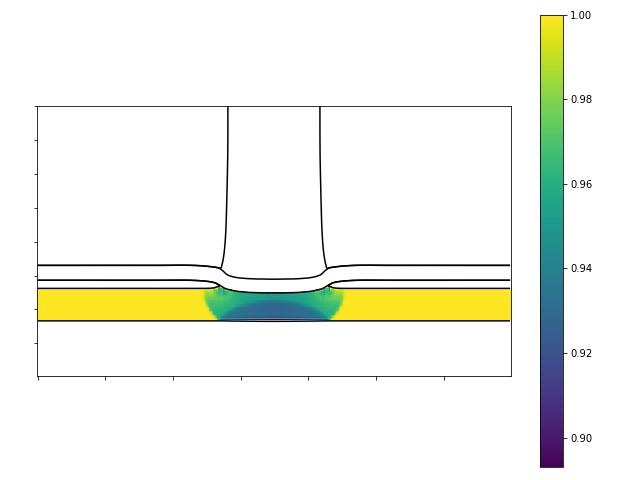}
\par\end{centering}
\caption{\label{fig:confined-detonation-airgap}Pressure (left) and reactant
concentration (right) for the confined detonation test (with air gap),
at times $2.4\mu s$ (top) and $4.9\mu s$ (bottom)}
\end{figure}

As can be seen in \prettyref{fig:confined-detonation-airgap}, the
rod displaces the air (with the numerical method coping with contact
of the region representing the plate with the region representing
the C4). The displacement of the air enables the kinetic energy of
the rod to be transmitted through the plate and into the C4, as before.
The earlier time of $t=2.4\times10^{-6}$ corresponds with the instant
after the plate makes contact with the C4. The concentration is depleted
to 0.9998 at $t=2.4\times10^{-6}$ and 0.894 at $t=4.9\times10^{-6}$.
The latter value is lower than the corresponding value in the previous
test, as the reaction has been delayed by the presence of the air
gap.

\subsection{Convergence Study}

In order to determine the order of convergence of the Riemann Ghost
Fluid Method presented in this study, the tests from \prettyref{subsec:Water-Air-Shock-Tube},
\prettyref{subsec:PBX9404-Copper-Shock-Tube}, and \prettyref{subsec:Aluminium-in-Vacuum}
were run with cell counts of $50,100,150,200,250,300$, and the error
as compared with the exact solutions were calculated at the interfaces.
The reason for choosing these two tests was to incorporate a range
of different interface types (gas-liquid, fluid-solid, and solid-vacuum).
The results are displayed in \prettyref{tab:ConvergenceRates}. As
can be seen, in all tests, the convergence rate is roughly first-order.

\begin{table}
\begin{centering}
\bigskip{}
\begin{tabular}{|c|c|c|c|c|c|c|}
\hline 
 &
\multicolumn{2}{c|}{Water-Air Test} &
\multicolumn{2}{c|}{PBX-Copper Test} &
\multicolumn{2}{c|}{Aluminium-Vacuum Test}\tabularnewline
\hline 
\hline 
\# Cells &
Error &
Rate &
Error &
Rate &
Error &
Rate\tabularnewline
\hline 
50 &
17.67419797 &
 &
20.27348796 &
 &
0.04232714 &
\tabularnewline
\hline 
100 &
7.98596866 &
1.146 &
10.14165497 &
0.999 &
0.02180830 &
0.957\tabularnewline
\hline 
150 &
5.20404021 &
1.056 &
6.57517319 &
1.069 &
0.01365700 &
1.154\tabularnewline
\hline 
200 &
4.19411427 &
0.750 &
4.99100631 &
0.958 &
0.01005346 &
1.065\tabularnewline
\hline 
250 &
3.32431075 &
1.042 &
3.73732754 &
1.296 &
0.00793378 &
1.061\tabularnewline
\hline 
300 &
2.77425968 &
0.992 &
3.10851219 &
1.010 &
0.00571356 &
1.800\tabularnewline
\hline 
\end{tabular}
\par\end{centering}
\caption{\label{tab:ConvergenceRates}Convergence Rates for the Water-Air Test,
PBX-Copper Test, and Aluminium-Vacuum Test}
\medskip{}
\end{table}

\section{Discussion\label{sec:Discussion}}

The Riemann Ghost Fluid Method presented in this study has been demonstrated
to be an effective way of accurately simulating the interfaces between
several different materials (in all three main phases of matter, plus
vacuum), described by the GPR model. Unlike in many existing implementations,
heat conduction across the interface was simulated accurately, leading
to the representation of physical phenomena that are often overlooked,
and to the redundancy of numerical techniques that are sometimes used
to enforce more empirically-accurate results (such as entropy and
temperature fixes).

The framework presented here greatly simplifies the conceptual framework
required for multimaterial interactions. Implementation should be
easier and quicker, and future work can be more focused on a single
model, rather than several fundamentally different frameworks. 

\subsection{Limitations}

Throughout this study, the various fluids have been assumed to be
immiscible. Whilst this is a common assumption in situations where
mixing is low or practically non-existent, there are many problems
which may require it. An area of further research would be the implementation
of a mixture model such as that proposed by Romenski et al. \citep{Romenski2007,Romenski2010},
which uses the same thermal conduction system as the GPR model.

The truncation of the Taylor series expansions \eqref{eq:q* Taylor Series}
and \eqref{eq:sigma* Taylor Series} used to find the star states
of the heat flux and the viscous stress tensor implicitly assumes
that there are only small differences between the side states and
the star states of the variables upon which $q_{1}^{*},\boldsymbol{\sigma_{1}^{*}}$
depend ($\rho$, $p$, $J_{1}$, and $\boldsymbol{A_{1}}$). If this
is the case, higher order terms of the expansion can be neglected.
If it is not, however, the method may fail. The linearised nature
of the GPR-RGFM solver also implicitly assumes that all waves of interest
present in the Riemann Problem are shocks. Thus, strong rarefactions
may cause the method to fail. 

\subsection{Potential Improvements\label{subsec:Potential-Improvements}}

As noted in \prettyref{sec:HeatConductionInAGas}, the GPR-RGFM method
does not necessarily ensure the continuity of the normal component
of heat flux across interfaces that feature discontinuous temperature
jumps. This is despite the method accurately modeling both the heat
conduction across the interface over time, and the corresponding evolution
to thermal equilibrium between the two materials. The reason for this
is that the star states produced by the linearised solver presented
in \prettyref{sec:SolvingTheRiemannProblem} represent the state of
the system at a time slightly beyond the point in time at which they
are applied to neighboring ghost cells. With more simple systems of
PDEs - such as the Euler equations - this often doesn't matter, as
the star states are constant in time, or their time evolution is easily
calculated. Owing to the source terms in the GPR equations, however,
the star states evolve in a manner for which an analytical solution
is not available. Thus, when the star states are applied to the ghost
cells, they contain higher heat fluxes and slightly different temperatures
to the actual values at the interface at that moment in time, leading
to the slight aberration apparent in the heat flux in \prettyref{fig:HeatFluxInAGas}.

A possible solution to this is as follows: Take materials $L$ and
$R$ either side of an interface. Given the states straddling the
interface, $Q_{L},Q_{R}$, derive $Q_{L}^{*}$ using the procedure
outlined in \prettyref{sec:SolvingTheRiemannProblem}. Then, by inverting
this procedure, derive a state $\tilde{Q_{R}}$ such that if $Q_{L},\tilde{Q_{R}}$
are both states for material $L$ (rather than for materials $L$
and $R$, respectively), following the procedure in \prettyref{sec:SolvingTheRiemannProblem}
obtains the same star state, $Q_{L}^{*}$. The derivation of such
an inverse mapping should be feasible, possibly with the addition
of some physical constraints. In this way, $\tilde{Q_{R}}$ represents
a state for the righthand cell at the current time, which - if cell
$R$ were occupied by material $L$ - would result in the same state
on the lefthand side of the interface at the end of the current timestep
as if the righthand cell were occupied by $Q_{R}$ and material $R$.

Another clear improvement to the GPR-RGFM method presented would be
to use a better Riemann solver than the iterative, linearised solver
devised in \prettyref{sec:SolvingTheRiemannProblem}. Let $L$ be
the matrix of left eigenvectors of the primitive system. As noted
previously, the solver relies upon the fact that each of the following
relations holds along the characteristic to which it corresponds:

\begin{equation}
L\cdot\frac{d\boldsymbol{P}}{dt}=L\cdot\boldsymbol{S}\label{eq:along-characteristics}
\end{equation}

Methods to accurately integrate \eqref{eq:along-characteristics}
from the left and right interface boundary states to their respective
star states warrant further research.

Alternatively, a completely different approximate Riemann solver could
be employed, such as the universal HLLEM solver of Dumbser et al.
\citep{Dumbser2016}. This path-conservative formulation of the HLLEM
solver works for general non-conservative systems (such as the GPR
model) and is simple to implement. It's based upon a new path-conservative
HLL method (building on the original method of Harten, Lax, and van
Leer \citep{Harten1983}) but is claimed to be able to represent linearly
degenerate intermediate waves ``with a minimum of smearing'' by
evaluating the eigenvalues and eigenvectors of the intermediate characteristic
fields (given in \prettyref{sec:Eigenstructure of the HPR Model}).

There are iterative exact Riemann solvers for the equations of non-linear
elasticity (to which the GPR model reduces as $\tau_{1}\rightarrow\infty$).
Thus, they will work for applications of the GPR model to solids problems
(and perhaps to very viscous fluids problems too). Although these
solvers are computationally expensive, they are only used once at
each material interface point at each time step, and thus the added
accuracy that they provide may be desirable. There are two ways to
formulate the equations of non-linear elasticity: one in which the
deformation tensor (the analogue of the inverse of the GPR model's
distortion tensor) is evolved in time, and one in which its inverse
(the analogue of $A$) is evolved instead. Miller's exact solver \citep{Miller2004}
uses the first formulation and the solver of Barton et al. \citep{Barton2009}
uses the second. The former can be used to evolve $A^{-1}$, from
which $A$ can be calculated. Unfortunately, both solvers critically
assume that the source terms of the system vanish, and so are unlikely
to produce the correct boundary conditions for the GPR-RGFM when modeling
relatively inviscid fluids. It should also be noted that they cannot
be used for problems involving heat conduction across material interfaces,
and they do not take the thermal conduction subsystem of the GPR model
into account.

\newpage{}

\section{References}

\bibliographystyle{siam}
\addcontentsline{toc}{section}{\refname}\bibliography{54_Users_hari_Git_public_phd_papers_working_A_u___multimaterial_continuum_mechanics_misc_refs,55_Users_hari_Git_public_phd_papers_working_A_u____for_multimaterial_continuum_mechanics_refs}

\begin{thebibliography}{10}

\bibitem{Alcrudo2001}
{\sc F.~Alcrudo and F.~Benkhaldoun}, {\em {Exact solutions to the Riemann
  problem of the shallow water equations with a bottom step}}, Computers {\&}
  Fluids, 30 (2001), pp.~643--671.

\bibitem{Barton2010}
{\sc P.~T. Barton and D.~Drikakis}, {\em {An Eulerian method for
  multi-component problems in non-linear elasticity with sliding interfaces}},
  Journal of Computational Physics, 229 (2010), pp.~5518--5540.

\bibitem{barton_eulerian_2010}
{\sc P.~T. Barton and D.~Drikakis}, {\em An {Eulerian} method for
  multi-component problems in non-linear elasticity with sliding interfaces},
  Journal of Computational Physics, 229 (2010), pp.~5518--5540.

\bibitem{barton_exact_2009}
{\sc P.~T. Barton, D.~Drikakis, E.~Romenski, and V.~a. Titarev}, {\em Exact and
  approximate solutions of {Riemann} problems in non-linear elasticity},
  Journal of Computational Physics, 228 (2009), pp.~7046--7068.

\bibitem{Barton2009}
\leavevmode\vrule height 2pt depth -1.6pt width 23pt, {\em {Exact and
  approximate solutions of Riemann problems in non-linear elasticity}}, Journal
  of Computational Physics, 228 (2009), pp.~7046--7068.

\bibitem{barton_eulerian_2011}
{\sc P.~T. Barton, D.~Drikakis, and E.~I. Romenski}, {\em An {Eulerian}
  finite-volume scheme for large elastoplastic deformations in solids},
  International Journal for Numerical Methods in Engineering, 81 (2011),
  pp.~453--484.

\bibitem{barton_conservative_2011}
{\sc P.~T. Barton, B.~Obadia, and D.~Drikakis}, {\em A conservative level-set
  based method for compressible solid/fluid problems on fixed grids}, Journal
  of Computational Physics, 230 (2011), pp.~7867--7890.

\bibitem{boettger_tabular_2012}
{\sc J.~Boettger, K.~G. Honnell, J.~H. Peterson, C.~Greeff, and S.~Crockett},
  {\em Tabular equation of state for gold}, AIP Conference Proceedings, 1426
  (2012), pp.~812--815.

\bibitem{boscheri_cell_2016}
{\sc W.~Boscheri, M.~Dumbser, and R.~Loubere}, {\em Cell centered direct
  {Arbitrary}-{Lagrangian}-{Eulerian} {ADER}-{WENO} finite volume schemes for
  nonlinear hyperelasticity}, Computers and Fluids, 134-135 (2016),
  pp.~111--129.

\bibitem{brauer_cartesian_2017}
{\sc A.~d. Brauer, A.~Iollo, and T.~Milcent}, {\em A {Cartesian} {Scheme} for
  {Compressible} {Multimaterial} {Hyperelastic} {Models} with {Plasticity}},
  Communications in Computational Physics, 22 (2017), pp.~1362--1384.

\bibitem{buyukcizmeci_tabulated_2014}
{\sc N.~Buyukcizmeci, A.~S. Botvina, and I.~N. Mishustin}, {\em Tabulated
  {Equation} of {State} for {Supernova} {Matter} {Including} {Full} {Nuclear}
  {Ensemble}}, The Astrophysical Journal, 789 (2014), p.~33.

\bibitem{chinnayya_modelling_2004}
{\sc A.~Chinnayya, E.~Daniel, and R.~Saurel}, {\em Modelling detonation waves
  in heterogeneous energetic materials}, Journal of Computational Physics, 196
  (2004), pp.~490--538.

\bibitem{de_brauer_cartesian_2016}
{\sc A.~de~Brauer, A.~Iollo, and T.~Milcent}, {\em A {Cartesian} scheme for
  compressible multimaterial models in 3d}, Journal of Computational Physics,
  313 (2016), pp.~121--143.

\bibitem{donea_arbitrary_1982}
{\sc J.~Donea, S.~Giuliani, and J.~P. Halleux}, {\em An arbitrary
  lagrangian-eulerian finite element method for transient dynamic
  fluid-structure interactions}, Computer Methods in Applied Mechanics and
  Engineering, 33 (1982), pp.~689--723.

\bibitem{Dumbser2016}
{\sc M.~Dumbser and D.~S. Balsara}, {\em {A new efficient formulation of the
  HLLEM Riemann solver for general conservative and non-conservative hyperbolic
  systems}}, Journal of Computational Physics, 304 (2016), pp.~275--319.

\bibitem{dumbser_high_2015}
{\sc M.~Dumbser, I.~Peshkov, E.~Romenski, and O.~Zanotti}, {\em High order
  {ADER} schemes for a unified first order hyperbolic formulation of continuum
  mechanics: viscous heat-conducting fluids and elastic solids}, Journal of
  Computational Physics, 314 (2015), pp.~824--862.

\bibitem{dumbser_universal_2011}
{\sc M.~Dumbser and E.~F. Toro}, {\em On universal {Osher}-type schemes for
  general nonlinear hyperbolic conservation laws}, Communications in
  Computational Physics, 10 (2011), pp.~635--671.

\bibitem{dumbser_simple_2011}
{\sc M.~Dumbser and E.~F. Toro}, {\em A simple extension of the {Osher}
  {Riemann} solver to non-conservative hyperbolic systems}, Journal of
  Scientific Computing, 48 (2011), pp.~70--88.

\bibitem{dumbser_ader-weno_2013}
{\sc M.~Dumbser, O.~Zanotti, A.~Hidalgo, and D.~S. Balsara}, {\em {ADER}-{WENO}
  finite volume schemes with space-time adaptive mesh refinement}, Journal of
  Computational Physics, 248 (2013), pp.~257--286.

\bibitem{favrie_diffuse_2012}
{\sc N.~Favrie and S.~L. Gavrilyuk}, {\em Diffuse interface model for
  compressible fluid - {Compressible} elastic-plastic solid interaction},
  Journal of Computational Physics, 231 (2012), pp.~2695--2723.

\bibitem{favrie_solid-fluid_2009}
{\sc N.~Favrie, S.~L. Gavrilyuk, and R.~Saurel}, {\em Solid-fluid diffuse
  interface model in cases of extreme deformations}, Journal of Computational
  Physics, 228 (2009), pp.~6037--6077.

\bibitem{Fedkiw1999}
{\sc R.~Fedkiw, T.~Aslam, B.~Merriman, and S.~Osher}, {\em {A Non-oscillatory
  Eulerian Approach to Interfaces in Multimaterial Flows (the Ghost Fluid
  Method)}}, Journal of Computational Physics, 152 (1999), pp.~457--492.

\bibitem{fedkiw_non-oscillatory_1999}
\leavevmode\vrule height 2pt depth -1.6pt width 23pt, {\em A {Non}-oscillatory
  {Eulerian} {Approach} to {Interfaces} in {Multimaterial} {Flows} (the {Ghost}
  {Fluid} {Method})}, Journal of Computational Physics, 152 (1999),
  pp.~457--492.

\bibitem{Fedkiw2002}
{\sc R.~P. Fedkiw}, {\em {Coupling an Eulerian Fluid Calculation to a
  Lagrangian Solid Calculation with the Ghost Fluid Method}}, Journal of
  Computational Physics, 175 (2002), pp.~200--224.

\bibitem{frenkel_kinetic_1947}
{\sc J.~Frenkel}, {\em Kinetic {Theory} of {Liquids}}, Oxford University Press,
  1947.

\bibitem{Glimm1981}
{\sc J.~Glimm and D.~Marchesin}, {\em {A Numerical Method for Two Phase Flow
  with an Unstable Interface}}, Journal of Computational Physics, 39 (1981),
  pp.~179--200.

\bibitem{Harten1983}
{\sc A.~Harten, P.~D. Lax, and B.~van Leer}, {\em {On Upstream Differencing and
  Godunov-Type Schemes for Hyperbolic Conservation Laws}}, SIAM Review, 25
  (1983), pp.~35--61.

\bibitem{hempert_simulation_2017}
{\sc F.~Hempert, S.~Boblest, T.~Ertl, F.~Sadlo, P.~Offenhauser, C.~W. Glass,
  M.~Hoffmann, A.~Beck, C.~D. Munz, and U.~Iben}, {\em Simulation of real gas
  effects in supersonic methane jets using a tabulated equation of state with a
  discontinuous {Galerkin} spectral element method}, Computers \& Fluids, 145
  (2017), pp.~167--179.

\bibitem{Hirt1981}
{\sc C.~W. Hirt and B.~D. Nichols}, {\em {Volume of fluid (VOF) method for the
  dynamics of free boundaries}}, Journal of Computational Physics, 39 (1981),
  pp.~201--225.

\bibitem{Hou2012}
{\sc G.~Hou, J.~Wang, and A.~Layton}, {\em {Numerical methods for
  fluid-structure interaction - A review}}, Communications in Computational
  Physics, 12 (2012), pp.~337--377.

\bibitem{howell_free_2002}
{\sc B.~Howell and G.~Ball}, {\em A {Free} {Lagrange} {Augmented} {Godunov}
  {Method} for the {Simulation} of {Elastic} {Plastic} {Solids}}, Journal of
  Computational Physics, 175 (2002), pp.~128--167.

\bibitem{jackson_fast_2017}
{\sc H.~Jackson}, {\em A {Fast} {Numerical} {Scheme} for the
  {Godunov}-{Peshkov}-{Romenski} {Model} of {Continuum} {Mechanics}}, Journal
  of Computational Physics, 348 (2017), pp.~514--533.

\bibitem{jackson_numerical_2019}
{\sc H.~Jackson and N.~Nikiforakis}, {\em A numerical scheme for
  non-{Newtonian} fluids and plastic solids under the {GPR} model}, Journal of
  Computational Physics, 387 (2019), pp.~410--429.

\bibitem{lee_phenomenological_1980}
{\sc E.~L. Lee and C.~M. Tarver}, {\em Phenomenological model of shock
  initiation in heterogeneous explosives}, Physics of Fluids, 23 (1980),
  p.~2362.

\bibitem{Legay2006}
{\sc A.~Legay, J.~Chessa, and T.~Belytschko}, {\em {An Eulerian-Lagrangian
  method for fluid-structure interaction based on level sets}}, Computer
  Methods in Applied Mechanics {\&} Engineering, 195 (2006), pp.~2070--2087.

\bibitem{levashov_tabular_2007}
{\sc P.~R. Levashov and K.~V. Khishchenko}, {\em Tabular multiphase equations
  of state for metals and their applications}, AIP Conference Proceedings, 955
  (2007), pp.~59--62.

\bibitem{Liu2003}
{\sc T.~G. Liu, B.~C. Khoo, and K.~S. Yeo}, {\em {Ghost fluid method for strong
  shock impacting on material interface}}, Journal of Computational Physics,
  190 (2003), pp.~651--681.

\bibitem{Liu1975}
{\sc T.-P. Liu}, {\em {The Riemann problem for general systems of conservation
  laws}}, Journal of Differential Equations, 18 (1975), pp.~218--234.

\bibitem{maire_nominally_2013}
{\sc P.~H. Maire, R.~Abgrall, J.~Breil, R.~Loubere, and B.~Rebourcet}, {\em A
  nominally second-order cell-centered {Lagrangian} scheme for simulating
  elastic-plastic flows on two-dimensional unstructured grids}, Journal of
  Computational Physics, 235 (2013), pp.~626--665.

\bibitem{malyshev_hyperbolic_1986}
{\sc A.~N. Malyshev and E.~I. Romenskii}, {\em Hyperbolic equations for heat
  transfer. {Global} solvability of the {Cauchy} problem}, Siberian
  Mathematical Journal, 27 (1986), pp.~734--740.

\bibitem{Michael}
{\sc L.~Michael and N.~Nikiforakis}, {\em {Coupling of elastoplastic solid
  models with condensed-phase explosives formulations}}, (submitted),  (2016).

\bibitem{michael_multi-physics_2018}
{\sc L.~Michael and N.~Nikiforakis}, {\em A multi-physics methodology for the
  simulation of reactive flow and elastoplastic structural response}, Journal
  of Computational Physics, 367 (2018), pp.~1--27.

\bibitem{Miller2004}
{\sc G.~H. Miller}, {\em {An iterative Riemann solver for systems of hyperbolic
  conservation laws, with application to hyperelastic solid mechanics}},
  Journal of Computational Physics, 193 (2004), pp.~198--225.

\bibitem{ndanou_multi-solid_2015}
{\sc S.~Ndanou, N.~Favrie, and S.~Gavrilyuk}, {\em Multi-solid and multi-fluid
  diffuse interface model: {Applications} to dynamic fracture and
  fragmentation}, Journal of Computational Physics, 295 (2015), pp.~523--555.

\bibitem{nichols_sola-vof:_1980}
{\sc B.~D. Nichols, C.~W. Hirt, and R.~S. Hotchkiss}, {\em {SOLA}-{VOF}: {A}
  solution algorithm for transient fluid flow with multiple free boundaries},
  NASA STI/Recon Technical Report N, 81 (1980).

\bibitem{Osher2002}
{\sc S.~Osher and R.~Fedkiw}, {\em {Level Set Methods and Dynamic Implicit
  Surfaces}}, Springer, 2002.

\bibitem{osher_level_2001}
{\sc S.~Osher and R.~P. Fedkiw}, {\em Level {Set} {Methods}: {An} {Overview}
  and {Some} {Recent} {Results}}, Journal of Computational Physics, 169 (2001),
  pp.~463--502.

\bibitem{peshkov_theoretical_2019}
{\sc I.~Peshkov, W.~Boscheri, R.~Loubère, E.~Romenski, and M.~Dumbser}, {\em
  Theoretical and numerical comparison of hyperelastic and hypoelastic
  formulations for {Eulerian} non-linear elastoplasticity}, Journal of
  Computational Physics, 387 (2019), pp.~481--521.

\bibitem{peshkov_hyperbolic_2016}
{\sc I.~Peshkov and E.~Romenski}, {\em A hyperbolic model for viscous
  {Newtonian} flows}, Continuum Mechanics and Thermodynamics, 28 (2016),
  pp.~85--104.

\bibitem{peterson_global_2012}
{\sc J.~H. Peterson, K.~G. Honnell, C.~Greeff, J.~D. Johnson, J.~Boettger, and
  S.~Crockett}, {\em Global equation of state for copper}, Chicago, Illinois,
  2012, pp.~763--766.

\bibitem{Pin2007}
{\sc F.~D. Pin, S.~Idelsohn, E.~Onate, and R.~Aubry}, {\em {The ALE/Lagrangian
  Particle Finite Element Method: A new approach to computation of free-surface
  flows and fluid-object interactions}}, Computers {\&} Fluids, 36 (2007),
  pp.~27--38.

\bibitem{rider_reconstructing_1998}
{\sc W.~J. Rider and D.~B. Kothe}, {\em Reconstructing {Volume} {Tracking}},
  Journal of Computational Physics, 141 (1998), pp.~112--152.

\bibitem{romenski_conservative_2010}
{\sc E.~Romenski, D.~Drikakis, and E.~Toro}, {\em Conservative models and
  numerical methods for compressible two-phase flow}, Journal of Scientific
  Computing, 42 (2010), pp.~68--95.

\bibitem{Romenski2010}
\leavevmode\vrule height 2pt depth -1.6pt width 23pt, {\em {Conservative models
  and numerical methods for compressible two-phase flow}}, Journal of
  Scientific Computing, 42 (2010), pp.~68--95.

\bibitem{Romenski2007}
{\sc E.~Romenski, A.~D. Resnyansky, and E.~F. Toro}, {\em {Conservative
  Hyperbolic Formulation for Compressible Two-Phase Flow with Different Phase
  Pressures and Temperatures}}, Quarterly of Applied Mathematics, 65 (2007),
  pp.~259--279.

\bibitem{romenski_conservative_2007}
{\sc E.~Romenski, A.~D. Resnyansky, and E.~F. Toro}, {\em Conservative
  hyperbolic model for compressible two-phase flow with different phase
  pressures and temperatures}, Quarterly of applied mathematics, 65(2) (2007),
  pp.~259--279.

\bibitem{romenski_hyperbolic_1989}
{\sc E.~I. Romenski}, {\em Hyperbolic equations of {Maxwell}'s nonlinear model
  of elastoplastic heat-conducting media}, Siberian Mathematical Journal, 30
  (1989), pp.~606--625.

\bibitem{Ryzhakov2010}
{\sc P.~B. Ryzhakov, R.~Rossi, S.~R. Idelsohn, and E.~Onate}, {\em {A
  monolithic Lagrangian approach for fluid-structure interaction problems}},
  Computational Mechanics, 46 (2010), pp.~883--899.

\bibitem{Sambasivan2009}
{\sc S.~K. Sambasivan and H.~S. UdayKumar}, {\em {Ghost Fluid Method for Strong
  Shock Interactions Part 1: Fluid-Fluid Interfaces}}, AIAA Journal, 47 (2009),
  pp.~2907--2922.

\bibitem{Sambasivan2009a}
\leavevmode\vrule height 2pt depth -1.6pt width 23pt, {\em {Ghost Fluid Method
  for Strong Shock Interactions Part 2: Immersed Solid Boundaries}}, AIAA
  Journal, 47 (2009), pp.~2923--2937.

\bibitem{saurel_multiphase_1999}
{\sc R.~Saurel and R.~Abgrall}, {\em A {Multiphase} {Godunov} {Method} for
  {Compressible} {Multifluid} and {Multiphase} {Flows}}, Journal of
  Computational Physics, 150 (1999), pp.~425--467.

\bibitem{scardovelli_direct_1999}
{\sc R.~Scardovelli and S.~Zaleski}, {\em Direct numerical simulation of
  free-surface and interfacial flow}, Annual Review of Fluid Mechanics, 31
  (1999), pp.~567--603.

\bibitem{Schoch2013}
{\sc S.~Schoch, K.~Nordin-Bates, and N.~Nikiforakis}, {\em {An Eulerian
  algorithm for coupled simulations of elastoplastic-solids and condensed-phase
  explosives}}, Journal of Computational Physics, 252 (2013), pp.~163--194.

\bibitem{toro_reimann_2009}
{\sc E.~F. Toro}, {\em Reimann {Solvers} and {Numerical} {Methods} for fluid
  dynamics}, vol.~40, 2009.

\bibitem{Toro2009}
{\sc E.~F. Toro}, {\em {Riemann Solvers and Numerical Methods for Fluid
  Dynamics: A Practical Introduction}}, Springer, 2009.

\bibitem{wang_thermodynamically_2004}
{\sc S.~P. Wang, M.~H. Anderson, J.~G. Oakley, M.~L. Corradini, and
  R.~Bonazza}, {\em A thermodynamically consistent and fully conservative
  treatment of contact discontinuities for compressible multicomponent flows},
  Journal of Computational Physics, 195 (2004), pp.~528--559.

\end{thebibliography}

\section{Acknowledgments}

The authors acknowledge financial support from the EPSRC Centre for
Doctoral Training in Computational Methods for Materials Science under
grant EP/L015552/1.

\section{Appendix\label{sec:Appendix}}

Taking the ordering $\boldsymbol{P}$ of primitive variables in \eqref{eq:Pvec},
note that \eqref{eq:EnergyEquation}, \eqref{eq:MomentumEquation},
\eqref{eq:DistortionEquation}, \eqref{eq:ThermalEquation} can be
stated as:

\begin{subequations}

\begin{align}
\frac{D\rho}{Dt}+\rho\frac{\partial v_{k}}{\partial x_{k}} & =0\\
\frac{Dv_{i}}{Dt}+\frac{1}{\rho}\frac{\partial\Sigma_{ik}}{\partial x_{k}} & =0\\
\frac{DA_{ij}}{Dt}+A_{ik}\frac{\partial v_{k}}{\partial x_{j}} & =-\frac{\psi_{ij}}{\theta_{1}}\\
\frac{DJ_{i}}{Dt}+\frac{1}{\rho}\frac{\partial T\delta_{ik}}{\partial x_{k}} & =-\frac{H_{i}}{\theta_{2}}\\
\frac{DE}{Dt}+\frac{1}{\rho}\frac{\partial\left(\Sigma_{ik}v_{i}+TH_{k}\right)}{\partial x_{k}} & =0
\end{align}

\end{subequations}

where the total stress tensor $\Sigma=pI+\rho A^{T}\psi$. Note that:

\begin{align}
\frac{DE}{Dt} & =\frac{\partial E}{\partial p}\frac{Dp}{Dt}+\frac{\partial E}{\partial\rho}\frac{D\rho}{Dt}+v_{i}\frac{Dv_{i}}{Dt}+\frac{\partial E}{\partial A_{ij}}\frac{DA_{ij}}{Dt}+H_{i}\frac{DJ_{i}}{Dt}\\
 & =\frac{\partial E}{\partial p}\frac{Dp}{Dt}-\rho\frac{\partial E}{\partial\rho}\frac{\partial v_{k}}{\partial x_{k}}-\frac{1}{\rho}v_{i}\frac{\partial\Sigma_{ik}}{\partial x_{k}}-\frac{\partial E}{\partial A_{ij}}\left(A_{ik}\frac{\partial v_{k}}{\partial x_{j}}+\frac{\psi_{ij}}{\theta_{1}}\right)-H_{i}\left(\frac{1}{\rho}\frac{\partial T\delta_{ik}}{\partial x_{k}}+\frac{H_{i}}{\theta_{2}}\right)\nonumber 
\end{align}

Thus, the energy equation becomes:

\begin{equation}
\frac{\partial E}{\partial p}\frac{Dp}{Dt}-\rho\frac{\partial E}{\partial\rho}\frac{\partial v_{k}}{\partial x_{k}}-\frac{1}{\rho}v_{i}\frac{\partial\Sigma_{ik}}{\partial x_{k}}-\frac{\partial E}{\partial A_{ij}}A_{ik}\frac{\partial v_{k}}{\partial x_{j}}-\frac{H_{k}}{\rho}\frac{\partial T}{\partial x_{k}}+\frac{1}{\rho}\frac{\partial\left(\Sigma_{ik}v_{i}+TH_{k}\right)}{\partial x_{k}}=\frac{\partial E}{\partial A_{ij}}\frac{\psi_{ij}}{\theta_{1}}+\frac{H_{i}H_{i}}{\theta_{2}}
\end{equation}

Simplifying:

\begin{equation}
\frac{Dp}{Dt}+\frac{1}{\rho E_{p}}\left(\Sigma_{ik}-\rho A_{ji}\frac{\partial E}{\partial A_{jk}}-\rho^{2}\frac{\partial E}{\partial\rho}\delta_{ik}\right)\frac{\partial v_{i}}{\partial x_{k}}+\frac{T}{\rho E_{p}}\frac{\partial H_{k}}{\partial x_{k}}=\frac{\partial E}{\partial A_{ij}}\frac{\psi_{ij}}{\theta_{1}E_{p}}+\frac{H_{i}H_{i}}{\theta_{2}E_{p}}
\end{equation}

We have\footnote{
\begin{align}
\frac{p-\rho^{2}E_{\rho}}{\rho E_{p}} & =\frac{\rho^{2}\left.E_{\rho}\right|_{s}-\rho^{2}\left.E_{\rho}\right|_{p}}{\rho\left.E_{p}\right|_{\rho}}=\rho\frac{\left.E_{\rho}\right|_{s}-\left(\left.E_{\rho}\right|_{s}+\left.E_{s}\right|_{\rho}\left.s_{\rho}\right|_{p}\right)}{\left.E_{s}\right|_{\rho}\left.s_{p}\right|_{\rho}}\\
 & =\rho\frac{-\left.s_{\rho}\right|_{p}}{\left.s_{p}\right|_{\rho}}=\rho\left.\frac{\partial p}{\partial\rho}\right|_{s}\nonumber 
\end{align}
}\footnote{
\begin{equation}
\frac{c_{t}^{2}T}{\rho E_{p}}=\frac{c_{t}^{2}T}{\rho c_{v}T_{p}}=\frac{\rho c_{h}^{2}}{T_{p}}
\end{equation}
}\footnote{
\begin{equation}
\left.\frac{\partial E}{\partial A}\right|_{\rho,p}=\left(c_{s}^{2}-\frac{\rho}{\Gamma}\frac{\partial c_{s}^{2}}{\partial\rho}\right)\frac{\psi}{c_{s}^{2}}=\left(1-2\frac{\rho^{2}}{\rho\Gamma}\frac{\partial\log\left(c_{s}\right)}{\partial\rho}\right)\psi
\end{equation}

\begin{align}
\frac{\partial\sigma}{\partial\rho} & =\frac{\partial}{\partial\rho}\left(-\rho c_{s}^{2}A^{T}\frac{\psi}{c_{s}^{2}}\right)=-c_{s}^{2}A^{T}\frac{\psi}{c_{s}^{2}}-\rho\frac{\partial c_{s}^{2}}{\partial\rho}A^{T}\frac{\psi}{c_{s}^{2}}\\
 & =\frac{\sigma}{\rho}+2\frac{\partial\log\left(c_{s}\right)}{\partial\rho}\sigma\nonumber 
\end{align}
}:

\begin{subequations}

\begin{align}
\frac{p-\rho^{2}E_{\rho}}{\rho E_{p}} & =\rho c_{0}^{2}\\
\frac{c_{t}^{2}T}{\rho E_{p}} & =\frac{\rho c_{h}^{2}}{T_{p}}\\
\left.\frac{\partial E}{\partial A}\right|_{\rho,p} & =\left(1-2\rho^{2}E_{p}\frac{\partial\log\left(c_{s}\right)}{\partial\rho}\right)\psi\\
-\rho A^{T}\left.\frac{\partial E}{\partial A}\right|_{\rho,p} & =\sigma+\rho^{2}E_{p}\left(\frac{\sigma}{\rho}-\frac{\partial\sigma}{\partial\rho}\right)
\end{align}

\end{subequations}

The full system then becomes:

\begin{subequations}

\begin{align}
\frac{D\rho}{Dt}+\rho\frac{\partial v_{k}}{\partial x_{k}} & =0\\
\frac{Dp}{Dt}+\rho c_{0}^{2}\frac{\partial v_{i}}{\partial x_{i}}+\left(\sigma_{ik}-\rho\frac{\partial\sigma_{ik}}{\partial\rho}\right)\frac{\partial v_{i}}{\partial x_{k}}+\frac{\rho c_{h}^{2}}{T_{p}}\frac{\partial J_{k}}{\partial x_{k}} & =\left(1-2\rho^{2}E_{p}\frac{\partial\log\left(c_{s}\right)}{\partial\rho}\right)\frac{\left\Vert \psi\right\Vert _{F}^{2}}{\theta_{1}E_{p}}+\frac{\left\Vert H\right\Vert ^{2}}{\theta_{2}E_{p}}\\
\frac{DA_{ij}}{Dt}+A_{ik}\frac{\partial v_{k}}{\partial x_{j}} & =-\frac{\psi_{ij}}{\theta_{1}}\\
\frac{Dv_{i}}{Dt}-\frac{1}{\rho}\frac{\partial\sigma_{ik}}{\partial\rho}\frac{\partial\rho}{\partial x_{k}}+\frac{1}{\rho}\frac{\partial p}{\partial x_{i}}-\frac{1}{\rho}\frac{\partial\sigma_{ik}}{\partial A_{mn}}\frac{\partial A_{mn}}{\partial x_{k}} & =0\\
\frac{DJ_{i}}{Dt}+\frac{T_{\rho}}{\rho}\frac{\partial\rho}{\partial x_{i}}+\frac{T_{p}}{\rho}\frac{\partial p}{\partial x_{i}} & =-\frac{H_{i}}{\theta_{2}}
\end{align}

\end{subequations}

Thus, the GPR system can be written in the following form:

\begin{equation}
\frac{\partial\boldsymbol{P}}{\partial t}+\boldsymbol{M}\cdot\nabla\boldsymbol{P}=\boldsymbol{S_{p}}
\end{equation}

where the first component of $M$ is given on \prettyref{eq:M1} for
illustrative purposes.

{\scriptsize{}
\begin{equation}
M_{1}=\left(\begin{array}{ccccccccccccccccc}
v_{1} & 0 & 0 & 0 & 0 & 0 & 0 & 0 & 0 & 0 & 0 & \rho & 0 & 0 & 0 & 0 & 0\\
0 & v_{1} & 0 & 0 & 0 & 0 & 0 & 0 & 0 & 0 & 0 & \Psi_{11}+\rho c_{0}^{2} & \Psi_{21} & \Psi_{31} & \frac{\rho c_{h}^{2}}{T_{p}} & 0 & 0\\
0 & 0 & v_{1} & 0 & 0 & 0 & 0 & 0 & 0 & 0 & 0 & A_{11} & A_{12} & A_{13} & 0 & 0 & 0\\
0 & 0 & 0 & v_{1} & 0 & 0 & 0 & 0 & 0 & 0 & 0 & A_{21} & A_{22} & A_{23} & 0 & 0 & 0\\
0 & 0 & 0 & 0 & v_{1} & 0 & 0 & 0 & 0 & 0 & 0 & A_{31} & A_{32} & A_{33} & 0 & 0 & 0\\
0 & 0 & 0 & 0 & 0 & v_{1} & 0 & 0 & 0 & 0 & 0 & 0 & 0 & 0 & 0 & 0 & 0\\
0 & 0 & 0 & 0 & 0 & 0 & v_{1} & 0 & 0 & 0 & 0 & 0 & 0 & 0 & 0 & 0 & 0\\
0 & 0 & 0 & 0 & 0 & 0 & 0 & v_{1} & 0 & 0 & 0 & 0 & 0 & 0 & 0 & 0 & 0\\
0 & 0 & 0 & 0 & 0 & 0 & 0 & 0 & v_{1} & 0 & 0 & 0 & 0 & 0 & 0 & 0 & 0\\
0 & 0 & 0 & 0 & 0 & 0 & 0 & 0 & 0 & v_{1} & 0 & 0 & 0 & 0 & 0 & 0 & 0\\
0 & 0 & 0 & 0 & 0 & 0 & 0 & 0 & 0 & 0 & v_{1} & 0 & 0 & 0 & 0 & 0 & 0\\
\Phi_{11} & \frac{1}{\rho} & \Upsilon_{11}^{11} & \Upsilon_{21}^{11} & \Upsilon_{31}^{11} & \Upsilon_{12}^{11} & \Upsilon_{22}^{11} & \Upsilon_{32}^{11} & \Upsilon_{13}^{11} & \Upsilon_{23}^{11} & \Upsilon_{33}^{11} & v_{1} & 0 & 0 & 0 & 0 & 0\\
\Phi_{21} & 0 & \Upsilon_{11}^{21} & \Upsilon_{21}^{21} & \Upsilon_{31}^{21} & \Upsilon_{12}^{21} & \Upsilon_{22}^{21} & \Upsilon_{32}^{21} & \Upsilon_{13}^{21} & \Upsilon_{23}^{21} & \Upsilon_{33}^{21} & 0 & v_{1} & 0 & 0 & 0 & 0\\
\Phi_{31} & 0 & \Upsilon_{11}^{31} & \Upsilon_{21}^{31} & \Upsilon_{31}^{31} & \Upsilon_{12}^{31} & \Upsilon_{22}^{31} & \Upsilon_{32}^{31} & \Upsilon_{13}^{31} & \Upsilon_{23}^{31} & \Upsilon_{33}^{31} & 0 & 0 & v_{1} & 0 & 0 & 0\\
\frac{T_{\rho}}{\rho} & \frac{T_{p}}{\rho} & 0 & 0 & 0 & 0 & 0 & 0 & 0 & 0 & 0 & 0 & 0 & 0 & v_{1} & 0 & 0\\
0 & 0 & 0 & 0 & 0 & 0 & 0 & 0 & 0 & 0 & 0 & 0 & 0 & 0 & 0 & v_{1} & 0\\
0 & 0 & 0 & 0 & 0 & 0 & 0 & 0 & 0 & 0 & 0 & 0 & 0 & 0 & 0 & 0 & v_{1}
\end{array}\right)\label{eq:M1}
\end{equation}
}{\scriptsize\par}

where we have:

\begin{subequations}

\begin{align}
\Psi_{ij} & =\sigma_{ij}-\rho\frac{\partial\sigma_{ij}}{\partial\rho}\\
\Upsilon_{mn}^{ij} & =-\frac{1}{\rho}\frac{\partial\sigma_{ij}}{\partial A_{mn}}\\
\Phi_{ij} & =-\frac{1}{\rho}\frac{\partial\sigma_{ij}}{\partial\rho}
\end{align}

\end{subequations}

\begin{equation}
\boldsymbol{P}=\left(\begin{array}{ccccccccccccccccc}
\rho & p & A_{11} & A_{21} & A_{31} & A_{12} & A_{22} & A_{32} & A_{13} & A_{23} & A_{33} & v_{1} & v_{2} & v_{3} & J_{1} & J_{2} & J_{3}\end{array}\right)^{T}\label{eq:Pvec}
\end{equation}

\begin{doublespace}
\begin{equation}
\boldsymbol{S_{p}}=\frac{1}{\theta_{1}}\left(\begin{array}{c}
0\\
\left(\frac{1}{E_{p}}-2\rho^{2}\frac{\partial\log\left(c_{s}\right)}{\partial\rho}\right)\left\Vert \psi\right\Vert _{F}^{2}\\
-\psi_{11}\\
-\psi_{21}\\
-\psi_{31}\\
-\psi_{12}\\
-\psi_{22}\\
-\psi_{32}\\
-\psi_{13}\\
-\psi_{23}\\
-\psi_{33}\\
0\\
0\\
0\\
0\\
0\\
0
\end{array}\right)+\frac{1}{\theta_{2}}\left(\begin{array}{c}
0\\
\frac{1}{E_{p}}\left\Vert \boldsymbol{H}\right\Vert ^{2}\\
0\\
0\\
0\\
0\\
0\\
0\\
0\\
0\\
0\\
0\\
0\\
0\\
-H_{1}\\
-H_{2}\\
-H_{3}
\end{array}\right)
\end{equation}
\end{doublespace}

\end{document}